\documentclass{aa}  

\usepackage{natbib,twoopt}
\usepackage[breaklinks=true, colorlinks=true, allcolors=blue]{hyperref} 
\usepackage{mathrsfs}
\usepackage{longfigure}
\makeatletter
\newcommandtwoopt{\citeads}[3][][]{\href{http://adsabs.harvard.edu/abs/#3}%
{\def\hyper@linkstart##1##2{}%
\let\hyper@linkend\@empty\citealp[#1][#2]{#3}}}
\newcommandtwoopt{\citepads}[3][][]{\href{http://adsabs.harvard.edu/abs/#3}%
{\def\hyper@linkstart##1##2{}%
\let\hyper@linkend\@empty\citep[#1][#2]{#3}}}
\newcommandtwoopt{\citetads}[3][][]{\href{http://adsabs.harvard.edu/abs/#3}%
{\def\hyper@linkstart##1##2{}%
\let\hyper@linkend\@empty\citet[#1][#2]{#3}}}
\newcommandtwoopt{\citeyearads}[3][][]%
{\href{http://adsabs.harvard.edu/abs/#3}
{\def\hyper@linkstart##1##2{}%
\let\hyper@linkend\@empty\citeyear[#1][#2]{#3}}}
\makeatother

\usepackage{graphicx}
\usepackage{txfonts}
\begin{document}

   \title{Investigating the interplay between the coronal properties and the hard X-ray variability of active galactic nuclei with {\it NuSTAR}}
   \titlerunning{{\it NuSTAR} variability}
   \authorrunning{R. Serafinelli et al.}


   \author{Roberto Serafinelli
          \inst{1}\fnmsep\thanks{\email{roberto.serafinelli@inaf.it}}
          \and
          Alessandra De Rosa\inst{2} 
          \and 
          Alessia Tortosa\inst{1}
          \and 
          Luigi Stella\inst{1}
          \and
          Fausto Vagnetti\inst{3,2}
          \and 
          Stefano Bianchi\inst{4}
          \and
          \\Claudio Ricci\inst{5,6}
          \and
          Elias Kammoun\inst{4,7}
          \and
          Pierre-Olivier Petrucci\inst{8}
          \and
          Riccardo Middei\inst{9,1}
          \and
          Giorgio Lanzuisi\inst{10}
          \and
          \\Andrea Marinucci\inst{11}
          \and
          Francesco Ursini\inst{4}
          \and
          Giorgio Matt\inst{4}
          }

   \institute{INAF - Osservatorio Astronomico di Roma, Via Frascati 33, 00078, Monte Porzio Catone (Roma), Italy
   \and
   INAF - Istituto di Astrofisica e Planetologia Spaziali, Via del Fosso del Cavaliere 100, 00133, Roma, Italy
   \and
   Dipartimento di Fisica, Universit\`a degli Studi di Roma “Tor Vergata”, via della Ricerca Scientifica 1, 00133 Roma, Italy
   \and
   Dipartimento di Matematica e Fisica, Universit\`a degli Studi Roma Tre, Via della Vasca Navale 84, 00146 Roma, Italy
   \and
   Instituto de Estudios Astrof\'isicos, Facultad de Ingenier\'ia y Ciencias, Universidad Diego Portales, Avenida Ejercito Libertador 441, Santiago, Chile
   \and
   Kavli Institute for Astronomy and Astrophysics, Peking University, Beijing 100871, People’s Republic of China
   \and
   INAF - Osservatorio Astrofisico di Arcetri, Largo Enrico Fermi 5, 50125 Firenze, Italy
   \and
   CNRS, IPAG, Universit\'e Grenoble Alpes, 38000 Grenoble, France
   \and
    Space Science Data Center, Agenzia Spaziale Italiana, Via del Politecnico snc, 00133 Roma, Italy
   \and
   INAF - Osservatorio di Astrofisica e Scienza dello Spazio di Bologna, Via Gobetti, 93/3, 40129 Bologna, Italy
   \and
   ASI - Agenzia Spaziale Italiana, Via del Politecnico snc, 00133 Roma, Italy}

   \date{Received XXX; accepted YYY}

 
  \abstract
   {Active galactic nuclei (AGN) are extremely variable in the X-ray band down to very short timescales. However, the driver behind the X-ray variability is still poorly understood. Previous results suggest that the hot corona responsible for the primary Comptonized emission observed in AGN is expected to play an important role in driving the X-ray variability. In this work, we investigate the connection between the X-ray amplitude variability and the coronal physical parameters; namely, the temperature ($kT$) and optical depth ($\tau$). We present the spectral and timing analysis of 46 {\it NuSTAR} observations corresponding to a sample of 20 AGN. For each source, we derived the coronal temperature and optical depth through X-ray spectroscopy and computed the normalized excess variance for different energy bands on a timescale of $10$ ks. We find a strong inverse correlation between $kT$ and $\tau$, with correlation coefficient of $r<-0.9$ and negligible null probability. No clear dependence was found among the temperature and physical properties, such as the black hole mass or the Eddington ratio. We also see that the observed X-ray variability is not correlated with either the coronal temperature or optical depth under the thermal equilibrium assumption, whereas it is anticorrelated with the black hole mass. These results can be interpreted through a scenario where the observed X-ray variability could primarily be driven by variations in the coronal physical properties on a timescale of less than $10$~ks; whereas we assume thermal equilibrium on such timescales in this work, given the capability of the currently available hard X-ray telescopes. Alternatively, it is also possible that the X-ray variability is mostly driven by the absolute size of the corona, which depends on the supermassive black hole mass, rather than resulting from any of its physical properties.}


   \maketitle
%

\section{Introduction}
\label{sec:intro}

Active galactic nuclei (AGN) are bright extragalactic sources powered by the accretion of matter onto a supermassive black hole (SMBH). In brief, AGN emit light at all wavelengths and they are characterized by a significantly loud X-ray emission \citep[e.g.,][]{padovani17r}. The X-ray emission of AGN is produced by inverse Compton scattering of UV seed photons, emitted by the accretion disk, off a hot relativistic electron plasma known as the corona \citep[e.g.,][]{haardt91,haardt93}. The typical shape of the X-ray spectrum of an AGN is that of a power law, characterized by a photon index $\Gamma$, up to a characteristic energy, $E_c$, known as the cut-off energy where the power law breaks. The relation between the cut-off energy and the temperature can be approximated with $E_c\sim2-3\;kT$ \citep[e.g.,][]{petrucci01}, depending on the geometry of the corona and the optical depth, while the photon index is dependent on both the coronal temperature and the optical depth. However, more complex relations with both the temperature and optical depth are needed when considering broader ranges of temperatures and optical depths \citep{middei19}.\\
\indent Many models have been proposed for the coronal geometry, including slab \citep[e.g.,][]{haardt91}, spherical \citep[e.g.,][]{frontera03} or lamp-post coronal geometry \citep[e.g.,][]{miniutti04}. Details of its shape, location and size are yet largely unknown, though, since spectroscopy alone is not able to distinguish among different geometries, which can be probed with polarimetry measurements \citep[e.g.,][]{ursini22}. Indeed, recent results with Imaging X-ray Polarimetry Explorer \citep[IXPE,][]{weisskopf16} are starting to unveil the geometrical properties of the AGN corona. \cite{gianolli23} measured the coronal X-ray polarization  for the first time in the Seyfert galaxy NGC 4151, strongly suggesting a wedge or a slab above the accretion disk \citep[e.g.,][]{poutanen18}. For IC4329A \citep{ingram23}, a marginal detection for the X-ray polarimetry also suggests a wedge coronal geometry. Only upper limits were found for MCG-5-23-16 \citep{marinucci22,tagliacozzo23}, although the results, combined with the inclination measurement obtained with {\it XMM-Newton} and {\it NuSTAR} \citep{serafinelli23b}, tentatively favors a wedge geometry as well.\\
\indent Several measurements of the cut-off energy have been undertaken using many hard X-ray instruments, like {\it Beppo}SAX \citep[e.g.,][]{dadina07}, INTEGRAL \citep[e.g.,][]{molina09,derosa12}, {\it Swift}-BAT \citep[e.g.,][]{ricci18} and {\it NuSTAR} \citep[e.g.,][]{fabian15,fabian17,tortosa18}, including obscured sources \citep[e.g.,][]{balokovic20,serafinelli23e}. This task is far from trivial, since the hard X-ray spectrum is also characterized by a reflection component, due to X-ray photons interacting with the surrounding environment, such as the accretion disk, the broad line region or the torus, whose parameters are often degenerate with those of the continuum \citep[see e.g. the review in][]{Reynolds2021}. The cut-off energy is found in a large energy range, from $E_c\sim23$ keV \citep{kammoun23} to $E_c\sim750$ keV \citep{matt15}, with average values around $\sim100-200$ keV \citep[e.g.,][]{ricci18,kamraj22}. Direct measurements of the coronal parameters such as the electron temperature $kT$ and the optical depth $\tau$ have also been extensively performed on many AGN, finding a tight correlation between the temperature and the optical depth \citep[e.g.,][]{tortosa18,kamraj22}.\\
\indent The X-ray emission of AGN is well-known to be variable on several timescales, both in amplitude \citep[e.g.,][]{markowitz04,ponti12,vagnetti16,middei17,serafinelli20} and spectral shape \citep[e.g.,][]{sobolewska09,serafinelli17}. Variability is found on very short timescales down to a few hours \citep[e.g.,][]{ponti12}, and this suggests that the X-ray emitting region is compact \citep[e.g.,][]{mushotzky93,demarco13}, with a typical radius of $R_c\sim10R_g$ \citep{ursini20l}, also supported by microlensing results \citep[e.g.,][]{chartas09,morgan12}. Moreover, the X-ray emission is variable in a wide range of energies, including very hard X-rays ($E>10$ keV) on both long \citep[e.g., years,][]{soldi14,akylas22,papadakis24} and shorter timescales \citep[e.g., hours,][]{rani19,akylas22}.\\
\indent The X-ray variability of AGN provides crucial insight on the size of the central source, but its main driver is still poorly understood. We aim here to investigate the possible relation between the variability of the X-ray emission coming from the corona and the physical properties of the corona itself with {\it NuSTAR}, which is able to study both coronal parameters and variability because of its high sensitivity at hard X-rays ($E=3-79$ keV). We present a study of a sample of 20 nearby ($z<0.2$) Seyfert galaxies for which we study the coronal parameters using {\it NuSTAR}, in order to investigate possible relations with the X-ray variability. In Sect.~\ref{sec:data}, we describe our sample, made up of sources with a wide range of coronal temperatures and optical depths, and the data reduction of the available X-ray data. In Sect.~\ref{sec:spectra} we describe the spectral analysis we performed. In Sect.~\ref{sec:var} we investigate the X-ray variability through the computation of the excess variance in different bands with {\it NuSTAR}. Finally our results are discussed in Sect.~\ref{sec:discussion} and present a summary in Sect. \ref{sec:end}. Throughout the paper, we adopt a standard $\Lambda$CDM cosmology, with $H_0=70$ km s$^-1$ Mpc$^{-1}$, $\Omega_m=0.3,$ and $\Omega_\Lambda=0.7.$

\section{Sample selection and data reduction}
\label{sec:data}

\begin{table*}[h!]
\centering
\caption{Sources analyzed in this paper.}
\label{tab:sample}
\begin{tabular}{lcccccccc}
\hline
Source & $z$ & Type & $M_{\rm BH}$ ($M_\odot$) & $\log L_{\rm bol}/{\rm (erg \; s}^{-1})$ & $\log\lambda_{\rm Edd}$ & \\
\hline
1H 0419-577 & $0.104$ & Sy1 & $2.2\times10^8$ & $45.68$ & $-0.84$\\ 
4C 50.55 & $0.02$ & RLSy1 & $9.3\times10^{7}$ & $44.91$ & $-1.24$\\
Ark 564$\dag$ & $0.0243$ & NLSy1 & $3.2\times10^6$ & $45.00$ & $0.39$\\ 
ESO 103-G35 & $0.00914$ & Sy1 & $2.3\times10^7$ & $44.54$ & $-1.01$\\
ESO 362-G18 & $0.01244$ & Sy1.5 & $1.3\times10^{7}$ & $44.07$ & $-1.22$ \\
ESO 383-G18 & $0.01241$ & Sy2 & $3.0\times10^5$ & $43.79$ & $0.15$\\
GRS 1734-292 & $0.021$ & Sy1 & $6.9\times10^7$ & $44.95$ & $-1.07$\\
HE 1143-1810 & $0.0329$ & Sy1 & $2.5\times10^7$ & $44.95$ & $-0.62$\\
IC4329A & $0.01613$ & Sy1 & $4.5\times10^7$ & $45.06$ & $-0.77$\\
MCG-5-23-16 & $0.00823$ & Sy1 & $4.5\times10^7$ & $44.34$ & $-1.49$\\
MCG+8-11-11 & $0.02$ & Sy1 & $4.1\times10^6$ & $44.93$ & $0.15$\\
Mrk 6 & $0.01951$ & Sy1.5 & $1.3\times10^8$ & $44.52$ & $-1.76$\\
Mrk 110 & $0.03552$ & Sy1 & $1.9\times10^7$ & $45.06$ & $-0.41$\\
Mrk 509 & $0.01951$ & Sy1 & $1.1\times10^8$ & $45.27$ & $-0.96$\\
NGC 3281 & $0.01073$ & Sy2 & $1.7\times10^8$ & $44.06$ & $-2.04$\\
NGC 5506 & $0.00589$ & NLSy1 & $1.7\times10^7$ & $44.14$ & $-1.24$\\
NGC 5728 & $0.00947$ & Sy2 & $1.8\times10^8$ & $44.14$ & $-2.28$\\
NGC 6814 & $0.00523$ & Sy1 & $1.1\times10^7$ & $43.52$ & $-1.7$\\
SWIFT J2127.4+5654 & $0.0144$ & NLSy1 & $1.4\times10^7$ & $44.11$ & $-1.22$\\
UGC 6728 & $0.00652$ & Sy1 & $3.5\times10^{5}$ & $43.16$ & $-0.57$\\
\hline
\hline
\end{tabular}
\tablefoot{The most common name, the redshift, the AGN type, the black hole mass ($M_{\rm BH}$), the bolometric luminosity ($L_{\rm bol}$), and the Eddington ratio ($\lambda_{\rm Edd}$) are listed in this table. All values of $M_{\rm BH}$, $L_{\rm bol}$ and $\lambda_{\rm Edd}$ are extracted from the BASS catalog published by \cite{koss22}. $\dag$ The only exception is Ark 564, for which we retrieved a reverberation mapping mass estimate from \cite{peterson04}, with $L_{\rm bol}$ estimated applying a bolometric correction to the X-ray luminosity.}
\end{table*}

We selected our sample of AGN from the $70$-Month {\it Swift}-BAT catalogue \citep{baumgartner13}. \citet{ricci17} computed many X-ray properties of the AGN in the catalog, such as the X-ray flux in several bands, the photon index, and the cut-off energy. We select all sources where the cut-off has been measured, namely, excluding the ones with only lower limits. Out of $836$ AGN of the whole {\it Swift}-BAT sample, $165$ satisfy this first condition. Among those AGN, we selected the ones with public {\it NuSTAR} observations as of 10th October 2021, for a total of 229 observations of 110 AGN. Not all {\it NuSTAR} observations have sufficient statistics to compute the coronal parameters; therefore, we selected only those with enough {\it NuSTAR} counts. To this end, we considered the value of the X-ray flux in the $20-50$ keV, as reported by \citet{ricci17}, and we converted such a flux to {\it NuSTAR} count rate in the same band using WebPimms\footnote{\url{https://heasarc.gsfc.nasa.gov/cgi-bin/Tools/w3pimms/w3pimms.pl}}, adopting an unabsorbed power law with $\Gamma=1.8$, which is typical for Seyfert galaxies \citep[e.g.,][]{serafinelli17}. This count rate was multiplied by the sum of the exposures of each observation of every source and we selected all sources with at least 1500 counts per FPM module. We note that this selection might exclude very variable sources in which the {\it NuSTAR} count rate may exceed the expected count rate from the BAT flux, not easily spotted with this criterion. A total of $34$ sources are selected with these criteria. Six of these sources (IC4329A, MCG-5-23-16, MCG+8-11-11, NGC 5506, NGC 6814, and SWIFT J2127.4+5654) were already present in the sample analyzed by \cite{tortosa18}. Out of completeness, we decided to include two more sources from \cite{tortosa18} that are not included in our selection, Ark 564 (not detected in BAT) and GRS 1734-292, to consider all non-jetted nearby ($z\leq0.2$) AGN from their sample. This resulted in a total of 36 AGN.\\ 
\indent Even though our sample might possibly be biased towards the low tail of the high energy cut-off distributions, this selection  provides the best spectral signal-to-noise ratio (S/N) needed to derive coronal physical properties with high confidence level. Furthermore, excluding the sources with only lower limits for the cut-off energy also excludes AGN with additional spectral complexities, such as multiple ionized absorption gas layers, which could introduce systematic errors in the measure of $kT$ and $\tau$ due to an inaccurate continuum fit. For these sources, a simultaneous spectroscopy with a low energy ($E<3$ keV) bandpass is recommended.\\
\indent We reduced the {\it NuSTAR} observations with {\tiny NUPIPELINE}, available through {\tiny HEASOFT} v6.30, which is part of the {\tiny NUSTARDAS} software package, calibration files ({\tiny CALDB}), updated as recently as March 31, 2022. We extracted the FPMA and FPMB source spectra and light curves from a circular region with $60"$ radius,  centered on the source, using the {\tiny NUPRODUCTS} command. The background spectra are extracted using two source-free regions of $40"$ each.

\begin{table*}
\centering
\caption{Coronal parameters of our sample derived with the models described in Sect.~\ref{sec:spectra}.\\
}
\label{tab:coropar}
\begin{tabular}{lcccccccccc}
\hline
Source & $kT_{\rm sl}$ (keV) & $\tau_{\rm sl}$ & $kT_{\rm sph}$ (keV) & $\tau_{\rm sph}$ & $L_{2-10\;{\rm keV}}$ & $F_{2-10\;{\rm keV}}$ & Model\\
\hline
1H0419-577 & $14^{+2}_{-1}$ & $2.5^{+0.2}_{-0.3}$ & $14^{+2}_{-1}$ & $5.7^{+0.4}_{-0.5}$ & $35\pm5$ & $1.3\pm0.2$ & A\\
4C50.55 & $18^{+5}_{-2}$ & $2.2^{+0.2}_{-0.3}$ & $18^{+3}_{-2}$ & $5.2^{+0.4}_{-0.5}$ & $10.0\pm0.1$ & $8.0\pm0.1$ & A\\
Ark 564 & $15\pm2$ & $1.4\pm0.1$ & $16\pm1$ & $3.3\pm0.3$ & $3.5\pm0.1$ & $6.4\pm0.1$ & A\\
ESO 103-G35 & $17^{+18}_{-3}$ & $2.2^{+0.3}_{-0.7}$ & $17^{+7}_{-3}$ & $5.2\pm0.7$ & $0.8\pm0.2$ & $1.9\pm0.5$ & C\\
ESO 362-G18 & $18^{+14}_{-4}$ & $2.5^{+0.4}_{-0.9}$ & $18^{+15}_{-4}$ & $5.8^{+0.8}_{-1.8}$ & $0.04\pm0.01$ & $1.1\pm0.1$ & B\\
ESO 383-G18 & $7.3^{+0.5}_{-0.4}$ & $4.7\pm0.3$ & $7.3^{+0.5}_{-0.4}$ & $10.1\pm0.6$ & $0.03\pm0.01$ & $0.5\pm0.1$ & C\\
GRS 1734-292 & $13^{+2}_{-1}$ & $2.9\pm0.2$ & $13_{-1}^{+2}$ & $6.5\pm0.4$ & $0.6\pm0.1$ & $2.9\pm0.1$ & A\\
HE 1143-1810 & $36^{+64}_{-19}$ & $1.2^{+1.1}_{-0.8}$ & $22^{+61}_{-5}$ & $4.4^{+0.9}_{-2.9}$ & $6.0\pm1.9$ & $2.7\pm0.9$ & A\\
IC 4329A & $44^{+20}_{-10}$ & $1.1\pm0.3$ & $44^{+17}_{-11}$ & $2.8\pm0.7$ & $5.6\pm0.4$ & $12.0\pm0.7$ & A\\
MCG-5-23-16 & $41\pm11$ & $0.9\pm0.2$ & $35^{+10}_{-11}$ & $2.7\pm0.3$ &  $1.80\pm0.02$ & $10.4\pm0.2$ & B\\
MCG+8-11-11 & $110^{+140}_{-40}$ & $0.3\pm0.2$ & $90^{+125}_{-45}$ & $1.2^{+0.7}_{-0.4}$ & $5.1\pm0.6$ & $5.6\pm0.7$ & B\\
Mrk 6 & $13^{+4}_{-2}$ & $3.4^{+0.4}_{-0.6}$ & $13^{+4}_{-2}$ & $7.5^{+0.7}_{-1.1}$ & $0.99\pm0.02$ & $0.70\pm0.01$ & C\\
Mrk 110 & $35^{+15}_{-10}$ & $1.2^{+0.5}_{-0.4}$ & $24^{+17}_{-5}$ & $4.0^{+0.6}_{-1.4}$ & $9\pm3$ & $3.1\pm0.8$ & B\\
Mrk 509 & $17^{+2}_{-1}$ & $2.2\pm0.1$ & $17^{+2}_{-1}$ & $5.2^{+0.2}_{-0.3}$ & $12.0\pm0.3$ & $4.5\pm0.1$ & A\\
NGC 3281 & $11^{+4}_{-2}$ & $3.7^{+0.8}_{-0.9}$ & $11^{+4}_{-2}$ & $8\pm2$ & $0.23\pm0.04$ & $0.4\pm0.1$ & C\\
NGC 5506 & $510^{+250}_{-150}$ & $0.02\pm0.01$ & $550\pm250$ & $0.09_{-0.05}^{+0.30}$ & $0.53\pm0.05$ & $6.2\pm0.5$ & A\\
NGC 5728 & $13\pm1$ & $5^{+2}_{-1}$ & $13\pm1$ & $10^{+6}_{-3}$ & $1.4\pm0.6$ & $0.16\pm0.6$ & C\\
NGC 6814 & $60^{+24}_{-20}$ & $0.8^{+0.7}_{-0.3}$ & $82^{+80}_{-10}$ & $1.5^{+0.8}_{-0.1}$ & $0.21\pm0.03$ & $3.8\pm0.5$ & B\\
SWIFTJ2127.4+5654 & $33^{+37}_{-15}$ & $1.0_{-0.6}^{+0.8}$ & $24_{-7}^{+24}$ & $3.5_{-1.6}^{+1.0}$ & $1.4\pm0.4$ & $2.9\pm0.8$ & A\\
UGC 6728 & $28^{+16}_{-18}$ & $2.0\pm1.2$ & $28^{+17}_{-15}$ & $5.0\pm2.0$ & $0.075\pm0.007$ & $1.1\pm0.1$ & B\\
\hline
\hline
\end{tabular}
\tablefoot{Intrinsic luminosities are listed in units of $10^{43}$ erg s$^{-1}$. Observed fluxes are in units of $10^{-11}$ erg s$^{-1}$ cm$^{-2}$. All values are obtained in this work with the spectral fits briefly described in Appendix~\ref{sec:specfits}. All spectra are shown in Fig.~\ref{fig:appB}.}
\end{table*}

\section{Spectral analysis}
\label{sec:spectra}

\begin{figure*}[h!]
\centering
\includegraphics[scale=0.7]{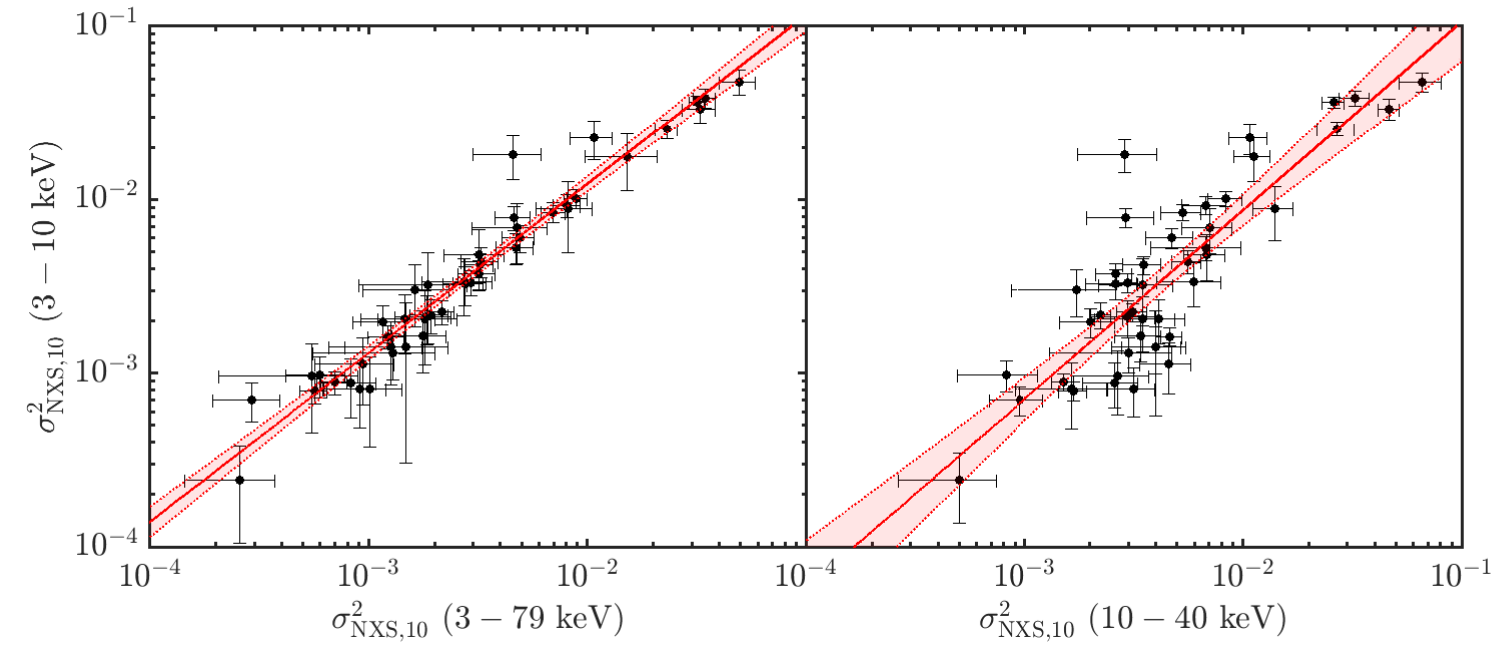} 
\caption{Excess variance in the soft ($3-10$ keV) versus the excess variance in the full ($3-79$ keV) {\it NuSTAR} bands (left). All excess variances are normalized at $10$ ks. The best-fit line is fully consistent with the bisector at $90\%$ confidence level, with an angular coefficient of $0.96\pm0.05$. The correlation coefficient is $r=0.95$ with a negligible or null probability. Excess variance in the soft band vs. excess variance in the hard ($10-40$ keV) X-ray band (right). The best-fit angular coefficient is $0.7\pm0.1$ with a correlation coefficient $r=0.90$ and a negligible probability of finding the correlation by chance.}
\label{fig:nxsfull}
\end{figure*}


We performed a quick analysis of 36 selected AGN to verify whether the coronal temperature is also constrained with {\it NuSTAR}. All fits included Galactic absorption extracted from \cite{hi4pi16} (see Appendix~\ref{sec:specfits} for all details). We considered three models. Model A consists of a continuum plus ionized reflection model, using {\tiny XILLVERCP}, which is the combination of the reflection model {\tiny XILLVER} \citep{garcia10,garcia11} and the continuum model {\tiny NTHCOMP} \citep{zdziarski96}. The {\tiny XSPEC} equation is: $${\tt ztbabs*xillvercp,}\;\;\; {\rm model\;A}.$$ For simplicity, the inclination and iron abundance are fixed to $\iota=30^\circ$, and $A_{\rm Fe}=1$, respectively. The reflection fraction $\mathcal{R}$ is left free to vary, as {\tiny XILLVERCP} models both continuum and reflection. Model B is adopted when the Fe K$\alpha$ region is not properly modeled by model A because of the presence of a broad emission line, interpreted as due to the presence of a relativistic reflection component \citep[e.g.,][]{Reynolds2021}. In such cases, we also included a second reflector, using {\tiny RELXILLCP}, which is the convolution of {\tiny RELLINE} \citep{dauser14} and {\tiny XILLVERCP} for the reflection component, and {\tiny NTHCOMP} for the continuum. In this scenario {\tiny XILLVERCP} models reflection from material located farther out from the black hole, while {\tiny RELLXILLCP} models reflection from the innermost parts of the accretion disk  \citep[e.g.,][]{serafinelli23b}. The {\tiny XSPEC} equation is: $${\tt ztbabs*(xillvercp+relxillcp),  }\;\;\; {\rm  model\;B}.$$ We  assumed a single emissivity index of $-3$ for the accretion disk and we assumed that {\tiny RELXILLCP} only models reflection ($\mathcal{R}=-1$). Unless otherwise specified (see Appendix~\ref{sec:specfits}), we fixed an inner radius of $R_{\rm in}=10~R_g$ and outer radius of $R_{\rm out}=400~R_g$. We assumed a Schwarzschild black hole ($a=0$), while the {\tiny XILLVERCP} parameters are the same as model A, including a free-to-vary reflection fraction. Finally, model C was adopted when the reflection component was due to a single neutral reflector, and the model with one or two ionized reflection components does not fit the data satisfactorily. In this case, the reflection component was modeled with {\tiny BORUS}, which models the reflection from spherical distribution of neutral material with polar cutouts\footnote{The BORUS grids are available at \url{https://sites.astro.caltech.edu/~mislavb/download}} \citep[see][for details]{balokovic18}. Thus, we added a {\tiny NTHCOMP} component for the continuum, since {\tiny BORUS} does not have an in-built continuum one and we tied all the reflection parameters in {\scriptsize BORUS} that describe the continuum to that of the {\scriptsize NTHCOMP}. The {\tiny XSPEC} equation is: $${\tt ztbabs*(nthcomp+borus12)}\;\;\; {\rm model\;C}.$$ Since the physical properties of the neutral reflector are not the main goal of this paper, we always assumed a Compton thick reflector ($\log N_{\rm H,refl}/{\rm cm^{-2}}=24.5$), with a covering factor $C_f=0.5$ and reflector assumed on the line of sight. All three models include a neutral absorption component modeled with {\tiny ZTBABS} whenever required (see Appendix~\ref{sec:specfits} for details).\\
\indent We did not find any relevant variability of the coronal parameters $kT$ or $\Gamma$, when considering different observation epochs of the same source. This is consistent with the recent results obtained by \cite{pal22}, where evidence of variations in the coronal parameters was found in less than $5\%$ of their AGN sample. Therefore, during the fit, we kept $kT$ and $\tau$ tied between different observations of the same AGN. We were able to constrain  the coronal parameters in 20 sources; however, we excluded the 16 sources for which we only obtained lower limits for either the coronal temperature or the optical depth. The final sample is shown in Table~\ref{tab:sample}, where  eblack hole masses ($M_{\rm BH}$), bolometric luminosities ($L_{\rm bol}$), and Eddington ratios ($\lambda_{\rm Edd}$) have been  taken from the catalog presented in \cite{koss22}; the only exception is Ark 564, which is absent from their list. \cite{koss22} selected masses preferably from reverberation mapping (when available)\ followed by single-epoch measurements from H$\alpha$ or H$\beta$ lines and, ultimately, from the host galaxy velocity dispersion. For Ark 564, we considered the reverberation mapping measurement in \cite{peterson04}, $M_{\rm BH}=3.2\times10^{6}$ $M_{\odot}$, with a bolometric luminosity computed by applying the bolometric correction proposed by \cite{marconi04} to the unabsorbed X-ray luminosity in the $2-10$ keV energy band, along with an Eddington ratio computed as $\lambda_{\rm Edd}=L_{\rm bol}/L_{\rm Edd}$, where $L_{\rm Edd}=1.26\times10^{38}M/M_\odot$ erg s$^{-1}$ is the Eddington luminosity. All the values are reported in Table~\ref{tab:sample}.\\
\indent As we are interested in exploring different coronal geometries and the interplay between temperature and optical depth, we separated the continuum and the reflection by including a Comptonization model as continuum, {\tiny COMPTT} \citep{magdziarz95}, which is capable of assuming slab-shaped and spherical-shaped coronae. In models A and B, we simply replaced {\tiny NTHCOMPT} with {\tiny COMPTT} by setting the reflection parameter to a fixed value of $-1$\footnote{We recall that {\tiny XILLVERCP} models both continuum and reflection, by fitting the reflection fraction $\mathcal{R}>0$ and assuming {\tiny NTHCOMP} as continuum. When the reflection fraction is assumed to be frozen to $\mathcal{R}=-1$, {\tiny XILLVERCP} will only model a pure reflection spectrum, which means that we then need a second component to model the continuum.}; whereas in model C, we explicitly removed the external {\tiny NTHCOMP} component and replaced it with {\tiny COMPTT}. We started with the baseline models used for the first round of fits. All the parameters of the reflection, with the exception of the normalization, are frozen to the results of the first round of fits, as we are mainly interested in the coronal parameters. We fix the parameter {\tt approx} in {\tiny COMPTT} to $0.5$ to model a slab geometry, and to $1.5$ to model a spherical coronal shape. We checked the consistency between the two continuum models by comparing the best-fit values of the temperature obtained before the addition of {\tiny COMPTT} and those obtained after, assuming a spherical geometry, the only one allowed by {\tiny NTHCOMP}. We find consistent $kT$ results at $90\%$ confidence level in all cases. Furthermore, we also simulated spectra in a large range of temperatures and exposures with {\tiny NTHCOMP}.We fit the simulated spectra with {\tiny COMPTT}, assuming a spherical geometry, such as the one considered in {\tiny NTHCOMP} model, also finding good agreement at at least $90\%$ confidence level.\\
\indent We report the values obtained for $kT$ and $\tau$ for both geometries in Table~\ref{tab:coropar}, while details on the fits are available in Appendix~\ref{sec:specfits}.

\section{Variability analysis}
\label{sec:var}

\begin{table*}
\centering
\caption{ {\it NuSTAR} excess variances in the energy bands $3-79$, $3-10,$ and $10-40$ keV.}
\label{tab:snxslist}
\begin{tabular}{lcccc}
\hline
Source & OBSID & $\sigma_{\rm NXS,10}^2$ ($3-79$ keV) & $\sigma_{\rm NXS,10}^2$ ($3-10$ keV) & $\sigma_{\rm NXS,10}^2$ ($10-40$ keV)\\
\hline
1H 0419-577 & 60101039002 & $1.2\pm0.2$ & $1.6\pm0.2$ & $4.6\pm0.6$\\
& 60402006002 & $2.7\pm0.7$ & $3.4\pm0.9$ & $6\pm2$\\
& 60402006004 & $1.8\pm0.6$ & $2.0\pm0.7$ & $4\pm1$\\
4C 50.55 & 60061305002 & $0.26\pm0.09$ & $0.2\pm0.1$ & $0.5\pm0.2$ \\
& 60301005002 & $0.29\pm0.08$ & $0.7\pm0.1$ & $1.0\pm0.3$\\
Ark 564 & 60101031002 & $49\pm7$ & $48\pm6$ & $66\pm15$\\
& 60401031002 & $23\pm2$ & $26\pm2$ & $27\pm5$\\
& 60401031004 & $33\pm4$ & $33\pm4$ & $47\pm5$\\
ESO 103-G35 & 60061288002 & $2.8\pm0.5$ & $3.3\pm0.6$ & $2.6\pm0.7$\\
& 60301004002 & $1.8\pm0.4$ & $1.6\pm0.5$ & $3.4\pm0.8$\\
ESO 362-G18 & 60201046002 & $5\pm1$ & $7\pm2$ & $7\pm2$\\
ESO 383-G18 & 60061243002 & $1.5\pm0.6$ & $1.4\pm0.9$ & $4\pm1$\\
& 60261002002 & $3.3\pm0.5$ & $4.4\pm0.7$ & $5.6\pm0.9$\\
GRS 1734-292 & 60061279002 & $0.6\pm0.1$ & $1.0\pm0.2$ & $0.8\pm0.3$\\
& 60301010002 & $1.1\pm0.2$ & $2.0\pm0.4$ & $2.0\pm0.6$\\
HE 1143-1810 & 60302002002 & $0.6\pm0.3$ & $0.4\pm0.3$ & $3\pm1$\\
& 60302002004 & $1.5\pm0.4$ & $2.1\pm0.6$ & $4\pm1$\\
& 60302002006 & $1.9\pm0.4$ & $2.1\pm0.5$ & $3\pm1$\\
& 60302002008 & $1.0\pm0.3$ & $0.8\pm0.3$ & $1.7\pm0.8$\\ 
& 60302002010 & $4.6\pm0.6$ & $8\pm1$ & $3\pm1$\\ 
IC 4329A & 60001045002 & $0.8\pm0.1$ & $0.9\pm0.3$ & $2.6\pm0.7$\\
MCG-5-23-16 & 10002019001 & $2.2\pm0.2$ & $2.3\pm0.3$ & $3.1\pm0.5$\\
& 60001046002 & $4.7\pm0.7$ & $5.3\pm0.8$ & $7\pm3$\\
& 60001046004 & $1.9\pm0.3$ & $2.2\pm0.4$ & $2.2\pm0.2$\\
& 60001046006 & $3.2\pm0.5$ & $3.7\pm0.5$ & $2.6\pm0.5$\\
& 60001046008 & $2.9\pm0.4$ & $3.3\pm0.4$ & $3.0\pm0.3$\\
MCG+8-11-11 & 60201027002 & $0.7\pm0.1$ & $0.9\pm0.1$ & $1.5\pm0.2$\\
Mrk6 & 60102044002 & $1.9\pm0.7$ & $3\pm1$ & $3\pm1$\\
& 60102044004 & $1.6\pm0.5$ & $3\pm1$ & $1.7\pm0.9$\\
Mrk 110 & 60201025002 & $1.3\pm0.6$ & $1.3\pm0.3$ & $3\pm2$\\
& 60502022002 & $0.9\pm0.3$ & $1.1\pm0.4$ & $5\pm1$\\
& 60502022004 & $1.3\pm0.4$ & $1.4\pm0.4$ & $4\pm1$\\
Mrk 509 & 60101043002 & $0.56\pm0.06$ & $0.8\pm0.1$ & $1.7\pm0.2$\\
& 60101043002 & $0.9\pm0.1$ & $0.8\pm0.3$ & $3.2\pm0.8$\\
NGC 3281 & 60061201002 & $8\pm2$ & $9\pm2$ & $14\pm3$\\
& 60662003002 & $3.2\pm0.8$ & $5\pm1$ & $7\pm1$\\
NGC 5506 & 60061323002 & $3.2\pm0.4$ & $4.2\pm0.5$ & $3.5\pm0.7$\\
NGC 5728 & 60061256002 & $5\pm1$ & $18\pm4$ & $3\pm1$\\
& 60662002002 & $11\pm2$ & $23\pm5$ & $11\pm2$\\
NGC 6814 & 60201028002 & $15\pm4$ & $18\pm5$ & $11\pm2$\\
SWIFT J2127.4+5654 & 60001110002 & $8.1\pm0.9$ & $9\pm1$ & $7\pm2$\\
& 60001110003 & $4.9\pm0.6$ & $6.0\pm0.8$ & $5\pm1$\\
& 60001110005 & $8.9\pm0.8$ & $10\pm1$ & $8\pm1$\\
& 60001110007 & $7.0\pm0.7$ & $8.4\pm0.9$ & $5\pm1$\\
UGC 6728 & 60160450002 & $34\pm3$ & $39\pm4$ & $33\pm5$\\
& 60376007002 & $32\pm2$ & $36\pm2$ & $26\pm3$\\
\hline
\hline
\end{tabular}
\tablefoot{All the excess variances are normalized to $10$ ks, as described in Sect.~\ref{sec:var}. All excess variances are in units of $10^{-3}$.}
\end{table*}


A straightforward estimator of the X-ray variability is the normalized excess variance \citep[e.g.,][]{vaughan03},  defined as: 
\begin{equation}
\sigma^2_{\rm NXS}=\frac{1}{N\mu^2}\sum_{i=1}^{N}\left[ \left( x_i-\mu \right )^2 - \sigma_i^2 \right ],
\label{eq:nxs}
\end{equation}
where $x_i$ are the values of the X-ray amplitude of every time bin, $\mu$ is the mean value of the amplitude, $N$ is the number of points, and $\sigma_i$ is the photometric error on the X-ray amplitude. The excess variance of a random process is the integral of the power spectral density (PSD) over all frequencies between 0 and infinity. However, with real data it is limited by $f_{\rm min}=1/t_{\rm max}$ and $f_{\rm max}=1/t_{\rm min}$, where $t_{\rm min}=2\Delta t$, with $\Delta t$ being the time bin of the light curves we use (in our case 1 ks), and $t_{\rm max}$ is the length of the observation segment (10 ks, see below). We note that the excess variance is not a good estimator when a large range of redshifts is considered, since same-length light curves in the observer frame represent different rest-frame lengths at different redshifts \citep{vagnetti16}; at the same time, we would be looking at a different energy range. However, our sample is limited at $z<0.2$, hence, we were able to avoid these biases among the excess variance.\\
\indent Since the excess variance is a quadratic sum over the number of points of a light curve, two conditions should be satisfied in order to properly compare these quantities over different sources. First of all, they must have the same binning, which is ensured in our case by how the light curves were prepared. Indeed, every {\it NuSTAR} light curve is binned at $1$ ks. The other condition is that the light curves should be equally long, which is not satisfied in our sample as the exposures are different in the observations of different sources. To avoid this bias, we normalize every excess variance to the same length. Given that the smallest {\it NuSTAR} exposure in our sample is $\Delta t\sim17$ ks, we decide to study the variability on a timescale lower than that, namely: $10$ ks. In order to normalize the excess variance, we divided the light curve in intervals of $10$ ks and discarded the leftover points up to a maximum of $9$ ks. We computed $\sigma^2_{\rm NXS}$ for each interval and we adopted the average value between all intervals as the excess variance of the whole light curve.\\
\indent Every excess variance value computed for each interval is associated with an error given by \citep{vaughan03}:

\begin{equation}
{\rm err}(\sigma^2_{\rm NXS}) = \sqrt{ \left( \sqrt{\frac{2}{N}} \frac{\overline{\sigma^2_{\rm err}}}{\mu^2}\right)^2 + \left( \sqrt{\frac{\overline{\sigma^2_{\rm err}}}{N}}\frac{2F_{\rm var}}{\mu}\right)^2 },
\label{eq:errnxs}
\end{equation}

where $\sigma^2_{\rm err}=\sum_i\sigma^2_i/N\mu^2$ and $F_{\rm var}=\sqrt{\sigma^2_{\rm NXS}}$ is the fractional variability.\\
\indent Following \cite{ponti12}, we considered the following three cases. When only one interval was present for the considered light curve ($\Delta t<20$ ks, one observation in our sample), the error associated with the excess variance is given by Eq.~\ref{eq:errnxs}, computed in the single interval available. If the number of intervals is between $2$ and $9$ (i.e. $20\leq\Delta t<100$ ks, 33 observations), we calculated Eq.~\ref{eq:errnxs} for each interval and took its average value as the global error on the whole light curve. Finally, when the number of intervals exceeded $10$ (i.e., $\Delta t\geq 100$ ks, 12 observations), the standard deviation of the excess variance is adopted as error. We caution that the distribution of $\sigma^2_{\rm NXS}$ is Gaussian for an ideal number of intervals $n\gtrsim20$ \citep[e.g.,][]{allevato13}. However, given that this condition is only satisfied for a total of four observations of two sources, we opted for the less conservative approach described above.\\
\indent We computed the excess variance in the full ($3-79$ keV), soft ($3-10$ keV) and hard ($10-40$ keV) {\it NuSTAR} bands, which are listed in Table~\ref{tab:snxslist}. As shown in Fig.~\ref{fig:nxsfull}, as expected, the broadband variability in the full band is dominated by the soft band variability, as most of the signal is detected in this energy band. Nearly all the points are located in the bisector line. Indeed, we fit the $\sigma^2_{\rm NXS}$ in the two bands in the logarithmic scale, we obtain a very tight correlation ($r=0.95$ with negligible or null probability) and we find a slope of $0.96\pm0.05$, indicating that the variability of the continuum in the $3-10$ keV band is dominant. We also find a tight correlation ($r=0.90$) between the excess variance between the  soft $3-10$ keV and the hard $10-40$ keV bands, which is however not consistent with the bisector at $90\%$ confidence level, as we find an angular coefficient of $0.7\pm0.1$ and an intercept of $-0.7\pm0.2$.

Since all three bands are so tightly correlated, in the following sections, we only used the excess variance derived from the light curves extracted in the full {\it NuSTAR} energy band ($E=3-79$ keV). Full band light curves of every observation for each AGN of the sample here presented are shown in Fig.~\ref{fig:appC}.

\section{Discussion}
\label{sec:discussion}

\subsection{Physics of the corona}
\label{sec:specdisc}

\begin{figure}
\centering
\includegraphics[scale=0.37]{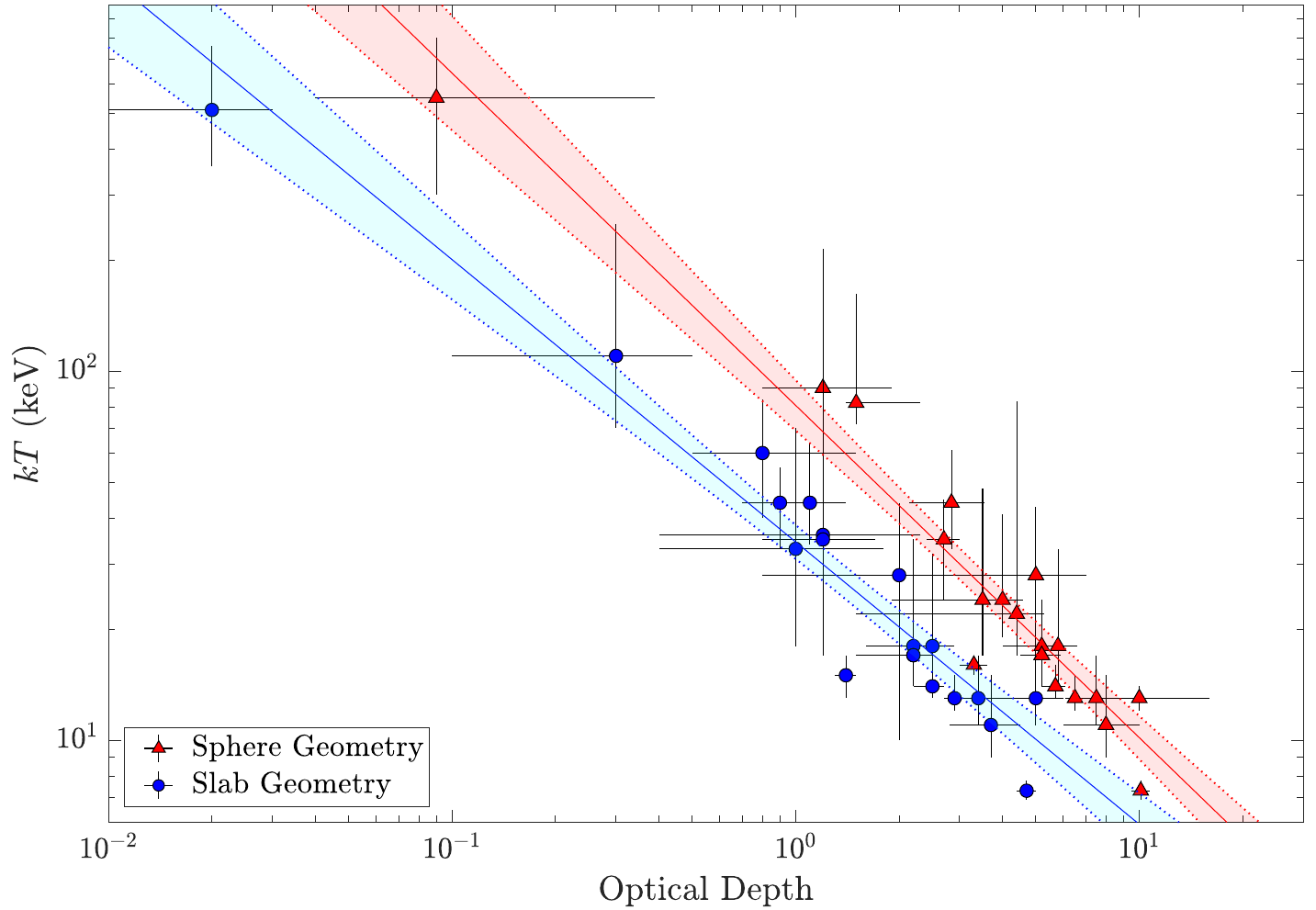}
\caption{Temperature versus optical depth plot. The $kT-\tau$ points obtained assuming a slab geometry are shown in blue, with the blue line indicating the best-fit line. The red points and line denote the points obtained with a sphere geometry and their best-fit line, respectively. Confidence intervals are shown at $90\%$ level.}
\label{fig:kTtau}
\end{figure}
\begin{figure*}
\centering
\includegraphics[scale=0.6]{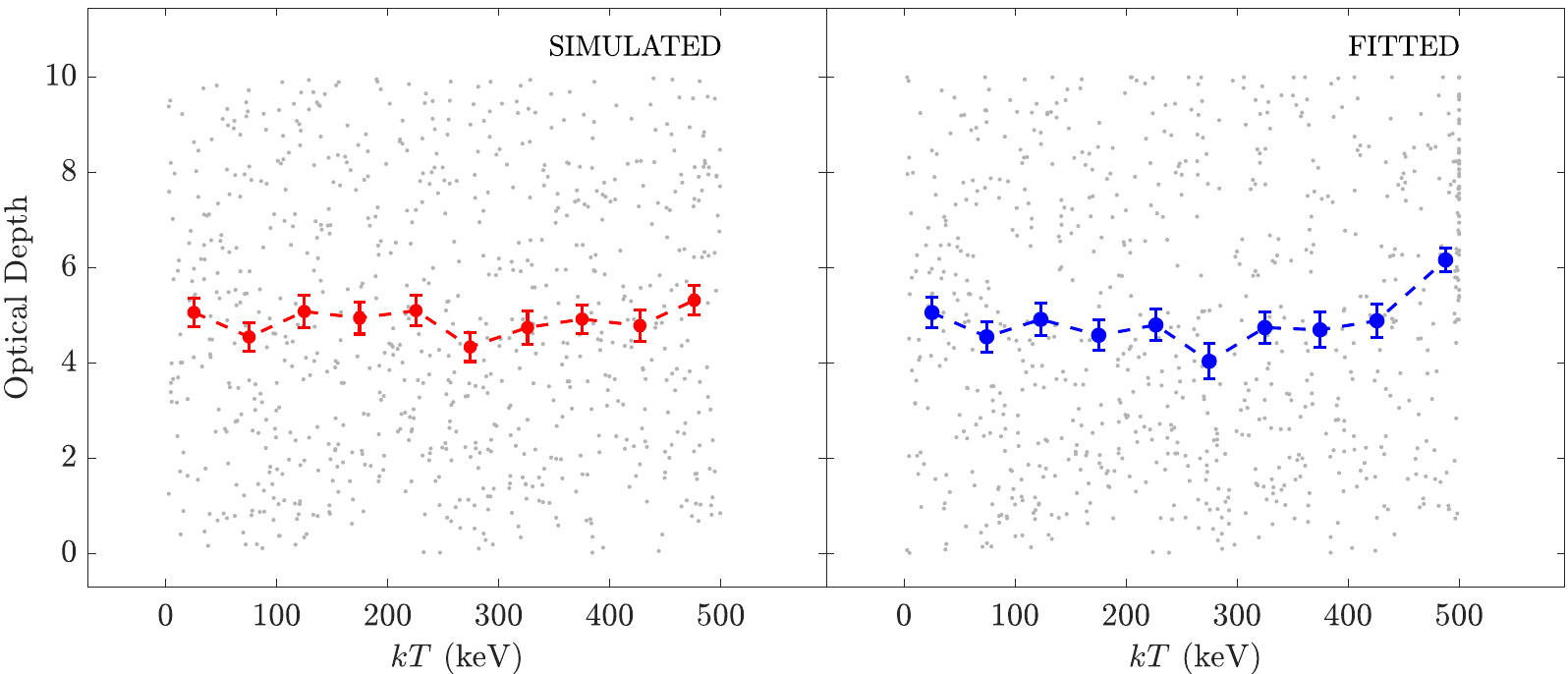}
\caption{Temperature and optical depth of the simulated spectra used to validate the correlation found in the {\it NuSTAR} data. A thousand {\it NuSTAR} spectra were simulated with random values of $kT$ and $\tau$ (see details in Sect.~\ref{sec:discussion}). The left panel show the values used to simulate the data with average values in bins of temperature (red), while the right panel shows the temperature and optical depth found when fitting the simulated spectra, with the average values in temperature bins.}
\label{fig:simul}
\end{figure*}
\begin{figure*}
\centering
\includegraphics[scale=0.7]{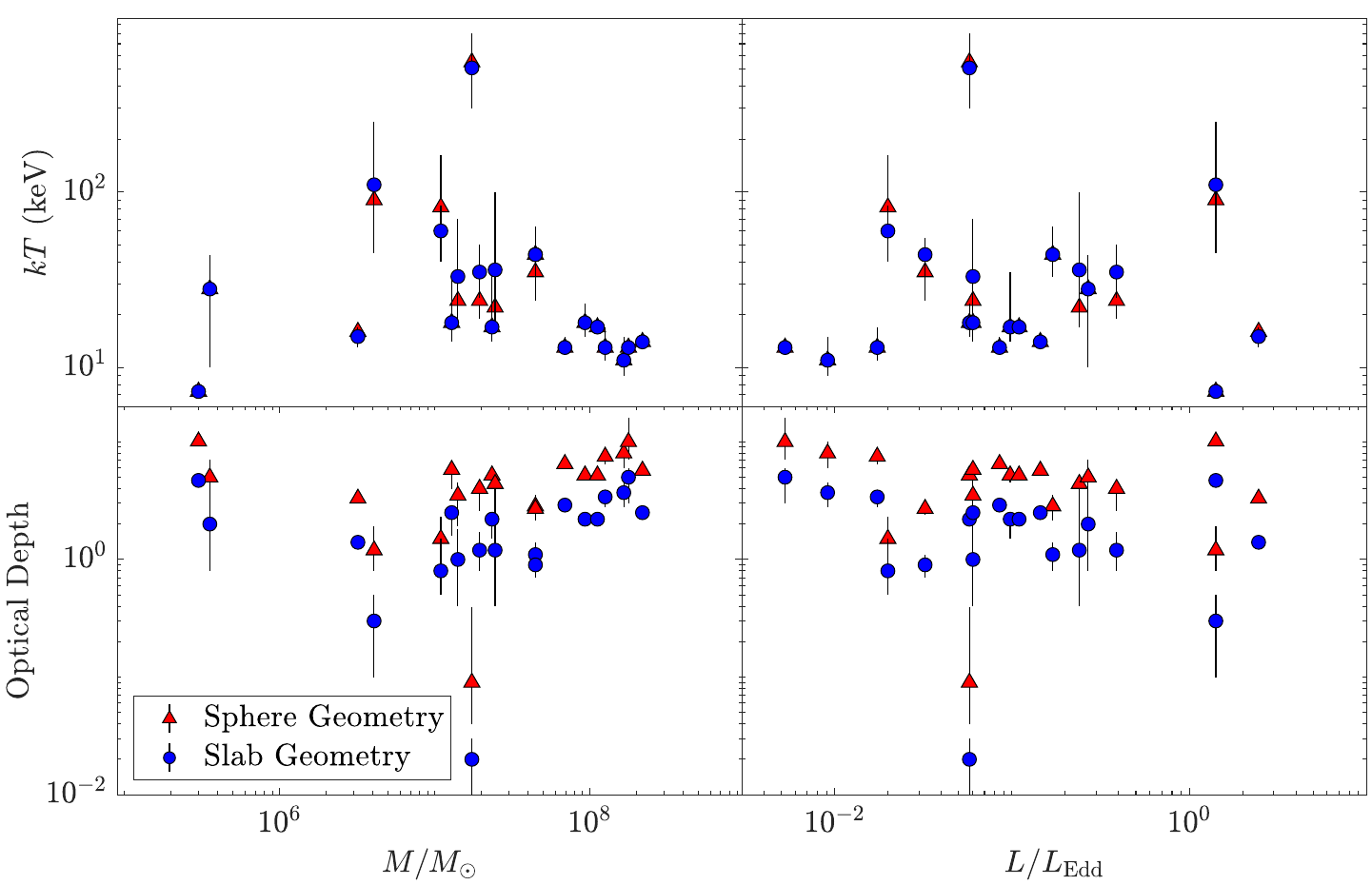}
\caption{We report here the coronal parameters (kT and $\tau$) $vs$ the black hole mass (left panels)  and the Eddington ratio (right panels) for the sources in our sample (see Table~\ref{tab:sample}) . The blue dots are the temperatures obtained assuming a slab corona, while the red dots are those obtained assuming a spherical one.}
\label{fig:kTpar}
\end{figure*}

\begin{figure*}
\centering
\includegraphics[scale=0.7]{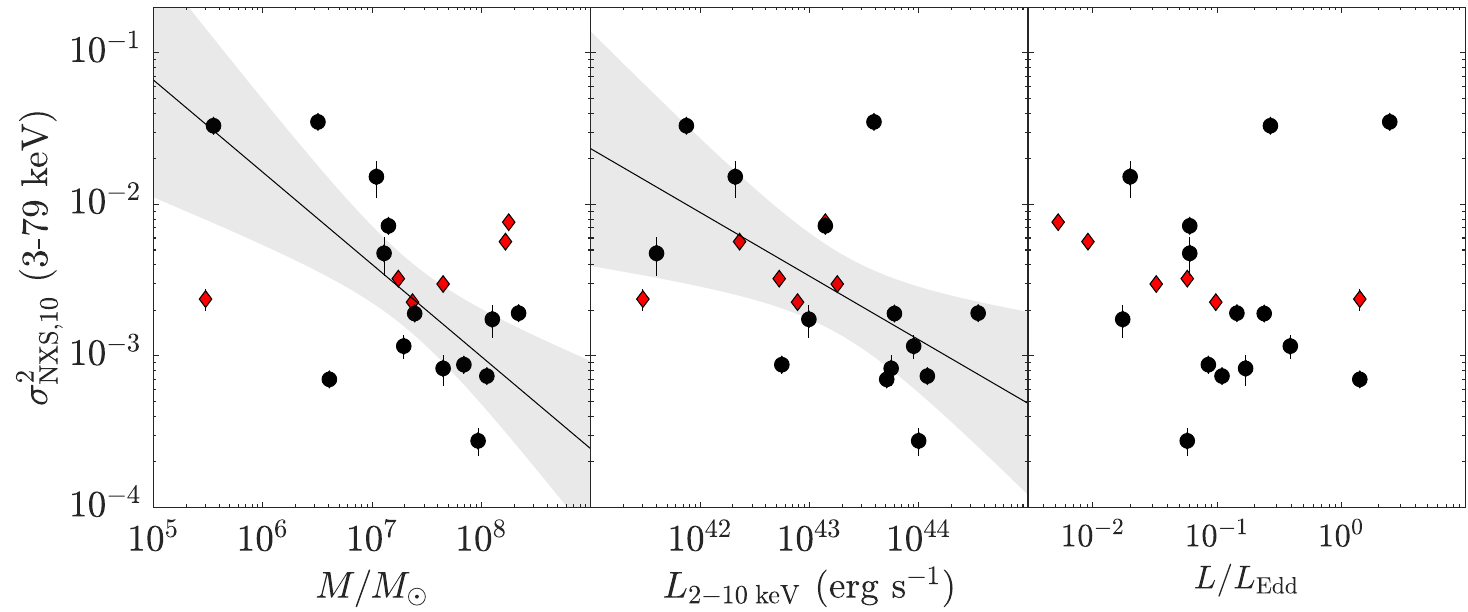}
\caption{ Excess variance in the $E=3-79$ keV energy band versus the black hole mass of the AGN (left). The excess variances shown here are averaged among the different epochs. A clear trend $\sigma^2_{\rm NXS}\propto M_{\rm BH}^{-0.6}$ is found with a strong anticorrelation ($r=-0.7$). Given their large uncertainties, masses obtained with the velocity dispersion method (red diamonds) are excluded from the fit. Middle panel gives The average excess variance vs. the unabsorbed X-ray luminosity $L_{\rm X}$ is shown. We find a trend $\sigma^2_{\rm NXS}\propto L_{\rm X}^{-0.4}$ with moderate anticorrelation ($r=-0.5$). Right panel shows the average excess variance vs. the Eddington ratio and no evident correlation is found. Also, in this panel, the sources with mass measurements obtained with the velocity dispersion method are drawn as red diamonds.}
\label{fig:nxsmass}
\end{figure*}
\begin{figure}
    \centering
    \includegraphics[scale=0.6]{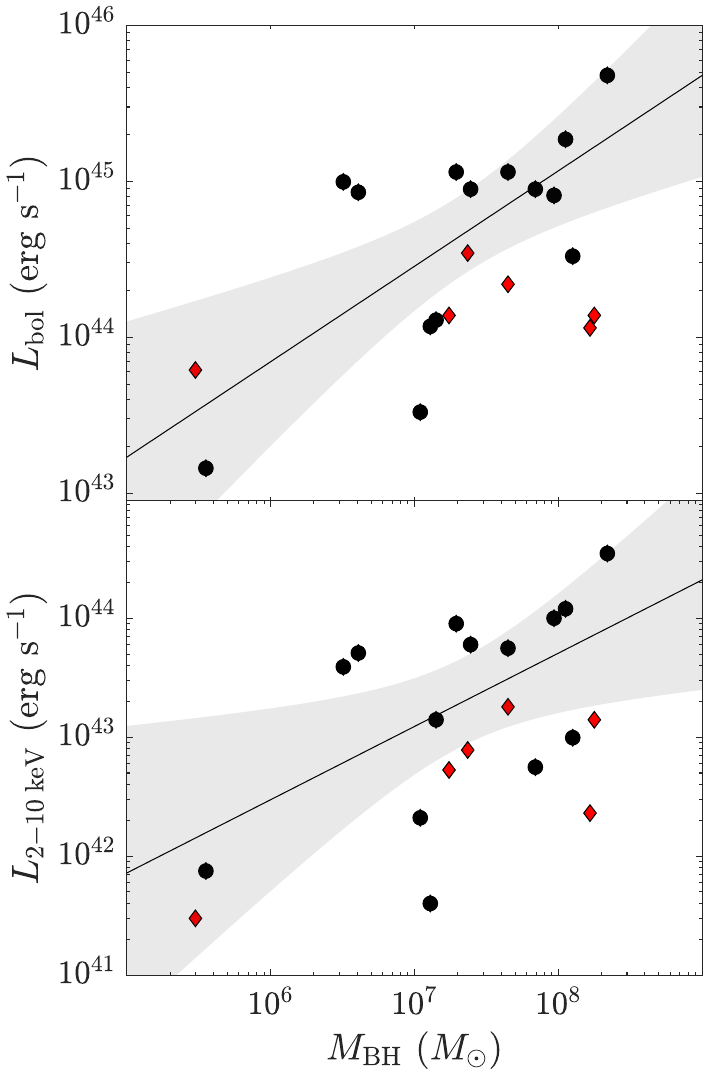}
    \caption{Bolometric (upper panel) and X-ray (lower panel) luminosities vs. black hole mass of the sources considered in this work. Both luminosities scale with mass as $M_{\rm BH}^{0.6}$, though the $L_{\rm bol}$ relation shows a larger correlation coefficient ($r\sim0.7$) than the $L_{\rm X}$ relation ($r\sim0.5$).}
    \label{fig:masslum}
\end{figure}
We considered the best-fit values of the temperature, $kT,$ and the optical depth, $\tau,$ for each source, as listed in Table~\ref{tab:coropar}. We find a strong correlation between the coronal temperature and the optical depth, as reported in Fig. \ref{fig:kTtau}. Indeed, the $\tau-kT$ relations have a correlation coefficient of $r=-0.96$ for the slab geometry, and $r=-0.97$ for the sphere geometry, both with a negligible probability of finding such correlations by chance. We fit the linear relation $\log kT=a\log \tau+b$ and found best-fit parameters of $a=-0.76\pm0.08$ and $b=1.54\pm0.05$, assuming a slab geometry. For the spherical corona, we obtained $a=-0.90\pm0.09$ and $b=1.91\pm0.07$. These results are consistent at $90\%$ confidence level with those obtained by \cite{tortosa18} with a smaller sample.\\
\indent It is very important to exclude any possible systematic effect when two quantities are found to be tightly correlated. To do that, we built a simple spectrum for an AGN, which is made of neutral absorption modeled with {\scriptsize TBABS}, a comptonizing continuum modeled with the physical model {\scriptsize COMPTT} and a ionized reflection component modeled with {\scriptsize XILLVER}, a mock version of model A. We simulate $1000$ {\it NuSTAR} spectra assuming random parameters in predetermined ranges. We assume that the column density may take any value between $N_{\rm H}=10^{20}$ cm$^{-2}$ to $N_{\rm H}=10^{23}$ cm$^{-2}$. We allowed for a wide range of photon indices in the reflection spectrum, from a flat spectrum with $\Gamma=1.45$ to a very steep one with $\Gamma=2.5$. The iron abundance is taken in the range $A_{\rm Fe}=\left[ 0.5\; 5 \right ]$, while the ionization describes a nearly neutral ($\log\xi/({\rm erg\;cm\;s^{-1}})=0$) to a highly ionized reflector ($\log\xi/({\rm erg\;cm\;s^{-1}})=4.7$). For simplicity, we tie the reflection normalization to the one of the continuum, assuming a reflection parameter in the range $\mathcal{R}=\left[ -0.1\; -1\right]$, where the negative values have been adopted in order to simulate pure reflection spectra. The adopted ranges of coronal temperature ($kT_e<500$ keV) and optical depth ($\tau<10$) are roughly based on the values of Table~\ref{tab:coropar}. Finally, the normalization of the continuum is chosen between $10^{-4}$ counts cm$^{-2}$ s$^{-1}$ keV$^{-1}$ and 5$\times10^{-2}$ counts cm$^{-2}$ s$^{-1}$ keV$^{-1}$, which corresponds to fluxes in the range $F_{\rm X}=5\times10^{-14}-10^{-10}$ erg cm$^{-2}$ s$^{-1}$. Each simulation was run assuming different exposures, between a minimum of $10$ ks to a maximum of $100$ ks, using simulated background spectra and the latest {\it NuSTAR} response and effective area. As shown in Fig.~\ref{fig:simul}, in the adopted large range of parameters, the average value of the best-fit parameters, $kT,$ and $\tau$ (right) are not much different than the simulated ones (left);  the exception is the last bin ($kT>450$ keV), where the optical depth tends to be slightly overestimated, due to the lower sensitivity of {\it NuSTAR} to high-temperature coronae. The simulations were designed with uncorrelated $kT-\tau$; therefore, if a spurious correlation was indeed present, it would also be expected to be present in the right panel of Fig.~\ref{fig:simul}. This suggests that from a statistical point of view, the correlation found in the data is not spurious. It is worth noting that the sample has been selected to be not strongly contaminated from complex, ionized, and multiple soft X-ray absorbers (see Sect. \ref{sec:data}); thus, we did not expect for the model systematics to contribute to affecting the continuum best-fit parameters.\\
\indent As discussed by \cite{tortosa18}, the observed $kT-\tau$ anticorrelation is not consistent with a global disk-corona configuration in radiative balance, which would imply the same heating-to-cooling ratio (HCR) for the coronae of all the AGN in the sample. One possibility to explain this anticorrelation is that colder coronae are more compact. In this case, a more compact corona would imply a larger optical depth, where the number of scatterings is larger. This will lead to a more efficient cooling (smaller HCR), and vice versa, a less compact corona would imply a smaller number of scattering, resulting in a larger HCR. Another possibility is that the disk-corona configuration is the same for all sources, but the sources might show different thermal emission due to viscous dissipation over the whole disk emission, which would also result in a larger cooling efficiency.\\
\indent We also investigate a possible dependence of the coronal temperature from the physical properties of the AGN, such as the black hole mass, $M_{\rm BH}$, and the Eddington ratio, $\lambda=L_{\rm bol}/L_{\rm Edd}$. As described in Sect.~\ref{sec:data}, the masses and bolometric luminosities were retrieved from the catalog published by \cite{koss22}, with   the sole exception of Ark 564, whose mass and bolometric luminosity were retrieved from \cite{peterson04} and adopting the bolometric correction from \cite{marconi04} to the X-ray luminosity obtained from the present data, respectively. As shown in Fig.~\ref{fig:kTpar}, there is no indication of a possible correlation between $kT$ or $\tau$ with either the black hole mass or the Eddington ratio. This is consistent with past studies of coronal parameters \citep[e.g.,][]{tortosa18,kamraj22}. However, we note that \cite{ricci18} found a trend between $\lambda_{\rm Edd}$ and the cut-off energy, $E_c$, when considering a very large sample of sources.

\subsection{X-ray variability and coronal properties}


\begin{figure*}
\centering
\includegraphics[scale=0.7]{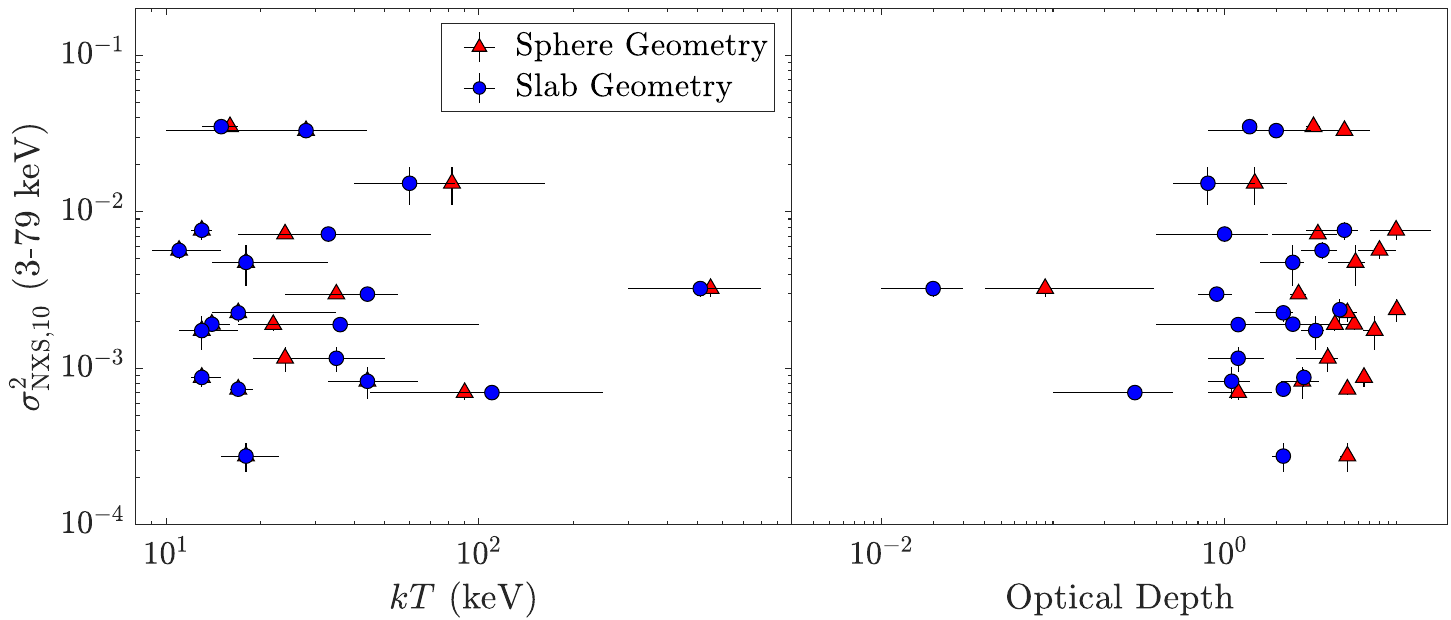}
\caption{Excess variance in the $E=3-79$ keV energy band versus the coronal temperature (left panel) and the optical depth (right panel) of the sample analyzed in this work. Blue circles identify temperature and optical depth values obtained assuming a slab geometry, while the red triangles indicate the values obtained assuming a spherical geometry.}
\label{fig:largesamplevar}
\end{figure*}


We investigate a possible correlation between the excess variance in the full {\it NuSTAR} band ($E=3-79$ keV) and the mass of the black hole (Fig.~\ref{fig:nxsmass}, left panel). For this purpose, we exclude the sources for which \cite{koss22} report the mass value from the velocity dispersion of the stars in the host galaxy, as they are affected by large uncertainties. We find that the two quantities are well correlated with a correlation coefficient of $r=-0.7$ and a null probability of $p_{\rm null}\simeq10^{-5}$. If we perform a linear fit on the logarithmic quantities, namely, $\log \sigma^2_{\rm NXS}=a \log (M_{\rm BH}/10^5M_\odot)+b$, we obtain $a=-0.6\pm0.2$ and $b=-1.2\pm0.7$. An anticorrelation was also found in previous variability analyses with {\it XMM-Newton} \citep[e.g.,][]{ponti12,tortosa23a}. We tested whether the excess variance is dependent on the X-ray luminosity, $L_{\rm X}$, and we found a moderate anticorrelation $r=-0.5$ ($p_{\rm null}\simeq0.02$), with the two quantities related as $\sigma^2_{\rm NXS}\propto L_{\rm X}^a$ (Fig.~\ref{fig:nxsmass}, middle panel) with $a=-0.4\pm0.3$. Despite not being highly significant, this relation is also found in other analyses with a much larger number of sources \citep[e.g.,][]{vagnetti16,prokhorenko24}. We also report an anticorrelation ($r\sim-0.7$ and $p_{\rm null}\sim10^{-5}$) with the bolometric luminosity ($\sigma^2_{\rm NXS}\propto L_{\rm bol}^a$ with $a=-0.6\pm0.3$). As shown in Fig.~\ref{fig:masslum}, though, both $L_{\rm X}$ and $L_{\rm bol}$ are correlated with the black hole mass, scaling as $\sim M_{\rm BH}^{0.6}$, with correlation coefficient $r\sim0.5$ ($p_{\rm null}\sim0.02$) and $r\sim0.7$ ($p_{\rm null}\sim10^{-5}$), respectively. Therefore it is likely that the two luminosity relations are degenerate with the black hole mass relation. Alternatively, the variability-luminosity relation has often been attributed to a superposition of small flares \citep[e.g.,][]{nandra97,almaini00}.\\
\indent We also test the possible relation between the excess variance and the Eddington ratio derived in Sect.~\ref{sec:specdisc} (see Fig.~\ref{fig:nxsmass}, right panel), finding a correlation coefficient $r=-0.03$, with a probability of finding such correlation by chance $p_{\rm null}=0.3$. This suggests a lack of correlation between the Eddington ratio and the X-ray variability. We also tried to remove the previously described dependency from the mass, checking a possible correlation between the quantity $\sigma^2_{\rm NXS}\times M_{\rm BH}^{0.6}$, but this quantity is also uncorrelated with the Eddington ratio values.\\
\indent We analyzed the possible correlation between the excess variance and the coronal parameters of the 20 sources analyzed here. We do not find a relevant dependence between the temperature obtained with both geometries and the excess variance in the $3-79$ keV energy band, as the correlation coefficient is $r=-0.02$, with null probability $p_{\rm null}=0.3$, and the angular coefficient is consistent with a flat line (see Fig.~\ref{fig:largesamplevar}, left panel). For completeness, we also show that the excess variance and the optical depth are not correlated ($r=0.01$, $p_{\rm null}=0.35$, see Fig.~\ref{fig:largesamplevar}, right panel). This result was expected once the $\sigma-kT$ relation is found to be absent, given the tight $kT-\tau$ correlations found in Sect.~\ref{sec:specdisc}.\\
\indent These results suggest that the X-ray variability on timescales of $10$ ks is not dependent on the coronal temperature nor the optical depth, raising questions about the origin of the X-ray variability. Indeed, the corona must introduce variability at timescales shorter than the ones observed for the UV radiation, which provides the seed photons for the Comptonization. Additionally, the corona must be responsible for the X-ray variability, because the UV light curves lag behind the X-ray ones \citep[e.g.,][]{kammoun21}.\\
\indent We note that, so far, we have considered the corona to be in thermal equilibrium, meaning that we have computed average values  of the temperature and optical depth over the whole observation, which is indeed larger than the timescale for which we have probed the variability. Therefore, one possibility is that the X-ray variability could be driven by changes of the temperature and optical depth at timescales that are consistent with the observed flux variations (i.e., below $10$ ks). However, measuring time-resolved temperatures is not yet possible with the currently most advanced hard X-ray telescope {\it NuSTAR}. Moreover, the physical environment could be much more complex than its simplification above. For instance, geometry is undoubtedly a parameter that could play a major role in driving the X-ray variability. In fact, variations in the intrinsic geometry or disk-corona geometry, such as the ones observed in X-ray binaries \citep[e.g.,][]{kara19}, could also drive the X-ray variability. It would be expected that, following a geometry variation, the temperature and optical depth would change accordingly, but these variations may also happen at shorter timescales than the ones probed by the spectral fits, possibly even shorter than 10 ks.\\
\indent An additional complexity to take into account is a possible spatial (likely radial) distribution of the temperature of the corona, while we have assumed an average single value for the whole electron plasma. The spatial and temporal average values of the temperature may spread out the link between the best-fit results and the calculations of the variability.\\
\indent Another possibility is that the variability is mainly driven by the observed anticorrelation with the black hole mass, which is likely proportional to the absolute size of the corona. If we consider, for instance, that all coronae of the AGN in the sample have a typical coronal radius of $R_c=10\;R_g$ \citep[e.g.,][]{dovciak16,ursini20l}, where $R_g=GM_{\rm BH}/c^2$ is the gravitational radius, the coronal size in physical units would be directly proportional to the supermassive black hole mass. However, we note that the size of the coronae in $R_g$ units could differ from source to source, as discussed in the previous section; furthermore, the relation between coronal size and black hole mass, although still increasing, it may be far from trivial to derive. In any case, even considering more complex relations between the coronal size and the black hole mass, more massive black holes correspond to larger coronae, which would imply a larger number of random scattering of the seed photons in the corona, resulting in a smaller X-ray variability amplitude.

\section{Summary and conclusions}
\label{sec:end}

We have presented a spectral and timing study of 20 bright AGN with the best signal-to-noise ratio available basedon {\it NuSTAR} data. We measured the temperature, $kT_e$, and optical depth, $\tau,$ of the X-ray emitting corona under two different geometries (sphere and slab) by modeling the spectrum continuum with the Comptonization model {\scriptsize COMPTT}. Additionally, we have studied the {\it NuSTAR} variability in the time range between 1 and 10 ks by means of the excess variance $\sigma^2_{\rm NXS}$. We note that the results of this paper are related to the sample at hand, which is the highest quality data to date. We summarize our results in the following

   \begin{itemize}
      \item We report that there is no correlation found between the X-ray variability and the electron temperature or the optical depth of the corona. This may imply that the X-ray variability is dependent on $kT$ and $\tau$ variations on timescales below $10$ ks, which is the timescale probed by the variability in this paper.
      \item We did find a strong anticorrelation between $kT$ and $\tau,$ adopting both slab and spherical geometries. In particular, we find that the optical depth is related to the temperature by a relation $kT\propto (\tau)^{-a}$ with $a\sim-0.7 \div -1$ (Fig.~\ref{fig:kTtau}), depending on the assumed geometry. Therefore, we confirm  the trend found by \cite{tortosa18} with our AGN  sample size increased by a factor of 2 with respect to the cited work.
      \item We did not find any dependence of $kT_e$ and $\tau$ with either the black hole mass or Eddington ratio. This is also consistent with previous results on other samples of AGN \citep[e.g.,][]{tortosa18,kamraj22}.
      \item There is a strong correlation between the $3-10$ keV and $3-79$ keV, which implies that the variability of the X-ray emission below $10$ keV is dominant over the rest of the spectrum. We also find a strong correlation between the variability of the {$10-40$} keV band and the one in {$3-10$ keV}. 
      \item We found a strong anticorrelation between the X-ray variability and the mass, following $\sigma^2_{\rm NXS}\propto M^{-0.6}$. This correlation is consistent to the one found by past variability studies with {\it XMM-Newton} \citep[e.g.,][]{ponti12,tortosa23a}. We also report a moderate correlation with the X-ray luminosity $\sigma^2_{\rm NXS}\propto L_{\rm X}^{-0.4}$, as well as no correlation seen with the Eddington ratio.
   \end{itemize}

The results of our study show that the main driver of the X-ray continuum variability produced in the hot-corona remains elusive; furthermore, it is not even clear whether there is a main driver for the observed variability at all and it may instead be the product of the superposition of several effects at work. We have shown here that the variability at $10$ ks does not depend on the physical properties of the corona, namely, electron temperature and optical depth. This then raises the question of what drives the X-ray variability. One possibility is that we might be probing different timescales, since we are studying relatively fast variability within $10$ ks; on the other hand, we have averaged the coronal temperature over days, months, and even years to reach a sufficient signal-to-noise ratio that would allow for an accurate measurement of the coronal temperature. Variations in the coronal geometry may also play an important role in producing the observed variability. In the future, detectors sensitive in a broadband energy range with much larger effective area, such as the Large Area Detector \citep[LAD; e.g.,][]{feroci22} proposed for the future enhanced X-ray Timing and Polarimetry mission \citep[eXTP;][]{zhang19} and the Spectroscopic Time-Resolving Observatory for Broadband Energy X-rays \citep[STROBE-X;][]{ray18}, will allow us to measure AGN coronal temperatures with high precision for exposures as short as 10 ks \citep{derosa19}. This would open up the possibility to probe both the coronal parameters and variability on the same timescale. Moreover, thanks to the extremely broadband $E=0.2-80$ keV, now only available with joint {\it NuSTAR} observations with other facilities such as {\it XMM-Newton}, the future X-ray telescope HEX-P \citep{kammoun24} will be able to measure optical depth and temperatures with much more accuracy than {\it NuSTAR}, with much shorter exposures.

\begin{acknowledgements}
      We thank the referee for their comments, which improved the quality of this paper. The authors thank Iossif Papadakis and Julien Malzac for useful discussions on the results of this paper. RS and ADR acknowledge support from the agreements ASI-INAF n.2017-14-H.0 "Science case study and scientific simulations for the enhanced X-ray Timing Polarimetry mission, eXTP", ASI-INAF eXTP Fase B-2020-3-HH.1-2021, Bando Ricerca Fondamentale INAF 2022 Large Grant "Dual and binary supermassive black holes in the multi-messenger era: from galaxy mergers to gravitational waves" and the INAF-PRIN grant “A Systematic Study of the largest reservoir of baryons and metals in the Universe: the circumgalactic medium of galaxies” (No. 1.05.01.85.10). AT acknowledges financial support from the Bando Ricerca Fondamentale INAF 2022 Large Grant "Toward an holistic view of the Titans: multi-band observations of z > 6 QSOs powered by greedy supermassive black holes". SB acknowledges funding from PRIN MUR 2022 SEAWIND 2022Y2T94C, supported by European Union - Next Generation EU, from INAF LG 2023 BLOSSOM, and from the EU grant AHEAD-2020 (GA no. 871158). CR acknowledges support from the Fondecyt Regular grant 1230345 and ANID BASAL project FB210003. POP acknowledges financial support from the French space agency (CNES) and the National High Energy Programme (PNHE) of CNRS. This research has made use of data and software provided by the High Energy Astrophysics Science Archive Research Center (HEASARC), which is a service of the Astrophysics Science Division at NASA/GSFC and the High Energy Astrophysics Division of the Smithsonian Astrophysical Observatory. This research has made use of the {\it NuSTAR} Data Analysis Software (NUSTARDAS) jointly developed by the ASI Space Science Data Center (SSDC, Italy) and the California Institute of Technology (Caltech, USA).
\end{acknowledgements}

%
   \bibliographystyle{aa} 
   \bibliography{biblio} 
%

\begin{appendix}
\section{Data}
\label{sec:datalist}
\begin{table}[h!]
\centering
\caption{{\it NuSTAR} spectra analyzed in this work.}
\begin{tabular}{lcccc}
\hline
Source & OBSID & Date & Exp. (s)\\
\hline
1H0419-577 & 60101039002 & 2015-06-03 & 169462\\
& 60402006002 & 2018-05-15 & 64216\\
& 60402006004 & 2018-11-13 & 48273\\
4C 50.55 & 60061305002 & 2014-12-13 & 24281\\
& 60301005002 & 2018-01-02 & 40338\\
Ark 564 & 60101031002 & 2015-05-22 & 211209\\
& 60401031002 & 2018-06-09 & 38090\\
& 60401031004 & 2018-11-28 & 408958\\
ESO 103-G35 & 60061288002 & 2013-02-24 & 27391\\
& 60301004002 & 2017-10-15 & 43834\\
ESO 362-G18 & 60201046002 & 2016-09-24 & 101905\\
ESO 383-G18 & 60061243002 & 2014-09-11 & 17342\\
& 60261002002 & 2016-01-20 & 106576\\
GRS 1734-292 & 60061279002 & 2014-09-16 & 20288\\
& 60301010002 & 2018-05-28 & 26020\\
HE 1143-1810 & 60302002002 & 2017-12-16 & 20960\\
& 60302002004 & 2017-12-18 & 20838\\
& 60302002006 & 2017-12-20 & 23096\\
& 60302002008 & 2017-12-22 & 20716\\
& 60302002010 & 2017-12-24 & 22378\\
IC 4329A & 60001045002 & 2012-08-12 & 162390\\
MCG-5-23-16 & 10002019001 & 2012-07-11 & 33925\\
& 60001046002 & 2013-06-03 & 160469\\
& 60001046004 & 2015-02-15 & 210887\\
& 60001046006 & 2015-02-21 & 98459\\
& 60001046008 & 2015-03-13 & 220835\\
MCG+8-11-11 & 60201027002 & 2016-08-16 & 97921\\
Mrk 6 & 60102044002 & 2015-04-21 & 62472\\
& 60102044004 & 2015-11-09 & 43816\\
Mrk 110 & 60201025002 & 2017-01-23 & 184563\\
& 60502022002 & 2019-11-16 & 86772\\
& 60502022004 & 2020-04-05 & 88674\\
Mrk 509 & 60101043002 & 2015-04-29 & 165885\\
& 60101043004 & 2015-06-02 & 36474\\
NGC 3281 & 60061201002 & 2016-01-22 & 22986\\
& 60662003002 & 2020-07-15 & 24616\\
NGC 5506 & 60061323002 & 2014-04-01 & 56585\\
NGC 5728 & 60061256002 & 2013-01-0 & 24357\\
& 60662002002   & 2020-07-13 & 24923\\
NGC 6814 & 60201028002 & 2016-07-04 & 148428\\
SWIFT\\
J2127.4+5654 & 60001110002 & 2012-11-04 & 49200\\
& 60001110003 & 2012-11-05 & 28764\\
& 60001110005 & 2012-11-06 & 74578\\
& 60001110007 & 2012-11-08 & 42106\\
UGC 6728 & 60160450002 & 2016-07-10 & 22615\\
& 60376007002 & 2017-10-13 & 58077\\
 \hline
 \hline
\end{tabular}
\tablefoot{The table lists source names with OBSID, observation date, and exposure.}
\end{table}

\section{Spectral fits}
\label{sec:specfits}

{\bf 1H0419-577}\\
This is a Seyfert 1 galaxy at $z=0.104$, with a black hole mass of $M_{\rm BH}=3.8\times10^8 M_\odot$. We binned the spectra at a minimum of $100$ counts per energy bin over the whole $3-79$ keV energy range. A Galactic absorption with column density $N_{\rm H,Gal}=1.14\times10^{20}$ cm$^{-2}$ is adopted. The spectra are well fitted by an absorbed continuum plus non-relativistic reflection (model A, {\tt ztbabs*(comptt+xillvercp)}). We let the intrinsic absorption column density $N_{\rm H}$ to vary between observations. We find $N_{\rm H}=(6\pm4)\times10^{21}$ cm$^{-2}$ for the 2015 observation and $N_{\rm H}=(1.0\pm0.5)\times10^{22}$ cm$^{-2}$ for both the 2018 observations. When we assume a slab geometry (parameter {\tt approx}$=0.5$) we obtain $kT=14^{+2}_{-1}$ keV and an optical depth $\tau=2.5^{+0.2}_{-0.3}$, while when a sphere geometry ({\tt approx}$=1.5$) is adopted we find the same temperature and an optical depth of $\tau=5.7^{+0.4}_{-0.5}$. We find a goodness of fit $\chi^2/{\rm dof}=1566/1588$. The results are consistent with the broadband analysis performed by \cite{turner18}, although we note that in that work the {\tiny OPTXAGNF} model \citep{done12} was used.\\
\\
{\bf 4C 50.55}\\
4C 50.55 is a radio-loud Seyfert 1 galaxy at redshift $z=0.02$, yielding a black hole mass of $6.4\times10^7$ $M_\odot$. We bin the {\it NuSTAR} spectra at a minimum of 100 counts per energy bin and consider the full $E=3-79$ keV energy band. We note that \cite{tazaki10} ruled out a significant role of the jet in the X-ray spectrum of this source using {\it Suzaku}. We consider a Galactic column of $N_{\rm H, Gal}=9.45\times10^{21}$ cm$^{-2}$. The source is well fitted by an absorbed continuum plus a single disk reflector, i.e. {\tt ztbabs*(comptt+xillvercp)} (model A). The column densities of the two observations are $N_{\rm H}=(3.0\pm0.4)\times10^{22}$ cm$^{-2}$ and $N_{\rm H}=(2.0\pm0.3)\times10^{22}$ cm$^{-2}$ for the 2014 and 2018 observations, respectively. We find a temperature of $kT=18^{+5}_{-2}$ keV ($kT=18^{+3}_{-2}$ keV) and optical depth of $\tau=2.2^{+0.2}_{-0.3}$ ($\tau=5.2^{+0.4}_{-0.5}$) assuming a slab (sphere) geometry. The statistic is $\chi^2/{\rm dof}=1325/1243$. These results are consistent with the analysis of \cite{buisson18}.\\
\\
{\bf Ark 564}\\
Ark 564 is a Narrow-Line Seyfert 1, with mass $M_{\rm BH}=3.2\times10^6 M_{\odot}$ \citep{peterson04}. The {\it NuSTAR} spectra have been binned at a minimum of $100$ counts per energy bin and only considered in the $E=3-30$ keV energy range. A Galactic column of $N_{\rm H,Gal}=5\times10^{20}$ cm$^{-2}$ is assumed. A simple power law model unveils a single narrow Fe K$\alpha$ component and a reflection component in the residual spectra. Therefore, the reflection is modeled with model A ({\tt compTT+xillvercp}), with no absorption needed. Assuming a slab (sphere) geometry we obtain $kT=15\pm2$ keV ($16\pm1$ keV) and an optical depth of $\tau=1.4\pm0.1$ ($3.3\pm0.3$). The statistic is $\chi^2/{\rm dof}=1267/1127$. A detailed broadband spectral analysis, which is consistent with the results presented here, can be found in \cite{kara17}.\\
\\
{\bf ESO 103-G35}\\
ESO 103-G35 is a Seyfert 2 galaxy at redshift $z=0.00914$ with a black hole mass of $M_{\rm BH}=1.3\times10^7$ $M_\odot$. The four FPM spectra were binned at a minimum of 100 counts per energy bin in the $E=3-50$ keV energy band. We consider a Galactic absorption of $N_{\rm H,Gal}=5.8\times10^{20}$ cm$^{-2}$. The source X-ray spectra appear reflection-dominated due to a heavy obscuration, therefore we adopt a model consisting of a distant absorber plus continuum, with intrinsic absorption on the line of sight, i.e. {\tt ztbabs*(comptt+borus12)} (model C). The two observations are characterized by nearly the same Compton-thin absorbing column density, which is consequently kept tied between epochs, of $N_{\rm H}=(1.65\pm0.08)\times10^{23}$ cm$^{-2}$. The temperature and optical depth assuming a slab geometry are $kT=17^{+18}_{-3}$ keV and $\tau=2.2^{+0.3}_{-0.7}$, respectively, while the two parameters assuming a sphere geometry are $kT=17^{+7}_{-3}$ keV and $\tau=5.2\pm0.7$. The goodness of fit is $\chi^2/{\rm dof}=1320/1302$. The results are consistent with the analysis of \cite{buisson18}, in which a different reflection model was used.\\
\\
{\bf ESO 362-G18}\\
ESO 362-G18 is a Seyfert 1.5 galaxy at redshift $z=0.01244$ with a black hole mass $M_{\rm BH}=4.5\times10^7$ $M_\odot$. The spectra were binned at a minimum of $50$ photon counts per energy bin in the full $E=3-79$ keV {\it NuSTAR} energy band. We consider a Galactic absorption of $N_{\rm H,Gal}=1.35\times10^{20}$ cm$^{-2}$. The X-ray spectrum has a typical Seyfert 1 shape with no evident intrinsic absorption. Following \cite{agis-gonzales14} we model the spectrum with two reflection components, i.e. {\tt comptt+xillvercp+relxillcp} (model B). For a slab (sphere) coronal geometry we obtain a temperature of $kT=18^{+14}_{-4}$ keV ($kT=18^{+15}_{-4}$ keV) and an optical depth of $\tau=2.5^{+0.4}_{-0.9}$ ($\tau=5.2\pm0.7$). The goodness of fit is given by $\chi^2/{\rm dof}=464/641$.\\
\\
{\bf ESO 383-G18}\\
ESO 383-G18 is a Seyfert 2 galaxy with $z=0.01241$, with $M_{\rm BH}=3\times10^5$ $M_\odot$. We bin the spectra of OBSID 60061243002 at a minimum of 50 counts per energy bin and the spectra of OBSID 60261002002 at a minimum of 100 counts per energy bin. We consider the $E=3-50$ keV energy band for both observations. We assume a Galactic absorption of $N_{\rm H,Gal}=3.8\times10^{20}$ cm$^{-2}$. The X-ray spectrum appears severely absorbed by a intervening cold material, which suggests that the reflector is best modeled by distant cold material, i.e. {\tt ztbabs*(comptt+borus12)} (model C). The absorbing column density of the two observations is consistent within the error at 90$\%$ confidence level, therefore we decided to keep them tied and obtain $N_{\rm H}=(1.2\pm0.2)\times10^{23}$ cm$^{-2}$. The temperature is identical assuming both assuming a slab or a spherical corona, i.e. $kT=7.3^{+0.5}_{-0.4}$ keV, while instead we find $\tau=4.7\pm0.3$ for a slab geometry and $\tau=10.1\pm0.6$ for a spherical one. The goodness of fit is $\chi^2/{\rm dof}=1035/1054$.\\
\\
{\bf GRS 1734-292}\\
GRS 1734-292 is a Seyfert 1 galaxy, having mass $M_{\rm BH}=2.5\times10^8 M_{\odot}$. We binned each FPM module of the two {\it NuSTAR} epochs at a minimum of 100 counts per energy bin, and considered the full $E=3-79$ keV energy band. We adopt a Galactic absorption of $N_{\rm H,Gal}=6.5\times10^{21}$ cm$^{-2}$. The source is also intrinsically moderately absorbed, with $N_{\rm H}=4^{+4}_{-3}\times10^{21}$ cm$^{-2}$ for OBSID 60061279002 and $N_{\rm H}=(15\pm4)\times10^{21}$ cm$^{-2}$ for OBSID 60301010002. An absorbed powerlaw only show narrow Fe K$\alpha$ residuals, therefore we adopt model A plus absorption ({\tt ztbabs(compTT+xillvercp)}). We obtain a coronal temperature of $kT=13^{+2}_{-1}$ keV for both slab and sphere geometries. An optical depth of $\tau=2.9\pm0.2$ ($6.5\pm0.4$) is recovered in a slab (sphere) geometry. We obtain a statistic of $\chi^2/{\rm dof}=1103/983$. These results are consistent with the broadband analysis presented in \cite{tortosa17}, in which OBSID 60061279002 was studied.\\
\\
{\bf HE 1143-1810}
\\
HE 1143-1810 is a Seyfert 1 galaxy at redshift $z=0.0329$ with an estimated black hole mass $M_{\rm BH}=4\times10^7$ $M_{\odot}$. The 10 FPM spectra were all binned at a minimum of 100 photon counts per energy bin in the $E=3-79$ keV energy band. The Galactic absorption is fixed at $N_{\rm H,Gal}=3\times10^{20}$ cm$^{-2}$. The spectrum has a typical unabsorbed shape, therefore no neutral absorption is needed. The source is nicely modeled by a single reflector plus continuum, i.e. {\tt comptt+xillvercp} (model A). Assuming a slab (spherical) coronal geometry we find a temperature of $kT=36^{+64}_{-19}$ keV ($kT=22^{+61}_{-5}$ keV) and an optical depth of $\tau=1.2^{+1.1}_{-0.8}$ ($\tau=4.4^{+0.9}_{-2.9}$). The goodness of fit is $\chi^2/{\rm dof}=2639/2665$. These values are largely in agreement with the broadband analysis presented in \cite{ursini20}.\\
\\
{\bf IC 4329A}\\
IC 4329A is a Seyfert 1 galaxy with mass $M_{\rm BH}=1.3\times10^8$. We bin both FPM modules of the analyzed observation (OBSID 60001045002) at a minimum of 100 counts per energy bin. We consider the full {\it NuSTAR} band, i.e. $E=3-79$ keV. The Galactic absorption is $N_{\rm H,Gal}=4\times10^{20}$ cm$^{-2}$. A modest absorption of $N_{\rm H}=(5\pm2)\times10^{21}$ cm$^{-2}$ is included. A single distant reflector is needed, therefore we adopt model A. The coronal temperature for a slab (sphere) coronal geometry is $kT=44^{+20}_{-10}$ keV ($44^{+17}_{-11}$ keV) and the optical depth is $\tau=1.1\pm0.3$ ($2.8\pm0.7$). The goodness of fit is $\chi^2/{\rm dof}=1411/1298$. A broadband spectral analysis is presented in \cite{brenneman14} and \cite{ingram23}, with consistent results with our {\it NuSTAR} fits.\\
\\
{\bf MCG-5-23-16}\\
MCG-5-23-16 is a Seyfert 1.9 galaxy with a mass of $M_{\rm BH}=1.8\times10^7 M_\odot$. All FPMA and FPMB spectra of the five available observations are binned at $100$ counts per energy bin and the $3-79$ keV energy range is considered. We assume a Galactic absorption of $N_{\rm H,Gal}=8\times10^{20}$ cm$^{-2}$ and we include cold intrinsic absorption with the model {\tt ztbabs}, finding a column density $N_{\rm H}=(1.4\pm0.1)\times10^{22}$ cm$^{-2}$. We the source is modeled with a single reflector plus continuum with {\tt xillvercp}, we find significant residuals on the Fe K$\alpha$ region, strongly hinting at the presence of a broad component, likely a reflection component from the accretion disk. Therefore, we adopt the absorbed model B, {\tt ztbabs(compTT+xillvercp+relxillcp)}. Following \cite{serafinelli23b}, we assume an emissivity index of $-3$, an inner radius $R_{\rm in}=40R_g$, inclination $45^\circ$, and a non-spinning black hole. Assuming a slab geometry, we find a coronal temperature $kT=44\pm{11}$ keV and an optical depth $\tau=0.9\pm0.2$, while $kT=35^{+10}_{-11}$ keV and $\tau=2.7\pm0.3$ are found if a spherical coronal geometry is assumed. The goodness of fit is $\chi^2/{\rm dof}=6995/5776$. These values are consistent at $99\%$ confidence level with the broadband analyses of \cite{balokovic15} and \cite{serafinelli23b} performed on this source.\\
\\
{\bf MCG+8-11-11}\\
MCG+8-11-11 is a Seyfert 1 galaxy with mass $M_{\rm BH}=1.6\times10^7 M_{\odot}$. The FPMA and FPMB spectra were binned at 100 counts per energy bin over the whole energy range ($E=3-79$ keV), and a Galactic column density of $N_{\rm H}=1.75\times10^{21}$ cm$^{-2}$ is assumed. No intrinsic cold absorption is present. Two reflection components are required in the model, as the Fe K$\alpha$ emission line is not adequately fitted by a single narrow component. We adopt therefore model B, {\tt compTT+xillvercp+relxillcp}. We assume a non-spinning black hole and a disk with inner radius $R_{\rm in}=2R_g$, where $R_g=GM/c^2$ is the gravitational radius, and emissivity index $-3$. We find a coronal temperature of $kT=110^{+140}_{-40}$ keV and an optical depth $\tau=0.3\pm0.2$ when assuming a slab coronal geometry. When a spherical geometry is assumed for the corona, we obtain $kT=90^{+125}_{-45}$ keV and $\tau=1.2^{+0.7}_{-0.4}$. The goodness of fit is $\chi^2/{\rm dof}=873/851$. These values are consistent with the results of \cite{tortosa18a}, where this data was analyzed with contemporaneous {\it Swift}-XRT data.\\
\\
{\bf Mrk 6}\\
Mrk 6 is a Seyfert 1.5 galaxy at $z=0.01951$ with mass $M_{\rm BH}=1.9\times10^7$ $M_{\odot}$. All four FPM spectra were binned at a minimum of 100 counts per energy bin. We consider the $E=3-50$ keV energy band. The X-ray spectrum appear severely absorbed as in previous analyses of this source \citep[e.g.,][]{molina19}. Therefore, once the Galactic absorption is modeled with a fixed $N_{\rm H,Gal}=7.6\times10^{20}$ cm$^{-2}$, we model the spectra with an absorbed continua and a distant neutral reflector, i.e. {\tt ztbabs*(comptt+borus12)} (model C). For OBSID 60102044002 and 60102044004 we find an intrinsic column density $N_{\rm H}=(1.2\pm0.1)\times10^{23}$ cm$^{-2}$ and $N_{\rm H}=(1.0\pm0.1)\times10^{23}$ cm$^{-2}$, respectively. The temperature is $kT=13^{+4}_{-2}$ keV both assuming a slab or a spherical coronal geometry, while the optical depth is $\tau=3.4^{+0.4}_{-0.6}$ and $\tau=7.5^{+0.7}_{-1.1}$, respectively. The statistic is $\chi^2/{\rm dof}=1283/1335$. A few moderate residuals are present below $5$ keV, possibly due to the presence of a warm absorber \citep{kayanoki23}, but given the low energy resolution of {\it NuSTAR} in the $3-5$ keV energy range, we did not model such a component.\\
\\
{\bf Mrk 110}\\
Mrk 110 is a Seyfert 1 galaxy with redshift $z=0.03552$ and mass $M_{\rm BH}=1.8\times10^7$ $M_\odot$. The FPM spectra are all binned at a minimum of 100 counts per energy bin in the full $E=3-65$ keV energy range. We adopt a Galactic absorption of $N_{\rm H,Gal}=1.3\times10^{20}$ cm$^{-2}$, while no intrinsic absorption is present confirming its nature as a "bare" AGN \citep{reeves21,porquet21}. We model the X-ray spectra with two reflectors, i.e. {\tt comptt+xillvercp+relxillcp} (model B) finding $kT=35^{+15}_{-10}$ keV ($kT=24^{+17}_{-5}$ keV) and $\tau=1.2^{+0.5}_{-0.4}$ ($\tau=4.0^{+0.6}_{-1.4}$) when assuming a slab (sphere) coronal geometry. The goodness of fit is $\chi^2/{\rm dof}=2600/2305$. The results agree with the broadband analysis presented in \cite{porquet21}.\\
\\
{\bf Mrk 509}\\
Mrk 509 is a Seyfert 1 galaxy at $z=0.01951$ with a black hole mass of $M_{\rm BH}=2\times10^{8}$ $M_\odot$. The FPMA and FPMB spectra for the two epochs were binned at a minimum of 100 photon counts per energy bin in $E=3-65$ keV. The Galactic absorption is fixed at $N_{\rm H,Gal}=3.9\times10^{20}$ cm$^{-2}$. We model the X-ray spectra with an unabsorbed continuum with a single ionized reflector, i.e. {\tt comptt*xillvercp} (model A). We find a temperature of $kT=17^{+2}_{-1}$ keV for both slab and sphere geometries, while the optical depth is $\tau=2.2\pm0.1$ ($\tau=5.2^{+0.2}_{-0.3}$) when assuming a slab (sphere) geometry. The statistic is $\chi^2/{\rm dof}=1949/1647$.\\
\\
{\bf NGC 3281}\\
NGC 3281 is a Seyfert 2 galaxy with redshift $z=0.01073$, with mass $M_{\rm BH}=1.7\times10^8$ $M_\odot$. We bin the FPM spectra with a minimum of 50 counts per energy bin in the energy band $E=3-60$ keV. We assume a Galactic absorption of $N_{\rm H,Gal}=6.6\times10^{20}$ cm$^{-2}$. The shape of the X-ray spectrum is that of a severely absorbed source, therefore we model the spectra with an absorbed continuum plus neutral reflection, i.e. {\tt ztbabs*(comptt+borus12)} (model C). We find that the intrinsic column density is moderately variable, by a factor of $\sim4$, since we find $N_{\rm H}=8^{+6}_{-5}\times10^{22}$ cm$^{-2}$ for OBSID 60061201002 and $N_{\rm H}=(3.1\pm0.5)\times10^{23}$ cm$^{-2}$ for OBSID 60662003002. Assuming either a slab or a spherical geometry we find a temperature $kT=11^{+4}_{-2}$ keV, with optical depth $\tau=3.7^{+0.8}_{-0.9}$ and $\tau=8\pm2$, respectively. The goodness of fit is $\chi^2/{\rm dof}=693/612$.\\
\\
{\bf NGC 5506}\\
NGC 5506 is a Seyfert 1 galaxy with mass $M=8.8\times10^7$ $M_\odot$. The {\it NuSTAR} Focal Plane Modules were both binned at a minimum of 100 counts per energy bin in the full {\it NuSTAR} band $E=3-79$ keV. The Galactic cold absorption is $N_{\rm H,Gal}=4.2\times10^{20}$ cm$^{-2}$, and we recover an intrinsic absorption of $N_{\rm H}=(2.4\pm0.3)\times10^{22}$ cm$^{-2}$. A single reflector is able to model the hard X-ray spectrum of this AGN, therefore we adopt model A with neutral absorption ({\tt ztbabs(compTT+xillvercp)}). When a slab geometry is assumed, we obtain $kT=510^{+250}_{-150}$ keV and $\tau=0.02\pm0.01$, while assuming a sphere geometry the best-fit parameters are $kT=550\pm250$ keV and $\tau=0.09^{+0.30}_{-0.05}$. The goodness of fit is $\chi^2/{\rm dof}=819/738$ and the best-fit values are largely in agreement with the broadband analysis performed in \cite{matt15}.\\
\\
{\bf NGC 5728}\\
NGC 5728 is a Seyfert 2 galaxy at $z=0.00947$ with an estimated black hole mass of $M_{\rm BH}=3.4\times10^7$ $M_\odot$. We bin all the X-ray spectra at a minimum of 50 counts per energy bin. We consider We adopt a fixed Galactic absorption column density of $N_{\rm H,Gal}=7.5\times10^{20}$ cm$^{-2}$. The spectrum appears as that of a typical absorbed source, therefore we model it with an absorbed continuum plus neutral reflection, i.e. {\tt ztbabs*(comptt+borus12)} (model C). The intrinsic absorption is not variable and therefore we keep it tied between the two epochs, finding $N_{\rm H}=4.6^{+1.6}_{-1.8}\times10^{23}$ cm$^{-2}$. We find a coronal temperature of $kT=13\pm1$ keV for both slab and spherical geometry and an optical depth of $\tau=5^{+2}_{-1}$ ($\tau=10^{+6}_{-3}$) for a slab (spherical) coronal geometry. The statistic is given by $\chi^2/{\rm dof}=480/524$.\\
\\
{\bf NGC 6814}\\
NGC 6814 is a Seyfert 1 galaxy with mass $M=2.7\times10^6$ $M_\odot$. We bin both FPMA and FPMB detectors at a minimum of 100 counts per energy bin in the $E=3-60$ keV energy band. We assume a Galactic absorption with column density $N_{\rm H,Gal}=8\times10^{20}$ cm$^{-2}$, while no intrinsic cold absorption is found in this source. The spectra need to be modeled with model B ({\tt comptt+xillvercp+relxillcp}), having two reflection components, since the Fe K$\alpha$ emission line is broadened. For simplicity, we assume a non-spinning black hole, and a disk with emissivity index $-3$ from a disk with inner radius $R_{\rm in}=2R_g$. The coronal temperature for a slab (spherical) coronal geometry is $kT=60^{+24}_{-20}$ keV ($kT=82^{+80}_{-10}$ keV) and the optical depth is $\tau=0.8^{+0.7}_{-0.3}$ ($1.5^{+0.8}_{-0.1}$). The goodness of fit is $\chi^2/{\rm dof}=897/846$. Though with larger errors due to the sole use of {\it NuSTAR}, the values are consistent with those found by \cite{tortosa18a} in the broad ($E=0.5-60$ keV) band.\\
\\
{\bf SWIFT J2127.4+5654}\\
The Narrow-line Seyfert 1 galaxy SWIFT J2127.4+5654 is characterized by a mass of $M=1.5\times10^7$ $M_\odot$ at redshift $z=0.014$. Both FPM spectra A and B are binned at 100 min counts per energy bin in the whole {\it NuSTAR} band $E=3-79$ keV. The Galactic absorption column density is $N_{\rm H,Gal}=7\times10^{20}$ cm$^{-2}$ and negligible intrinsic absorption. The source is well modeled by model A with a single non-relativistic reflector, i.e. {\tt compTT+xillvercp}, we obtain for a slab (spherical) coronal geometry a temperature of $kT=33^{+37}_{-15}$ keV ($kT=24^{+24}_{-7}$ keV) and an optical depth of $\tau=1.0^{+0.8}_{-0.6}$ ($\tau=3.5^{+1.0}_{-1.6}$). The statistic is $\chi^2/{\rm dof}=1751/1775$, with values consistent at $99\%$ confidence level with those of the broadband study by \cite{marinucci14}.\\
\\
{\bf UGC 6728}\\
UGC 6728 is a Seyfert 1 galaxy with redshift $z=0.00652$ with a black hole mass of $M_{\rm BH}=7.1\times10^5$ $M_{\odot}$. The four FPM spectra were binned at a minimum of 50 photon counts per energy bin. We consider data in the $E=3-40$ keV energy band. We assume a Galactic absorption of $N_{\rm H}=4.5\times10^{20}$ cm$^{-2}$. No evident intrinsic absorption is present. The spectra are best modeled with continuum and two reflectors, i.e. {\tt comptt+xillvercp+relxillcp} (model B). We find a temperature of $kT=28^{+16}_{-18}$ keV ($kT=28^{+17}_{-15}$ keV) and an optical depth of $\tau=2.0\pm1.2$ ($\tau=5.0\pm2.0$) when assuming a slab (sphere) coronal geometry. The goodness of fit is $\chi^2/{\rm dof}=1321/1183$.\\

\onecolumn
\section{Spectra}
\label{sec:specfig}
\begin{longfigure}{c}
\includegraphics[scale=0.35]{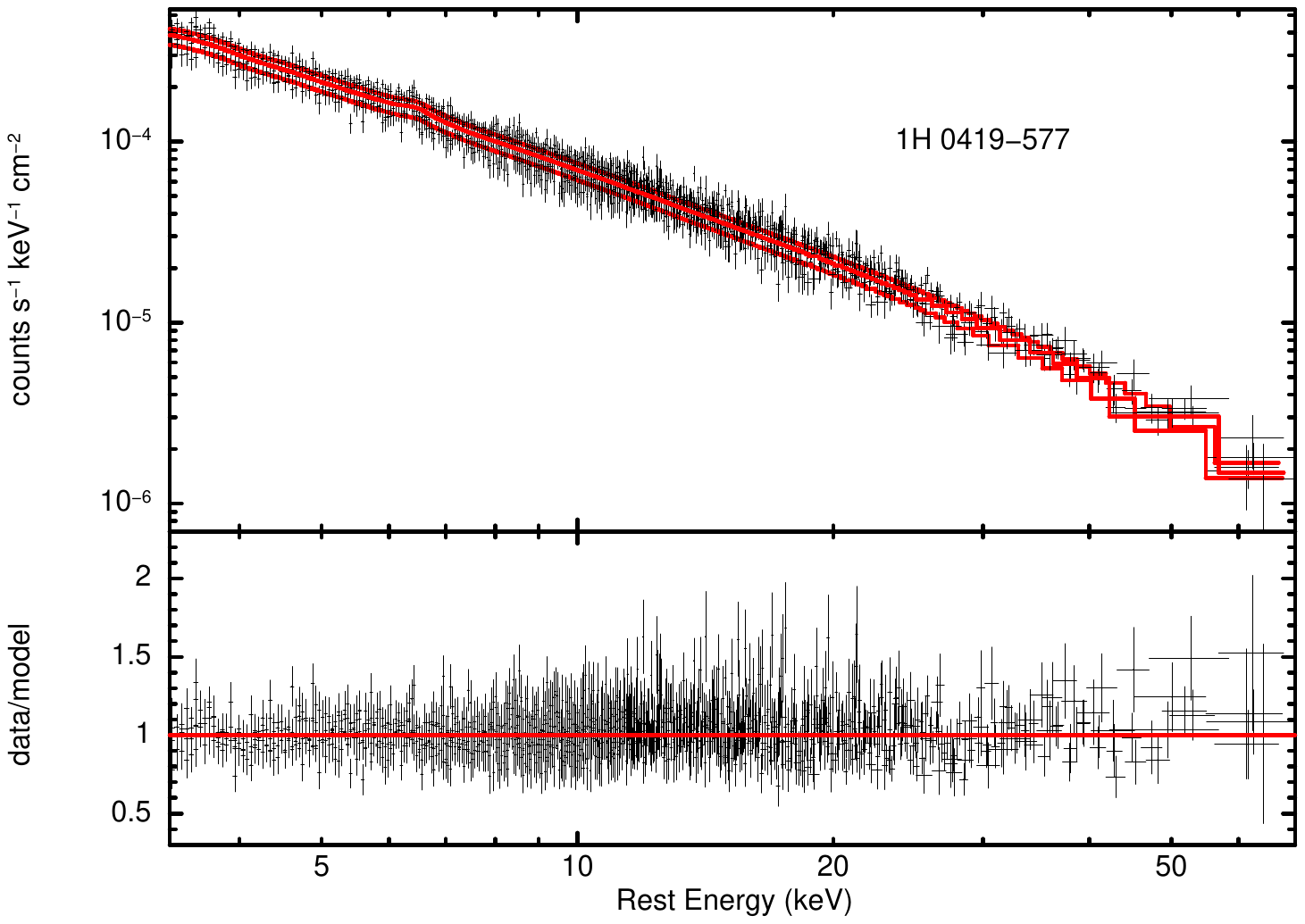}\includegraphics[scale=0.35]{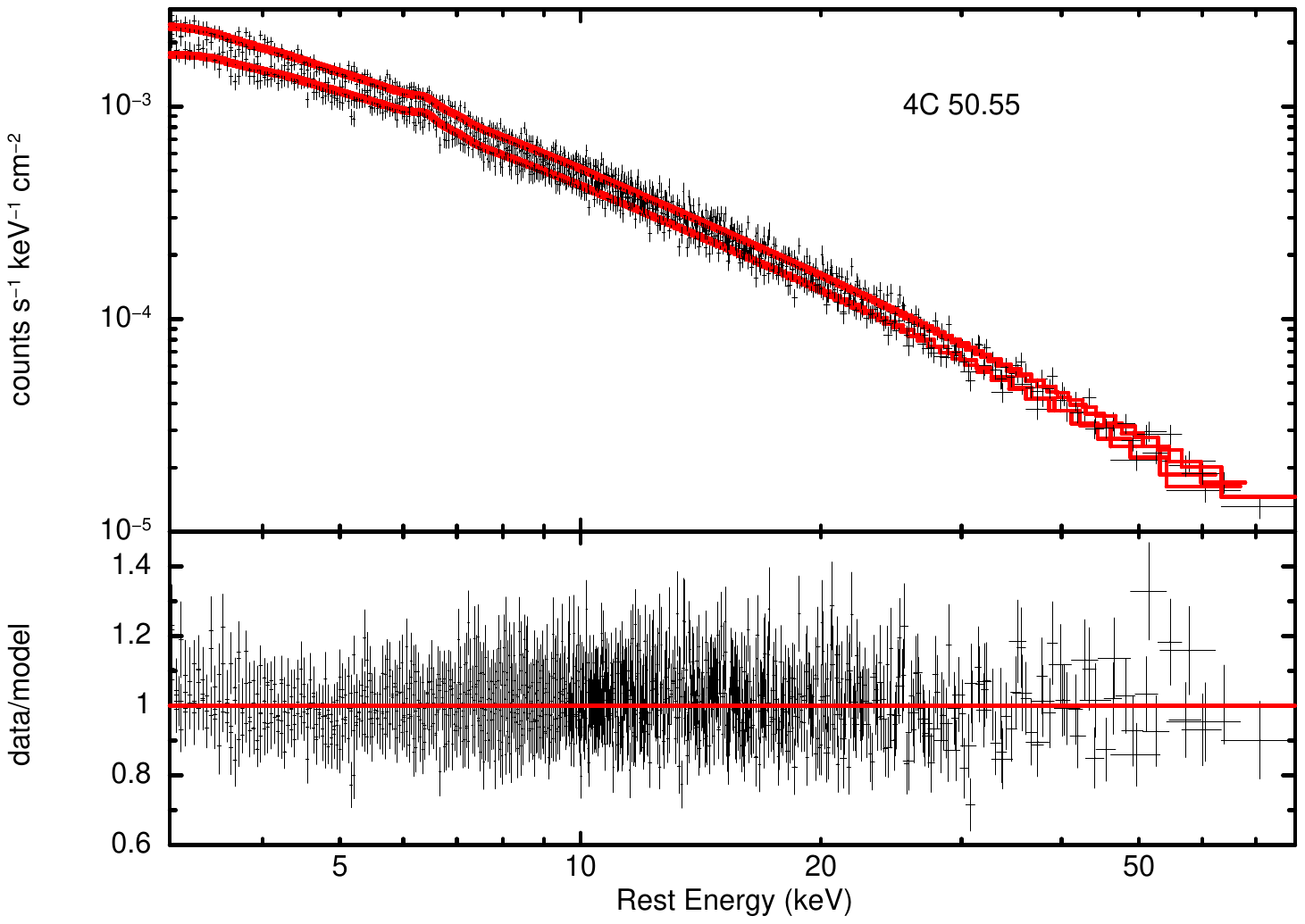}\\\includegraphics[scale=0.35]{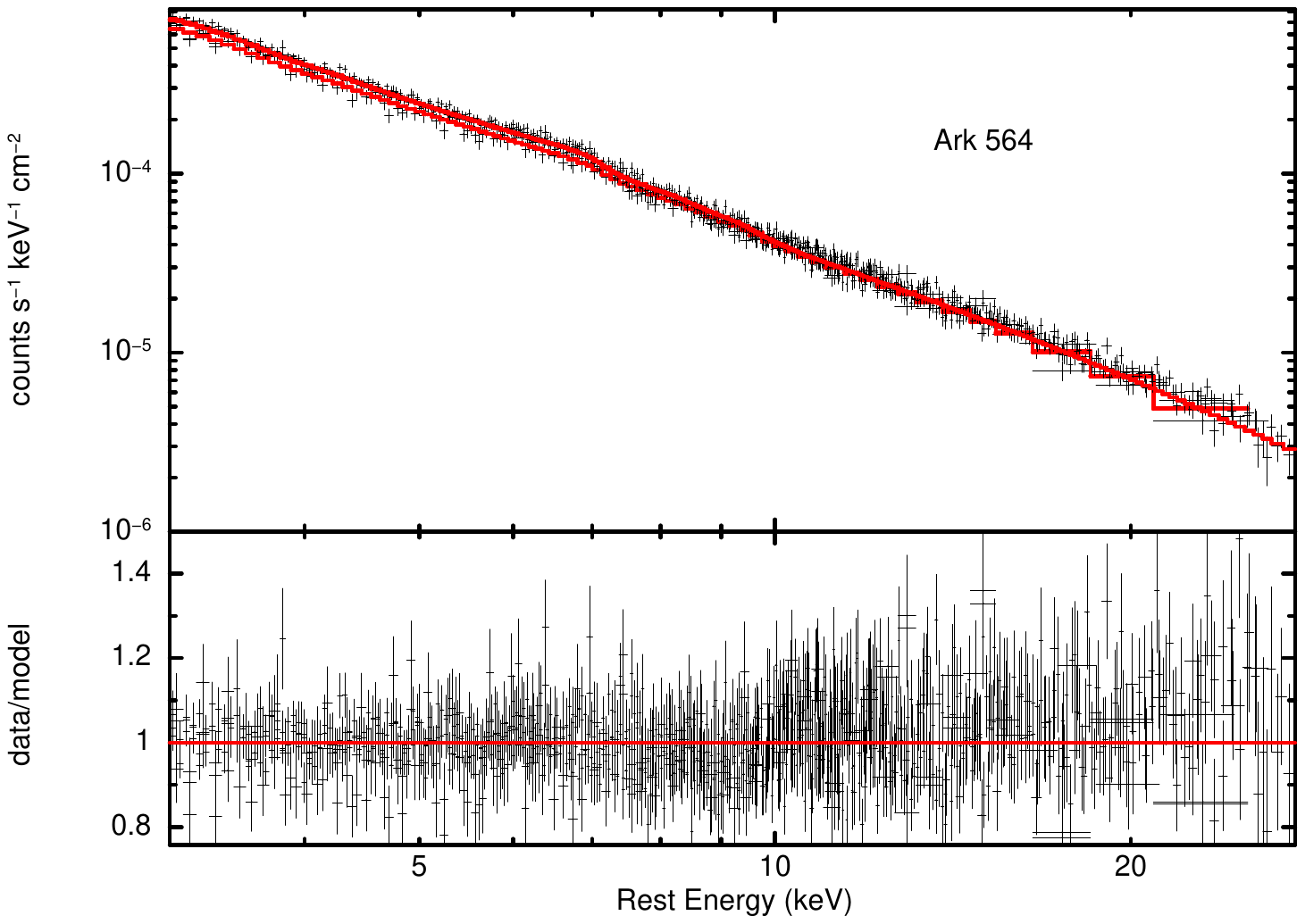}
\includegraphics[scale=0.35]{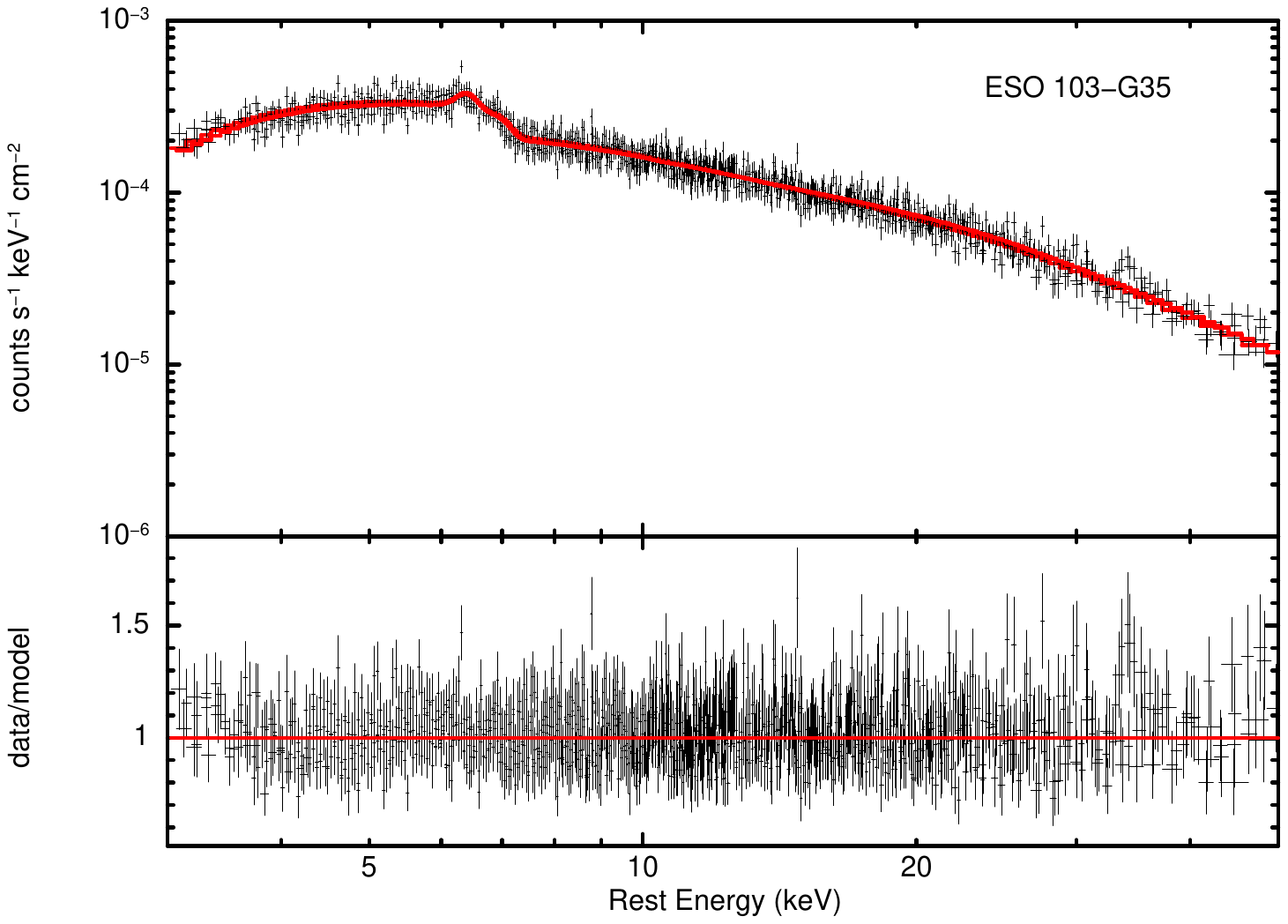}\\
\includegraphics[scale=0.35]{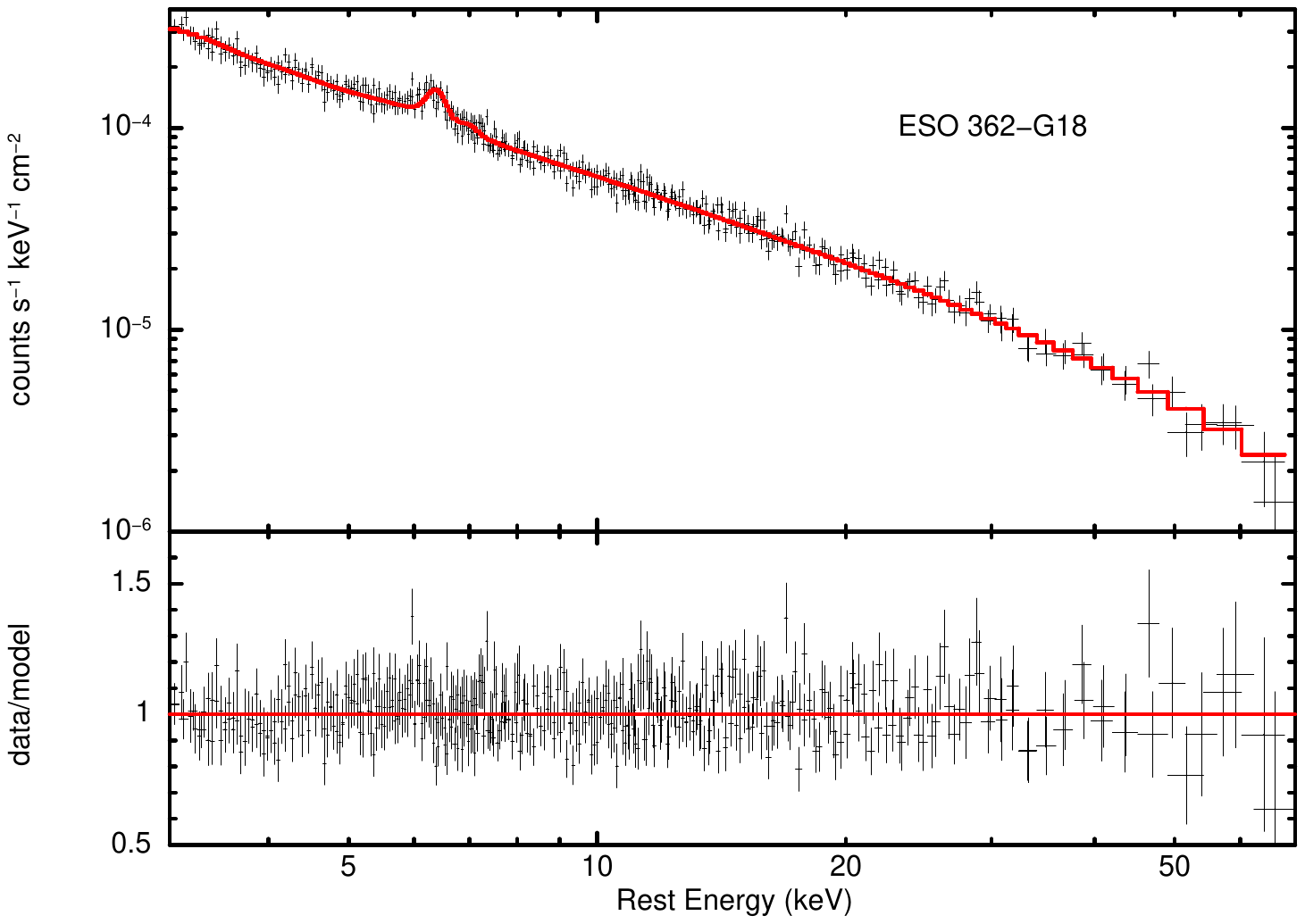}\includegraphics[scale=0.35]{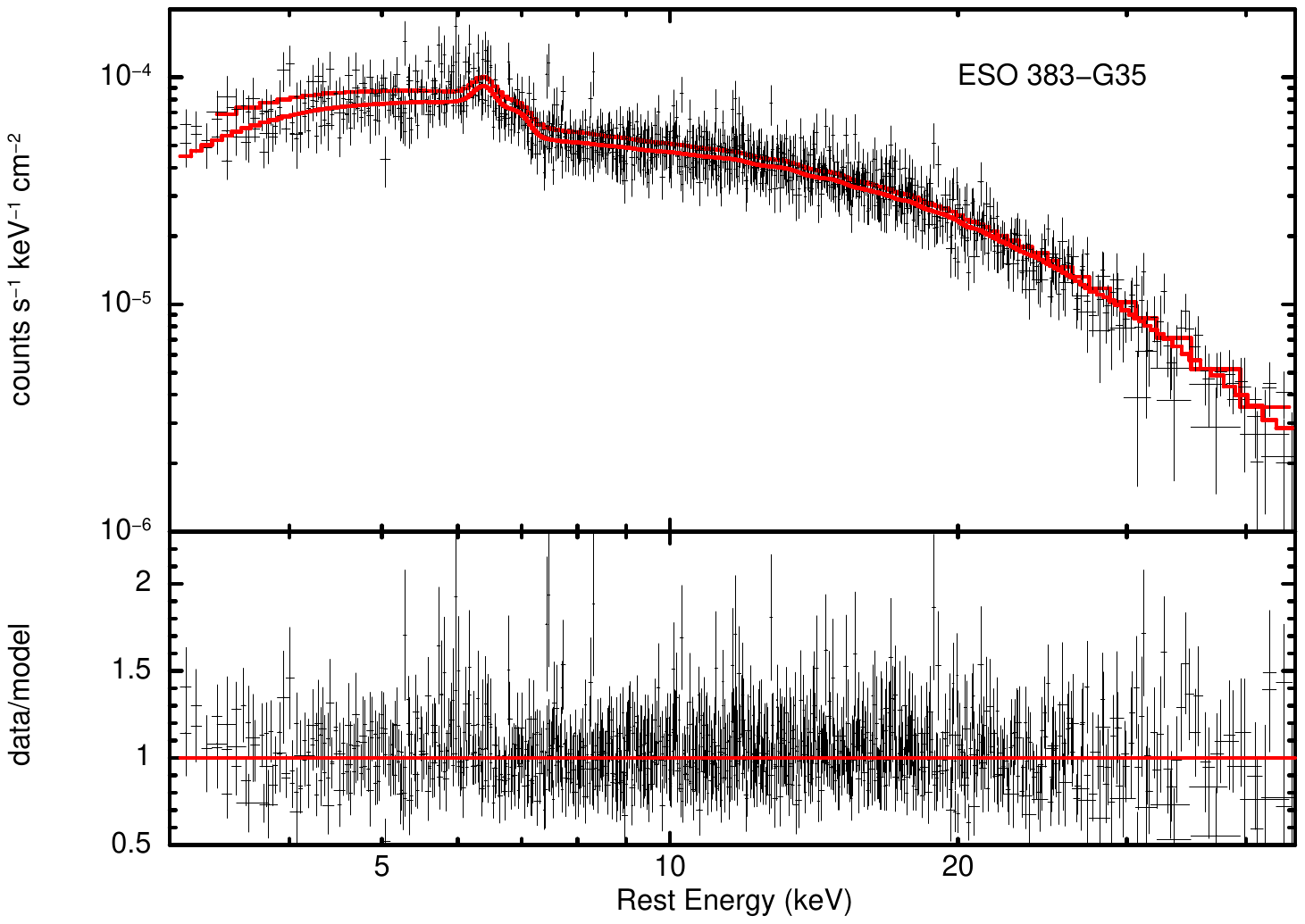}\\
\includegraphics[scale=0.35]{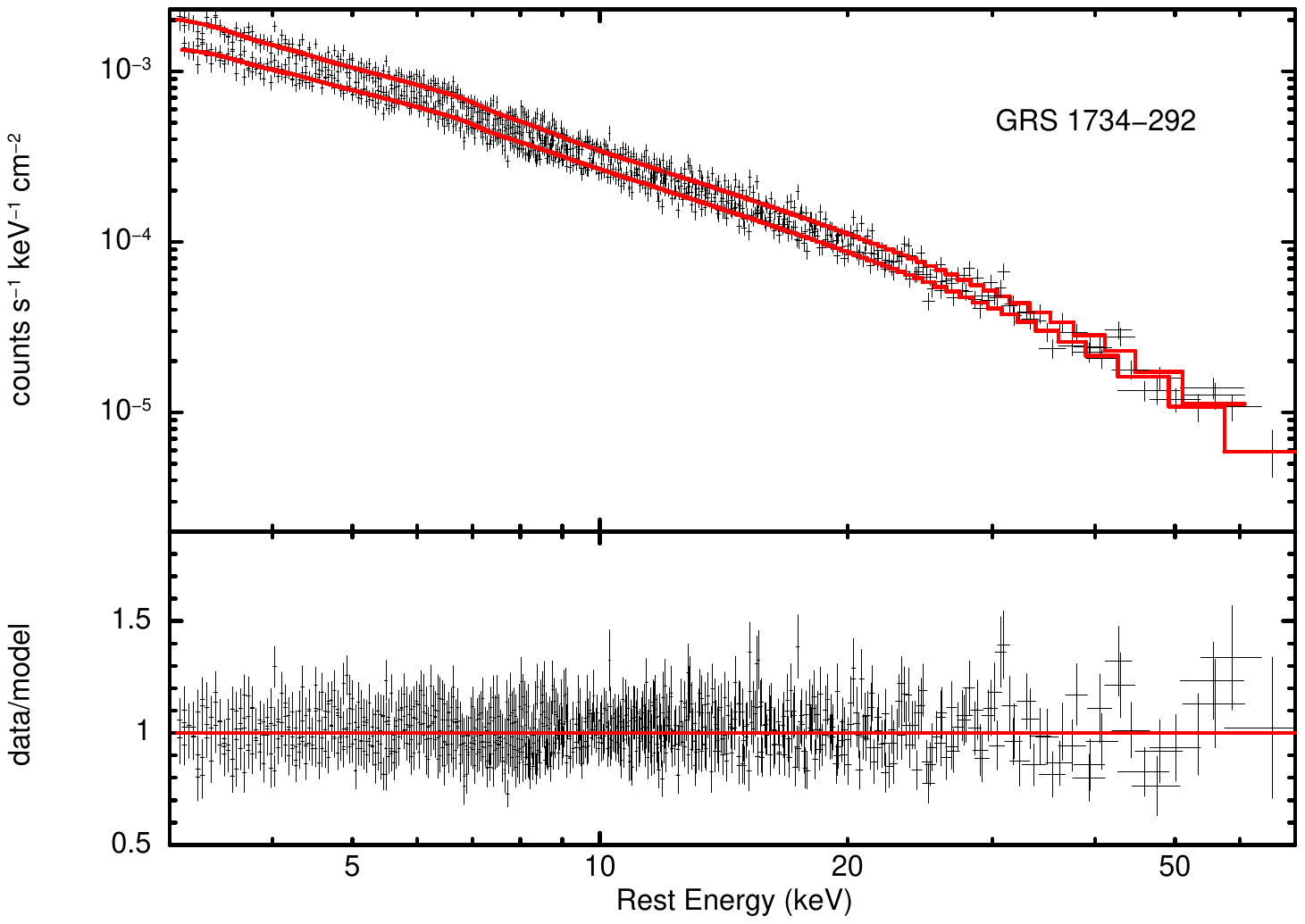}\includegraphics[scale=0.35]{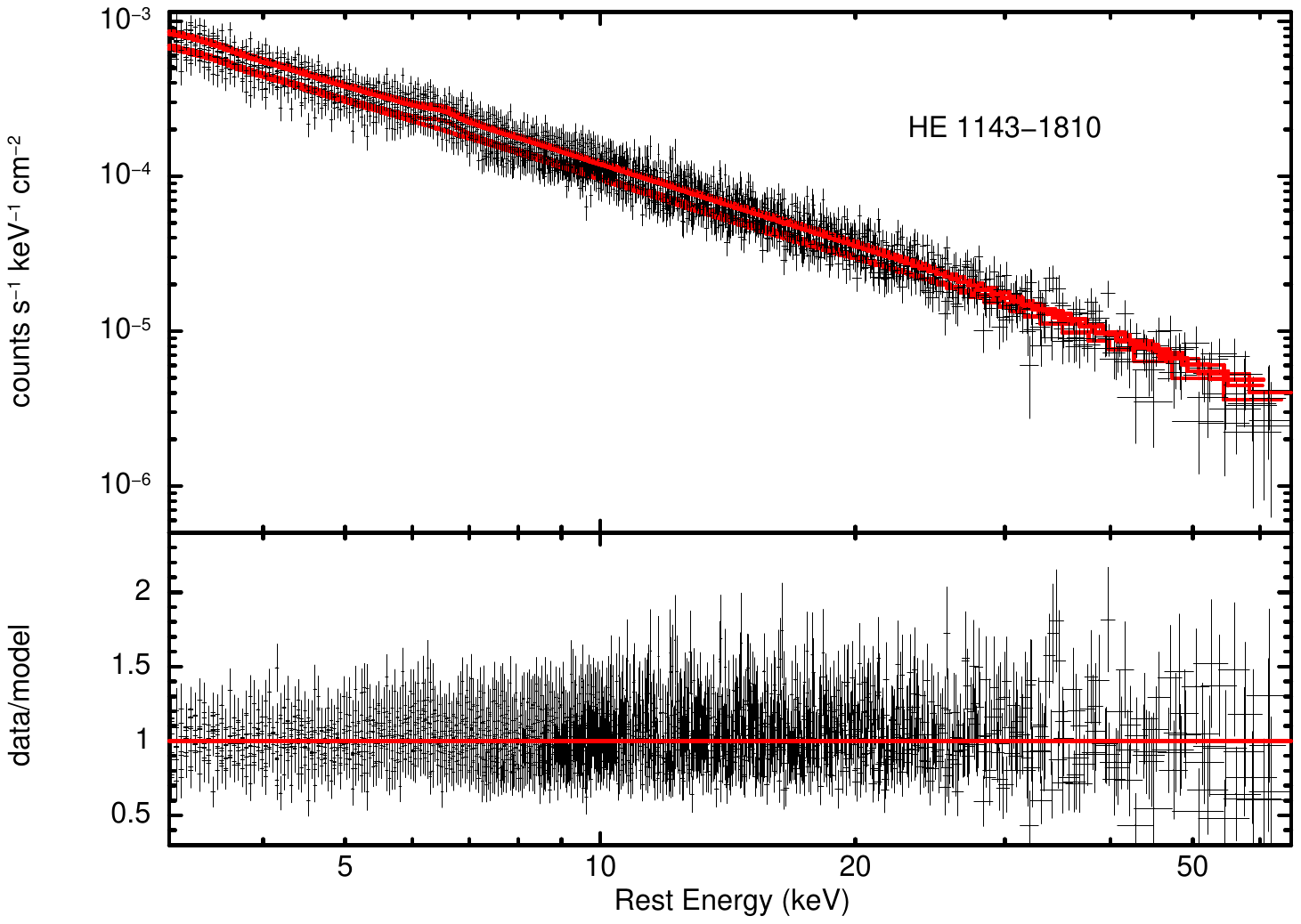}\\
\includegraphics[scale=0.35]{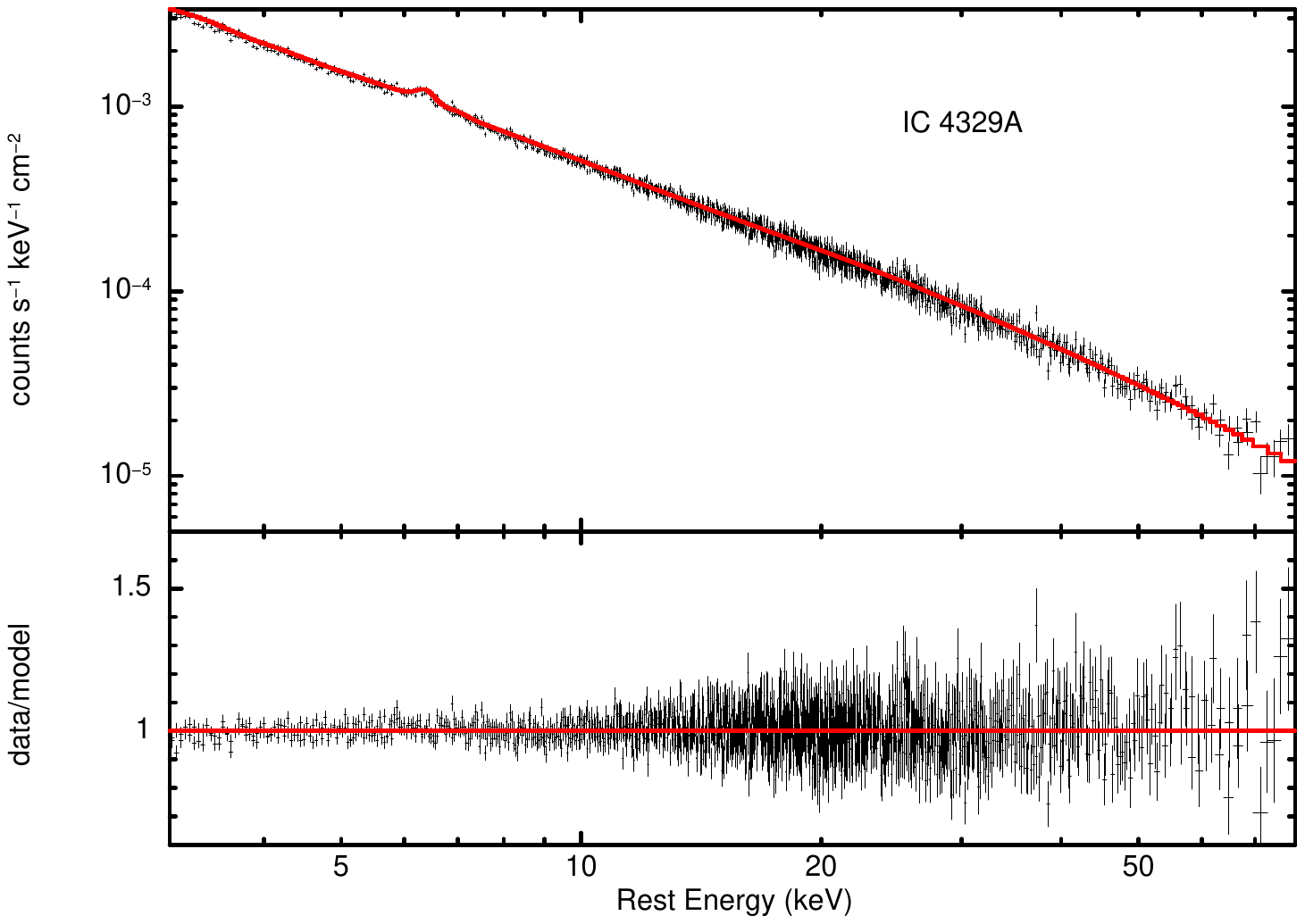}
\includegraphics[scale=0.35]{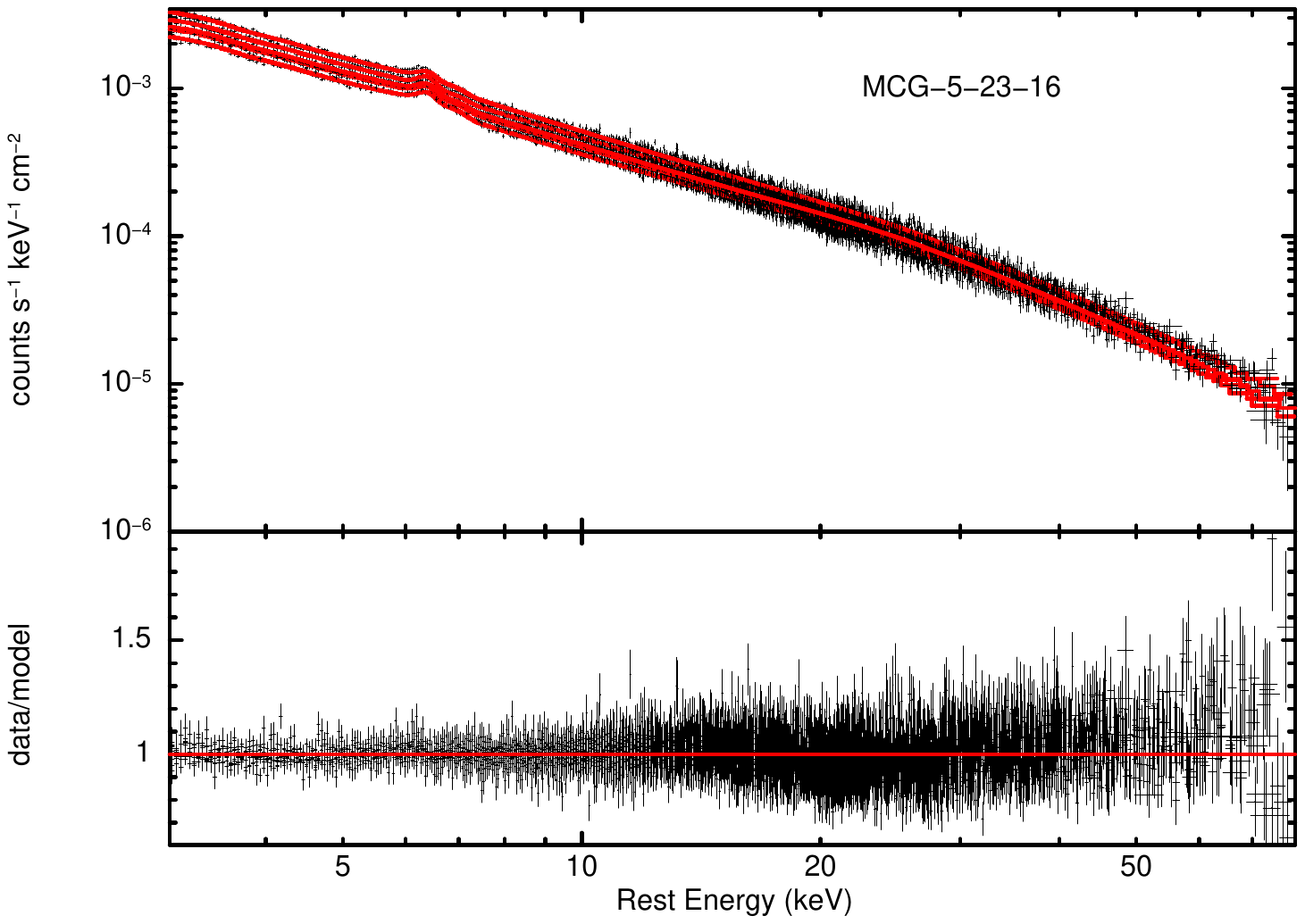}\\\includegraphics[scale=0.35]{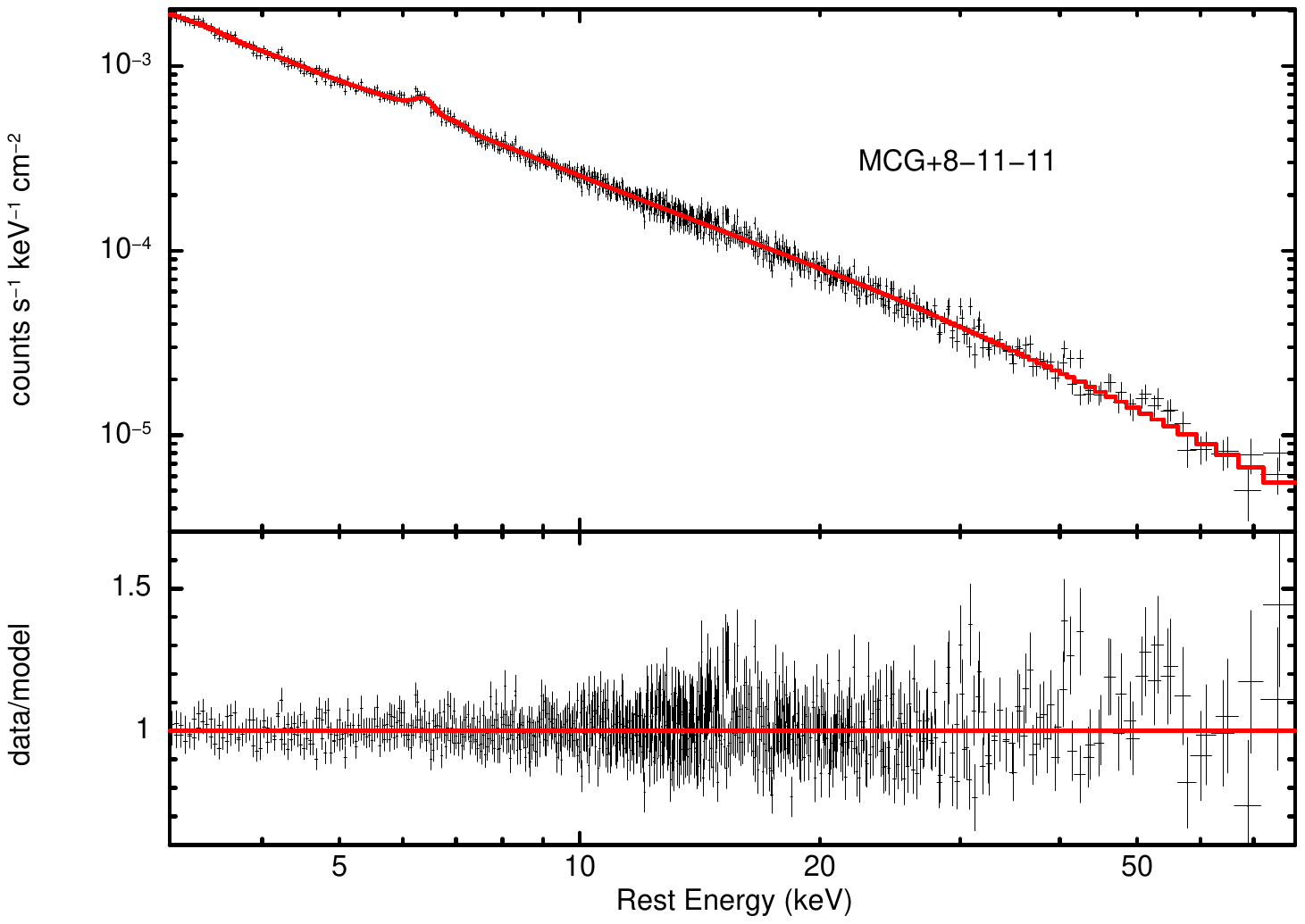}\includegraphics[scale=0.35]{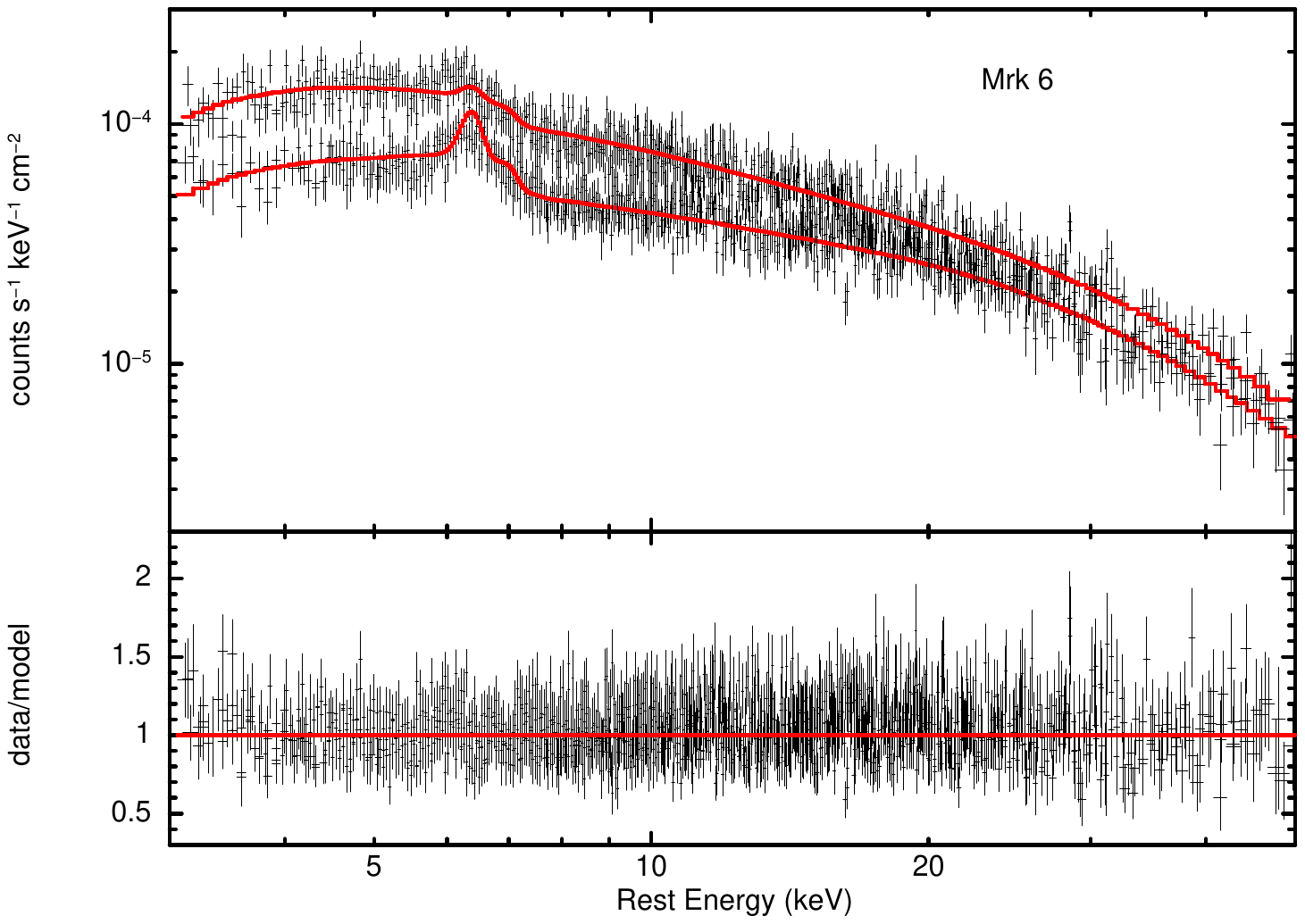}\\
\includegraphics[scale=0.35]{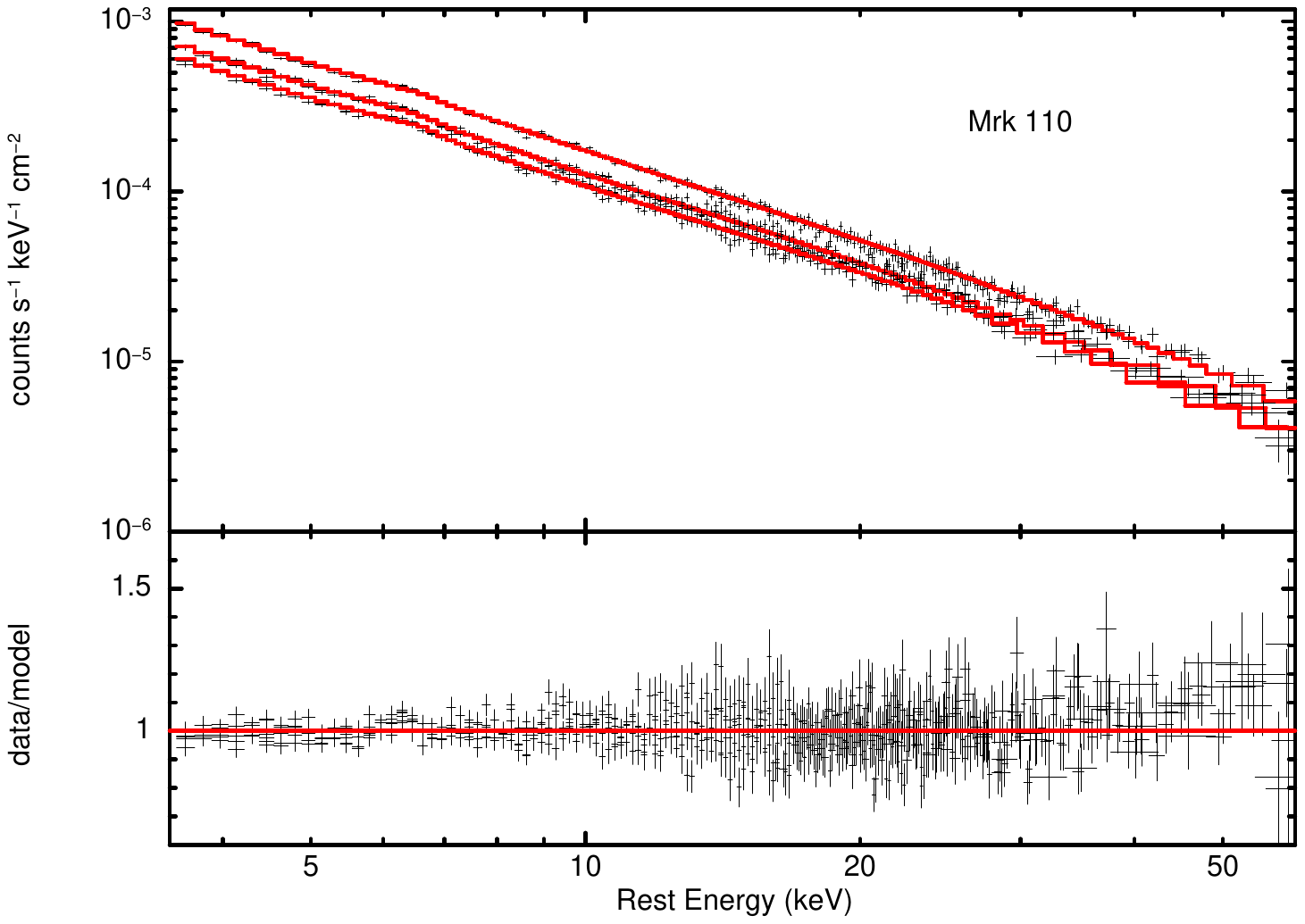}
\includegraphics[scale=0.35]{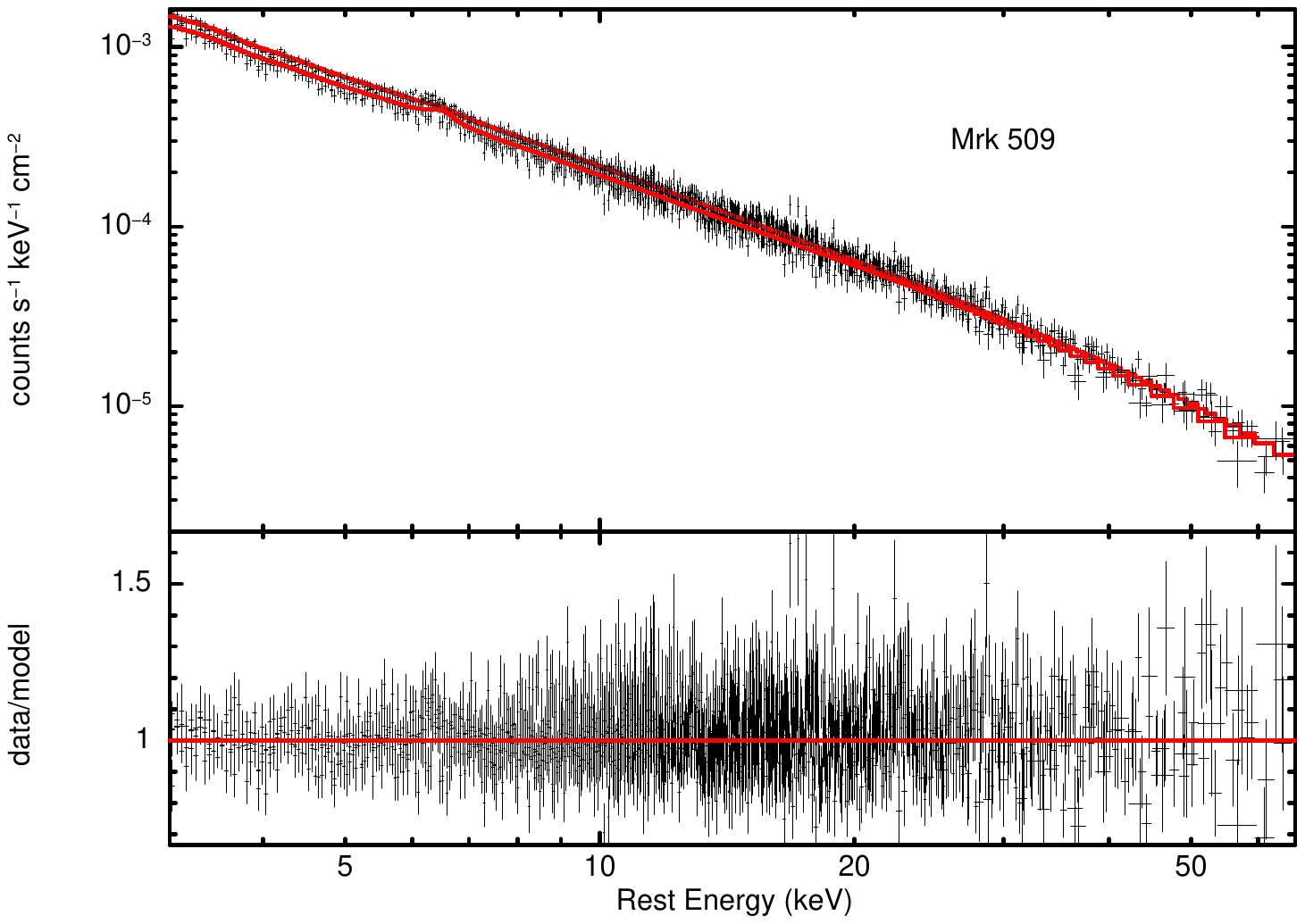}\\\includegraphics[scale=0.35]{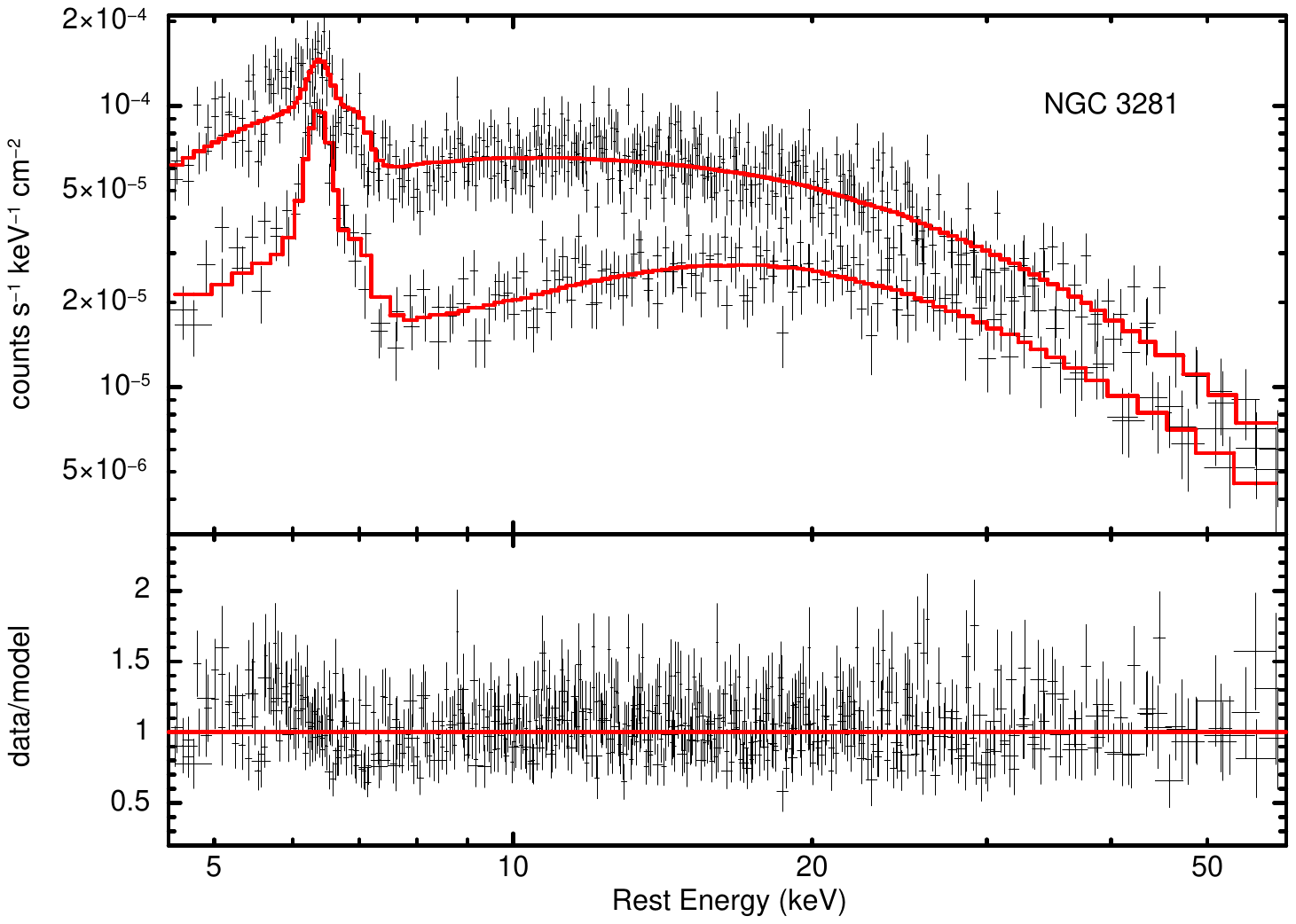}
\includegraphics[scale=0.35]{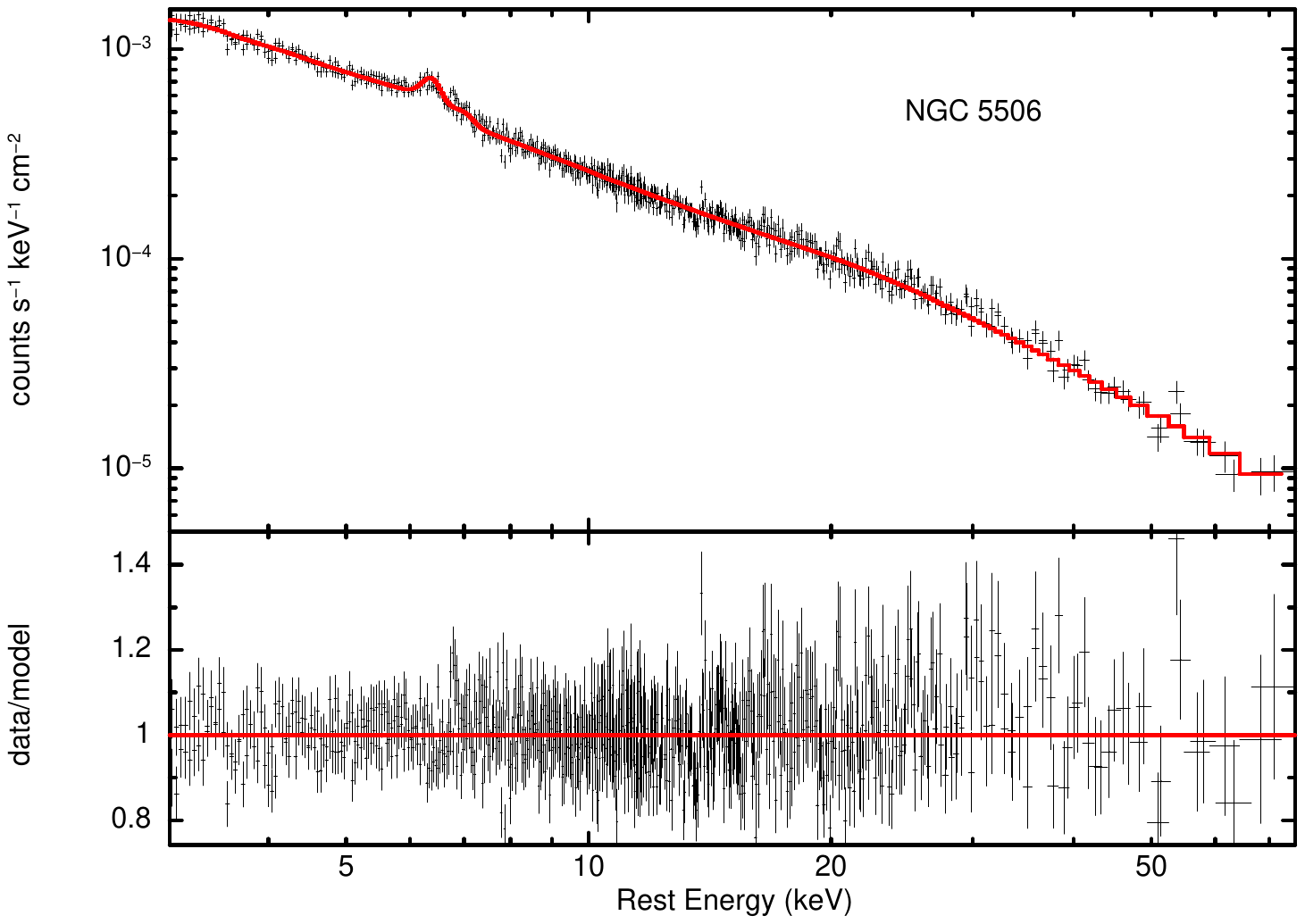}\\\includegraphics[scale=0.35]{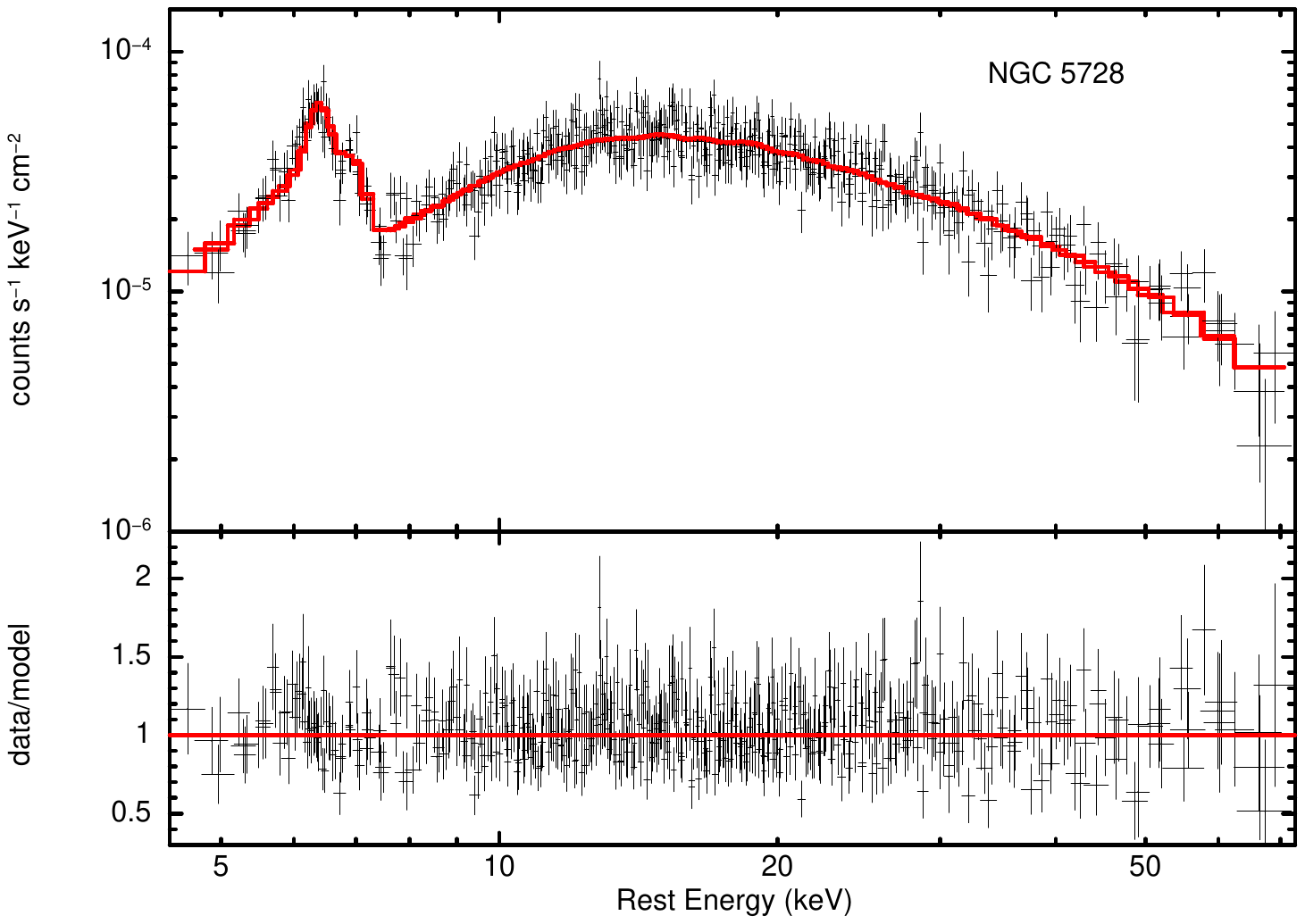}\includegraphics[scale=0.35]{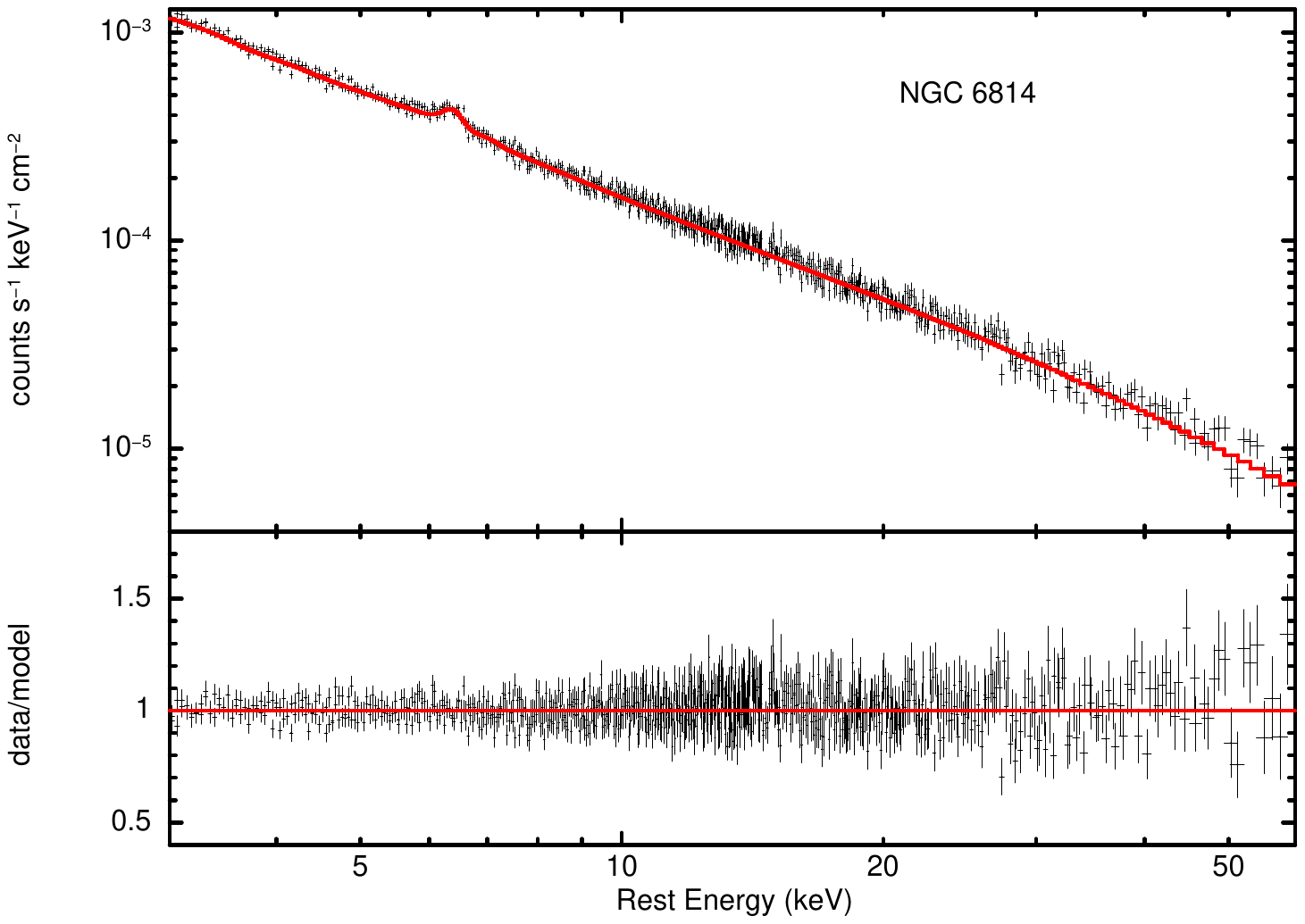}\\
\includegraphics[scale=0.35]{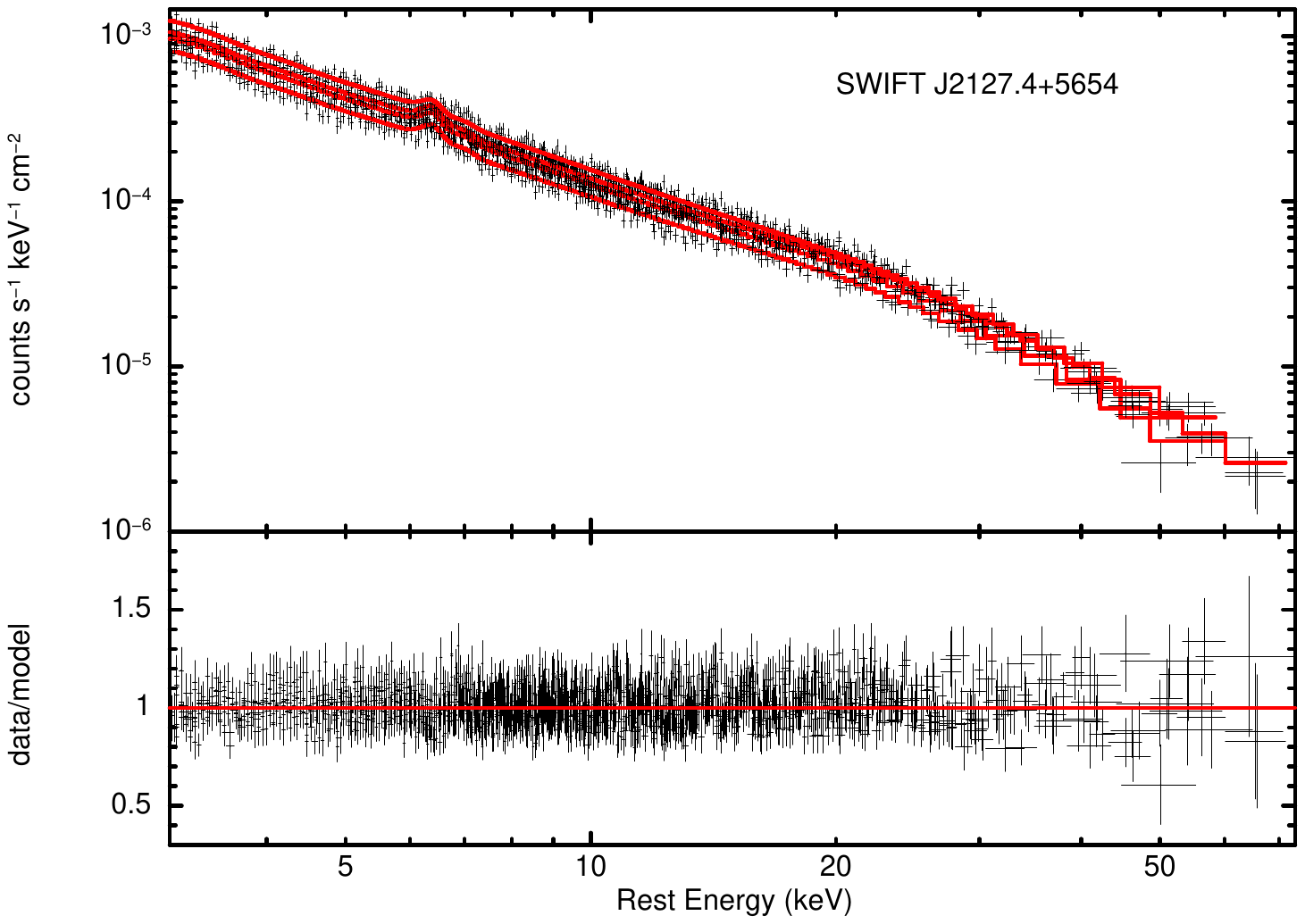}
\includegraphics[scale=0.35]{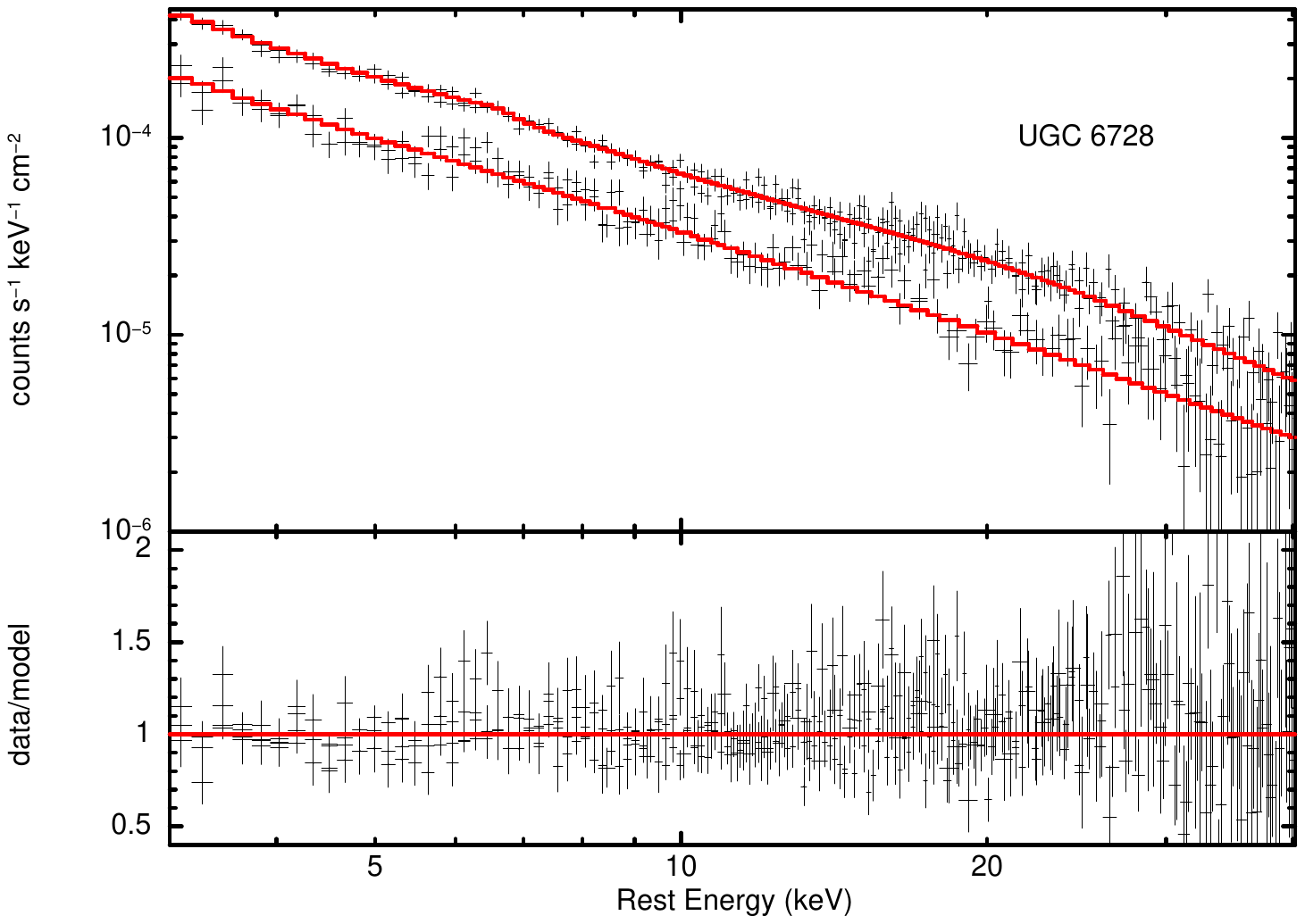}\\
\caption{\footnotesize Spectra and data-to-model ratios of the sources analyzed in this work. Multiple observation spectra are also shown together. FPMA and FPMB are plotted together.}
\label{fig:appB}
\end{longfigure}

\section{Light curves}
\label{sec:lcrepo}

\begin{longfigure}{c}
\includegraphics[scale=0.32]{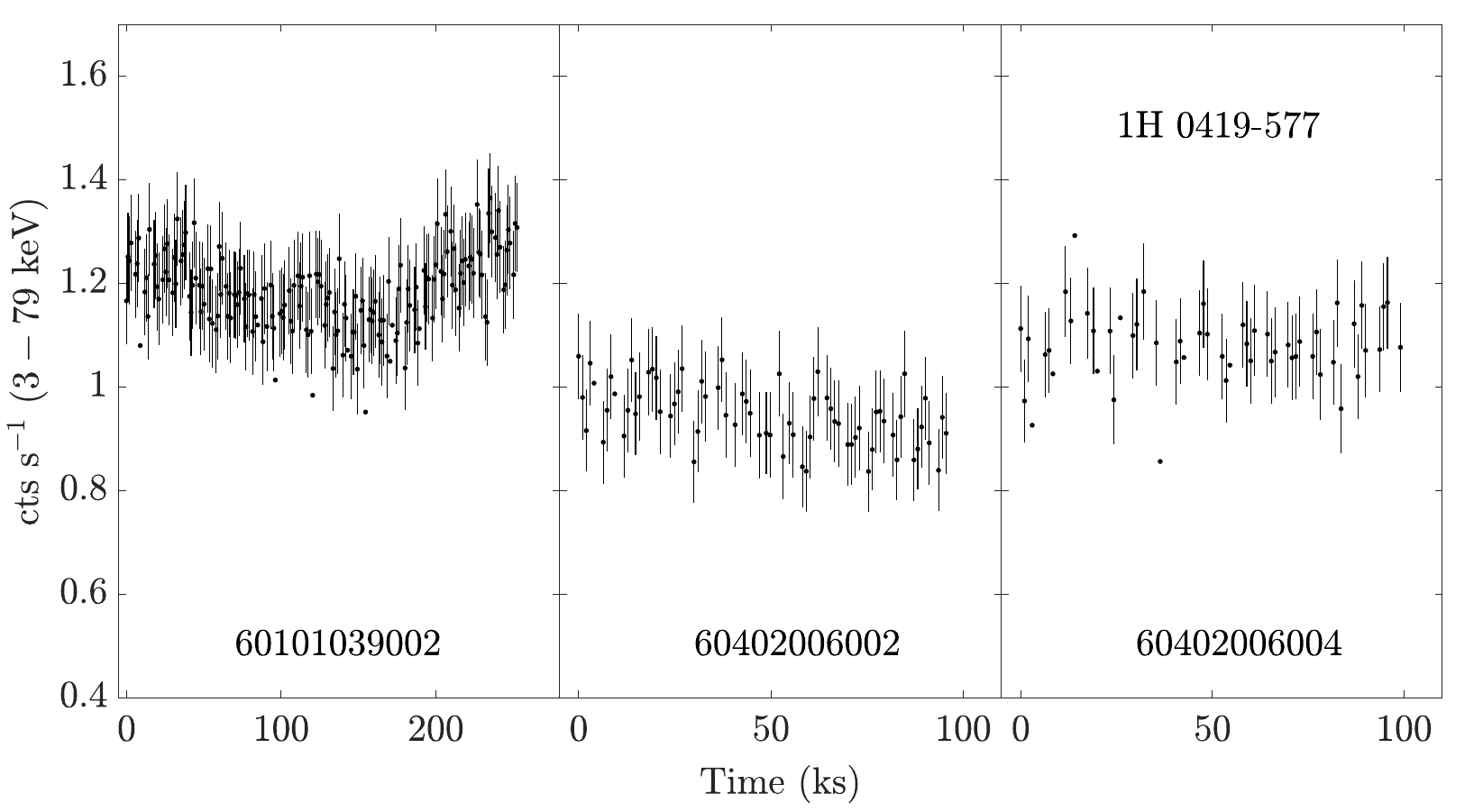}
\includegraphics[scale=0.32]{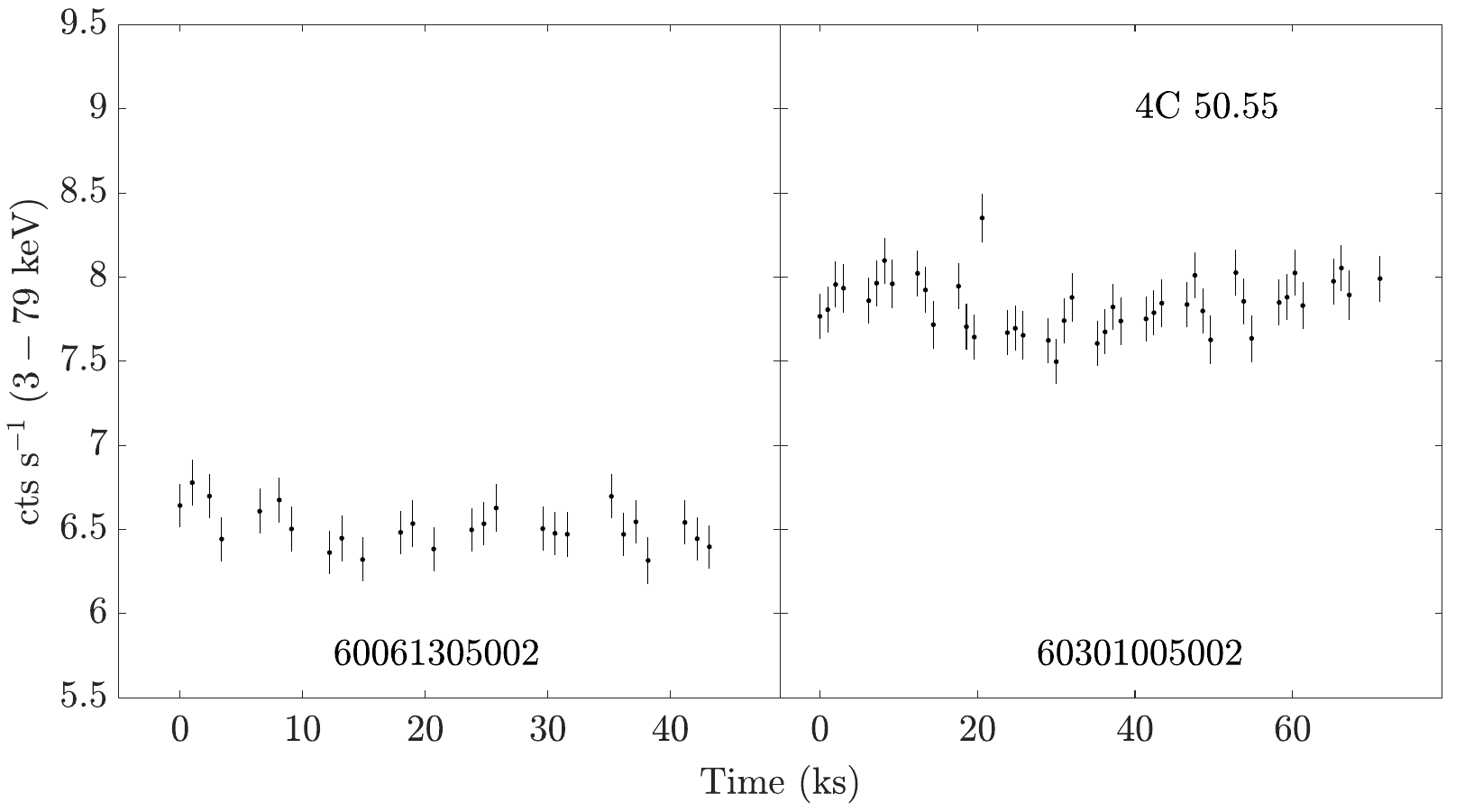}\\
\includegraphics[scale=0.32]{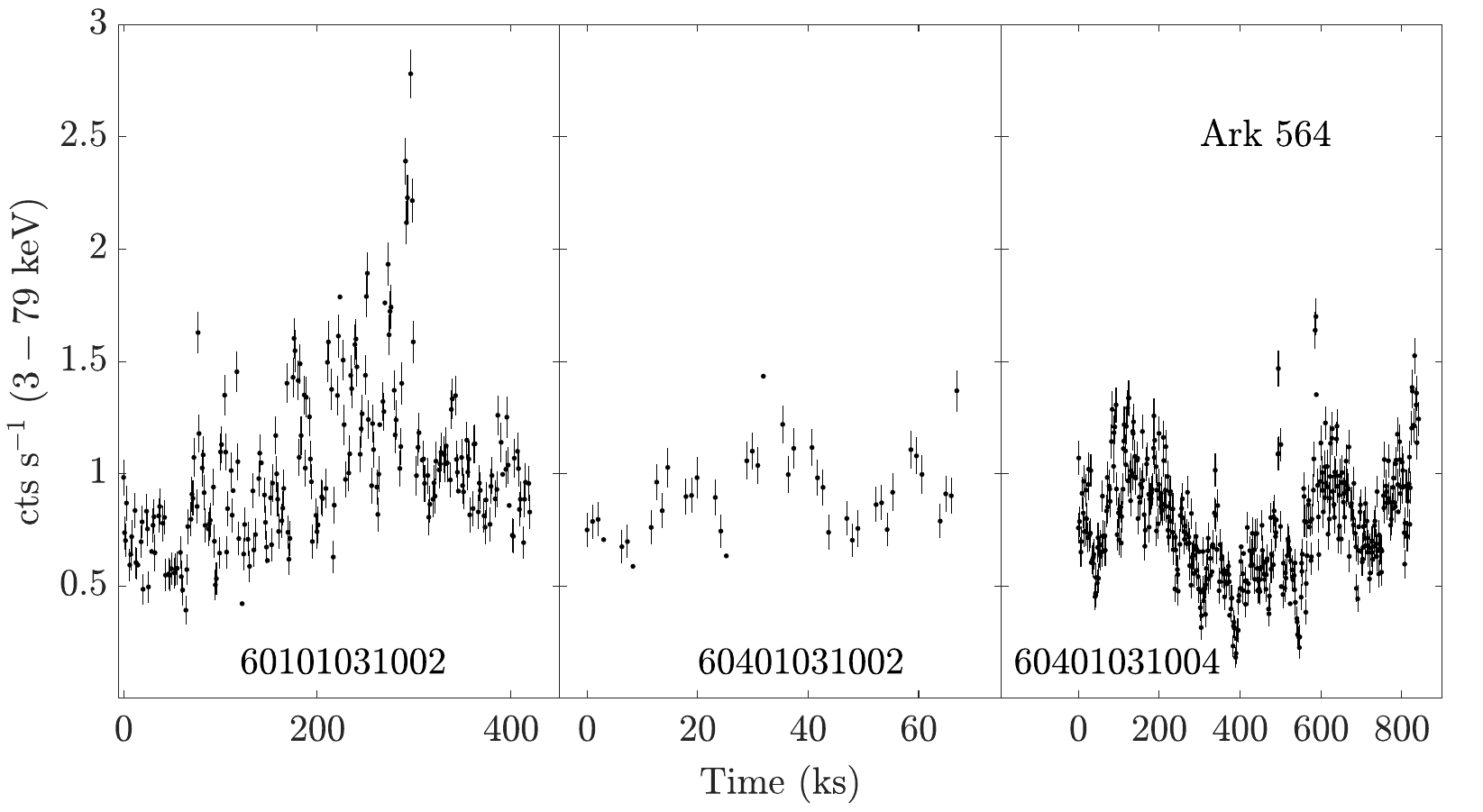}
\includegraphics[scale=0.32]{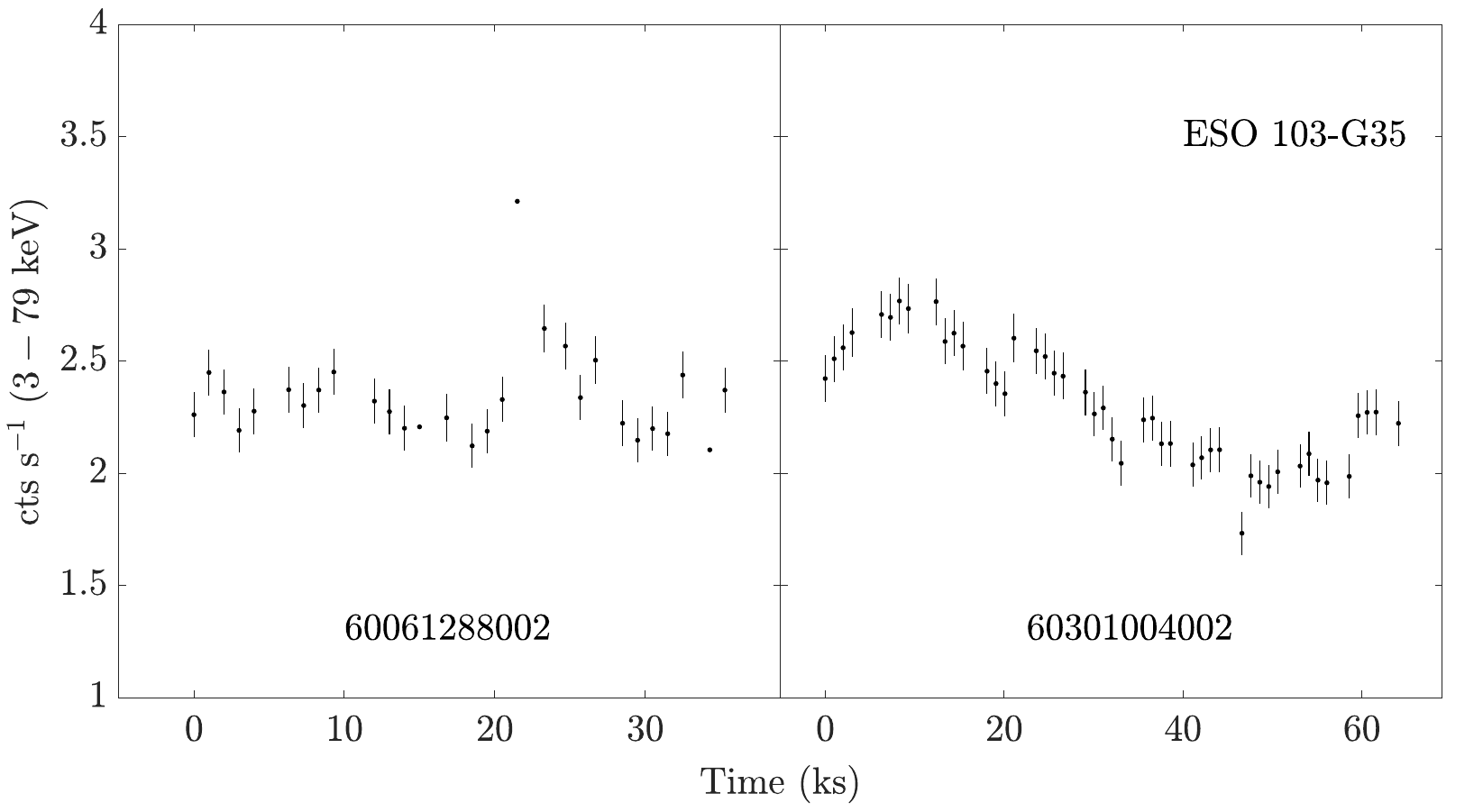}\\
\includegraphics[scale=0.32]{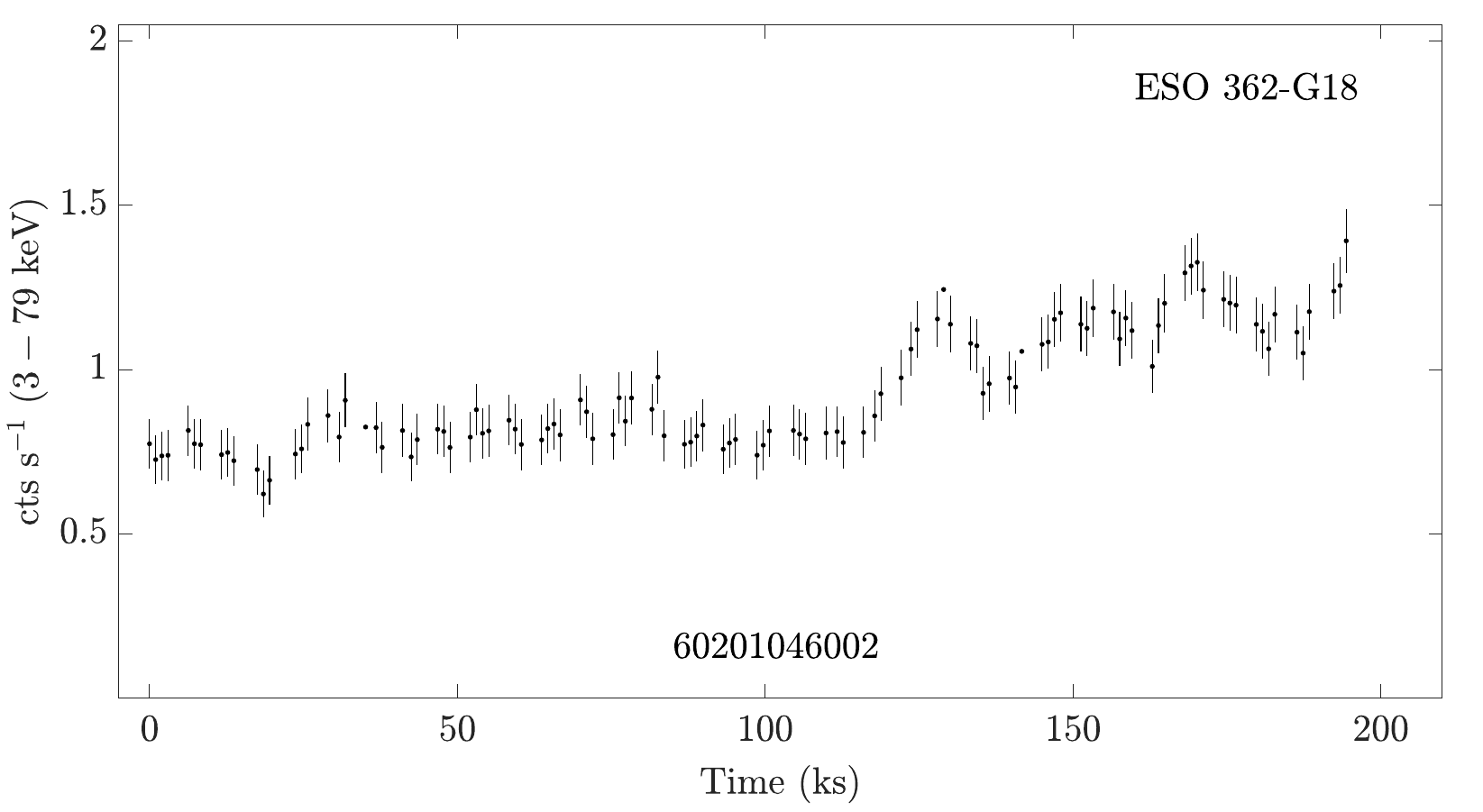}
\includegraphics[scale=0.32]{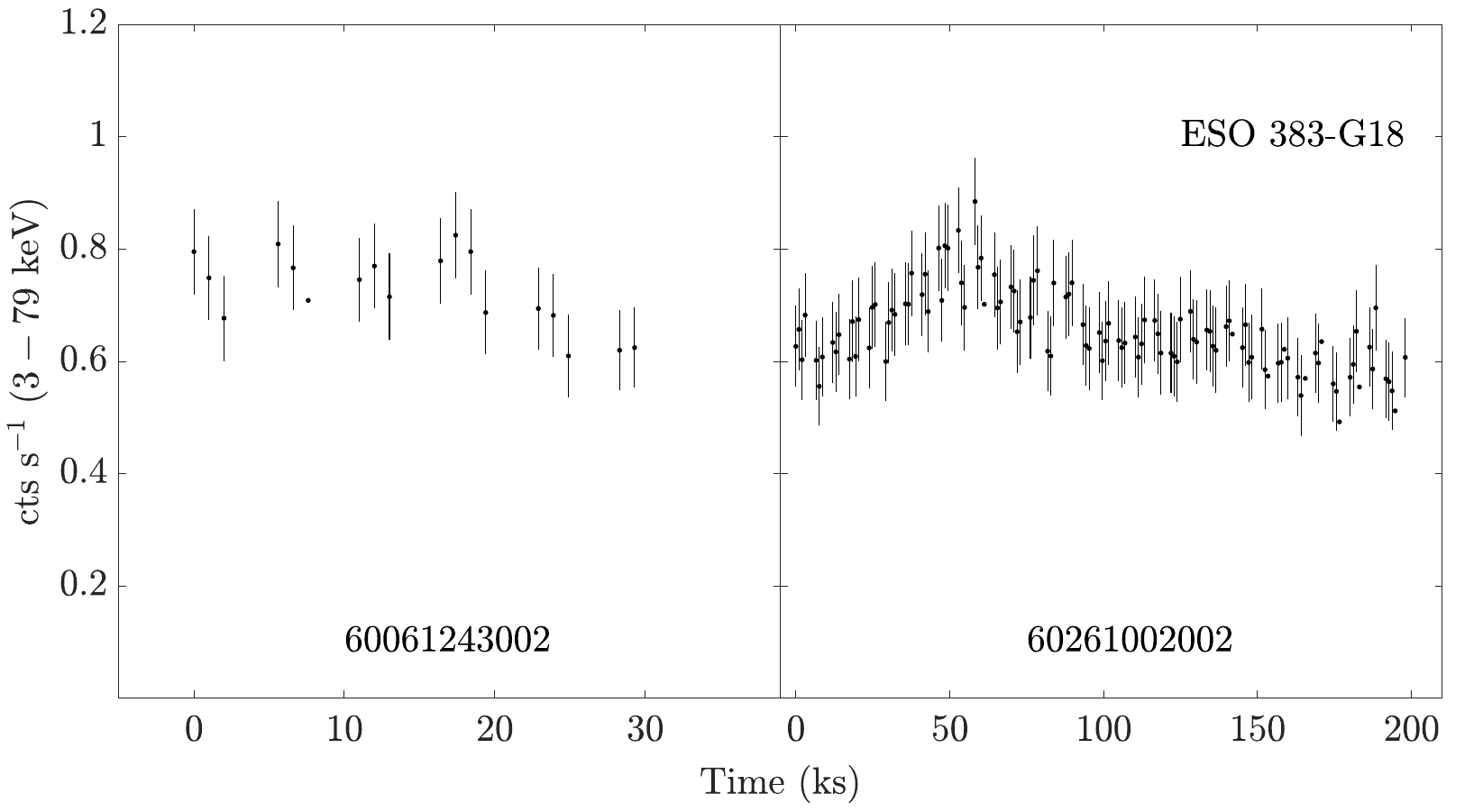}\\
\includegraphics[scale=0.32]{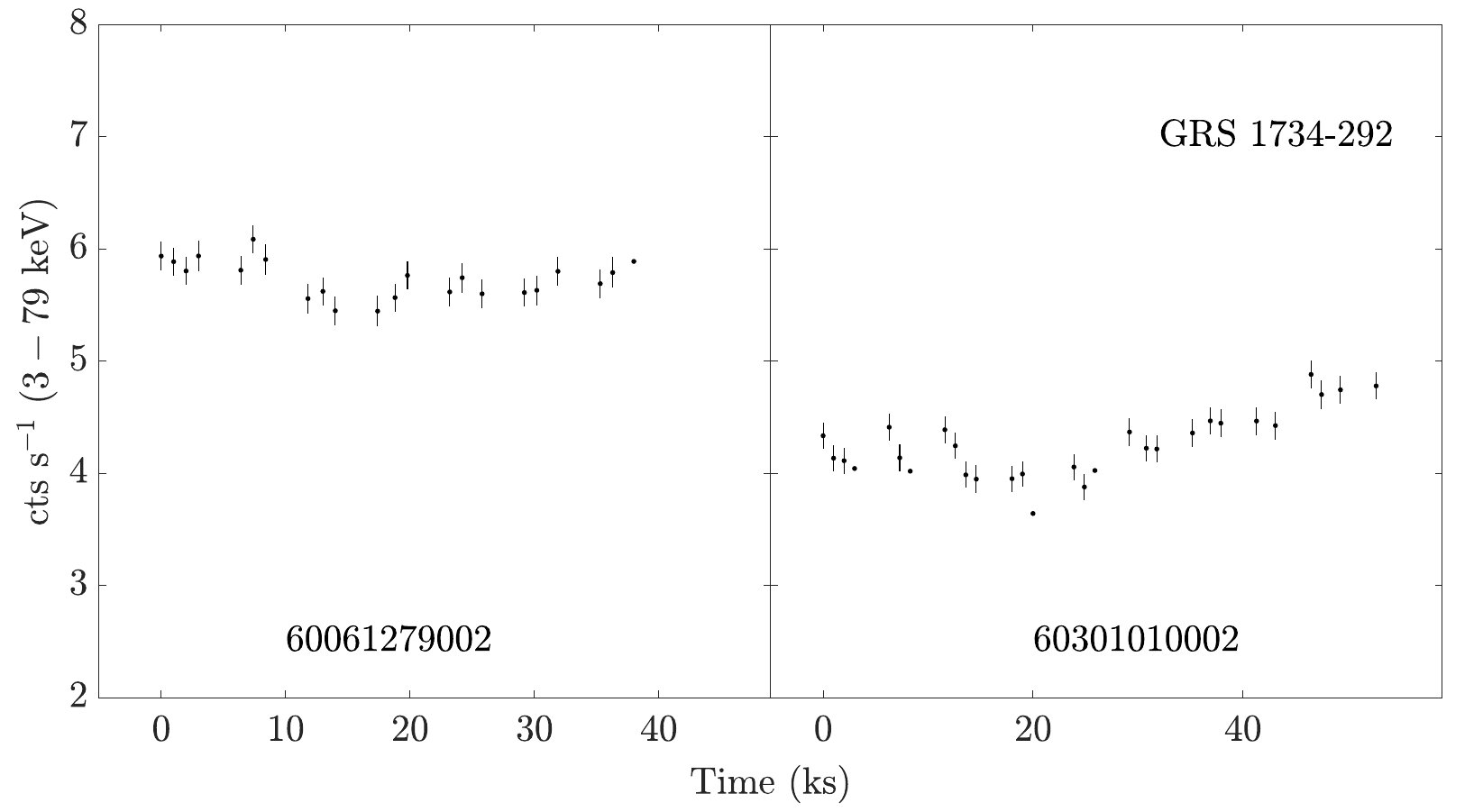}
\includegraphics[scale=0.32]{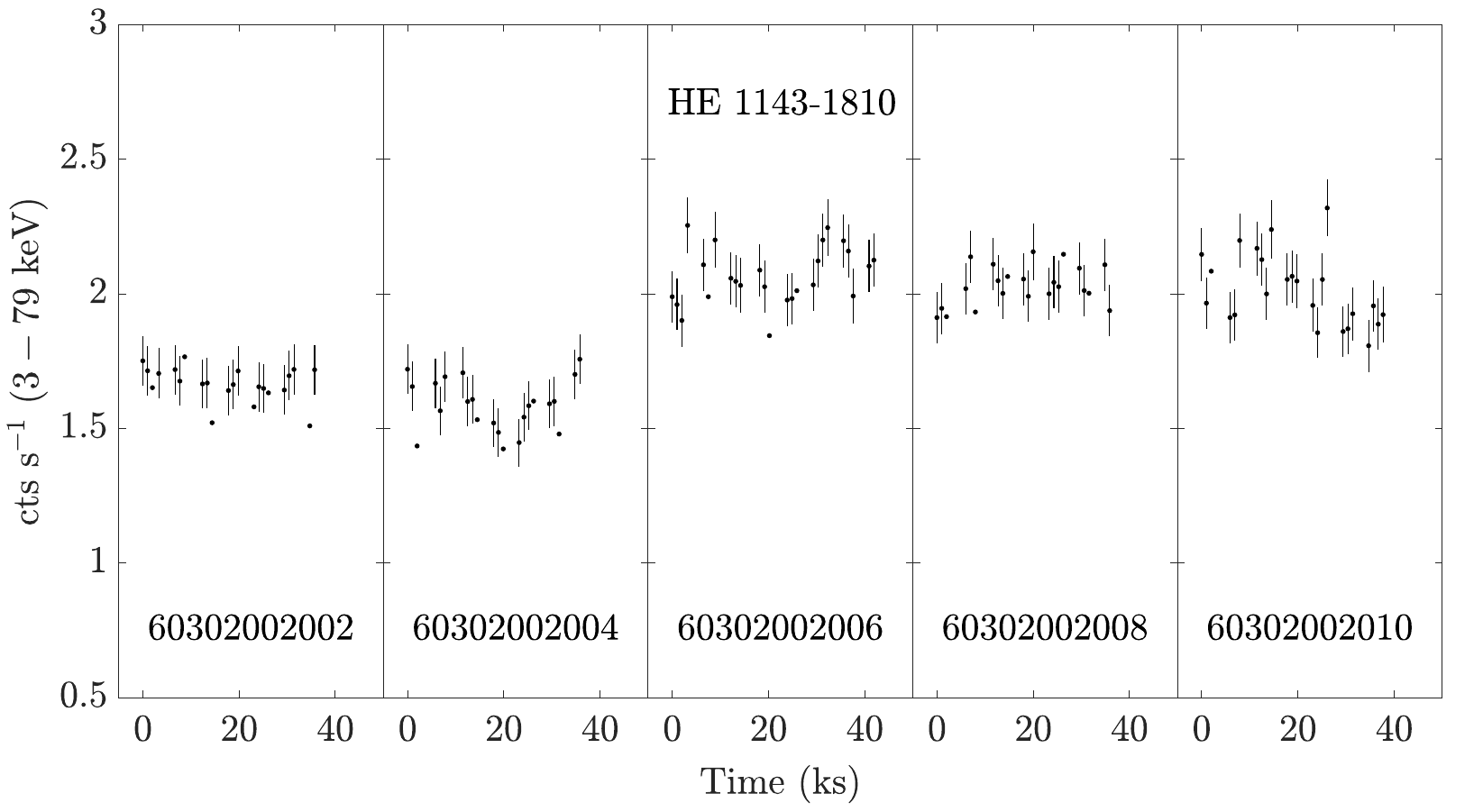}\\
\includegraphics[scale=0.32]{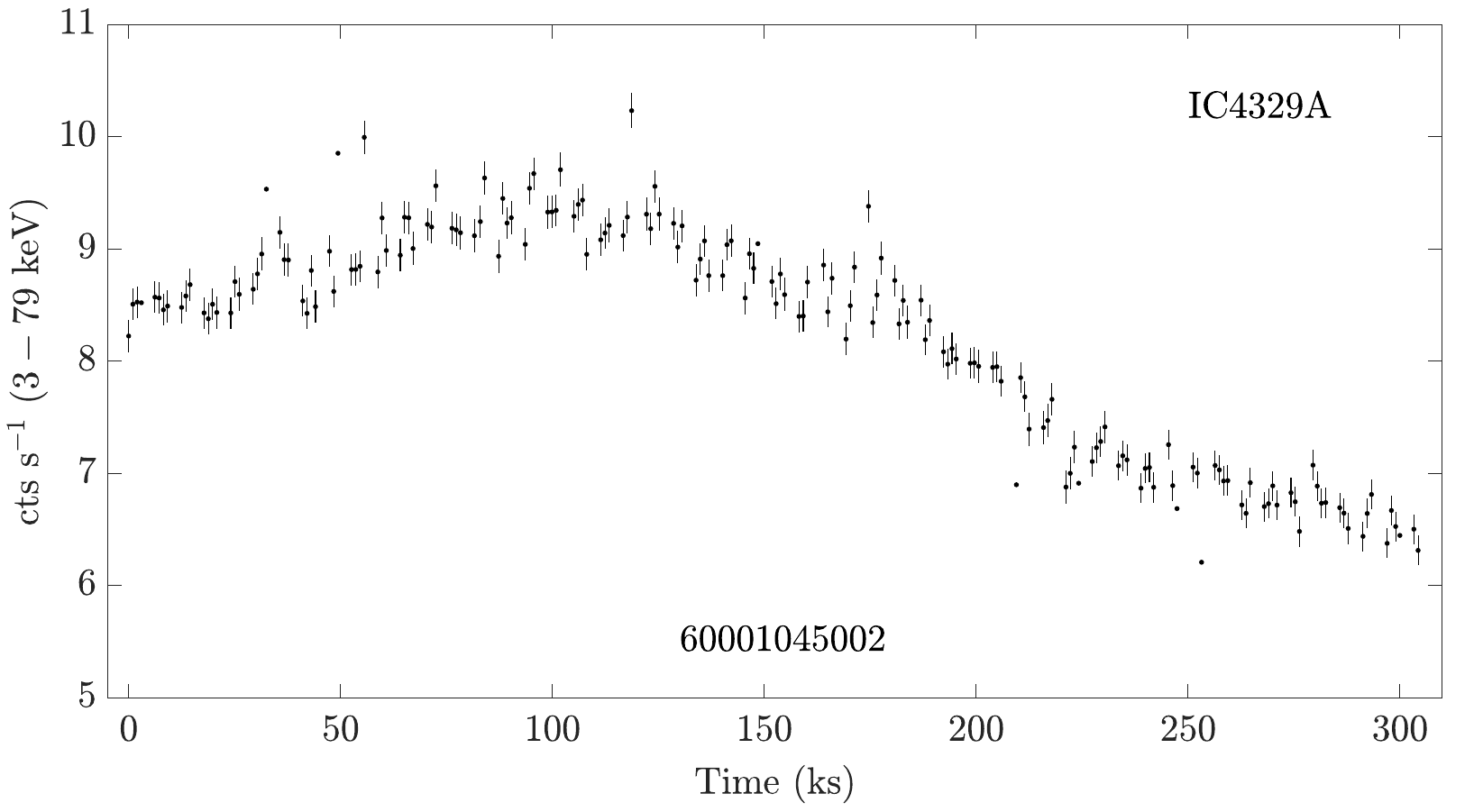}
\includegraphics[scale=0.32]{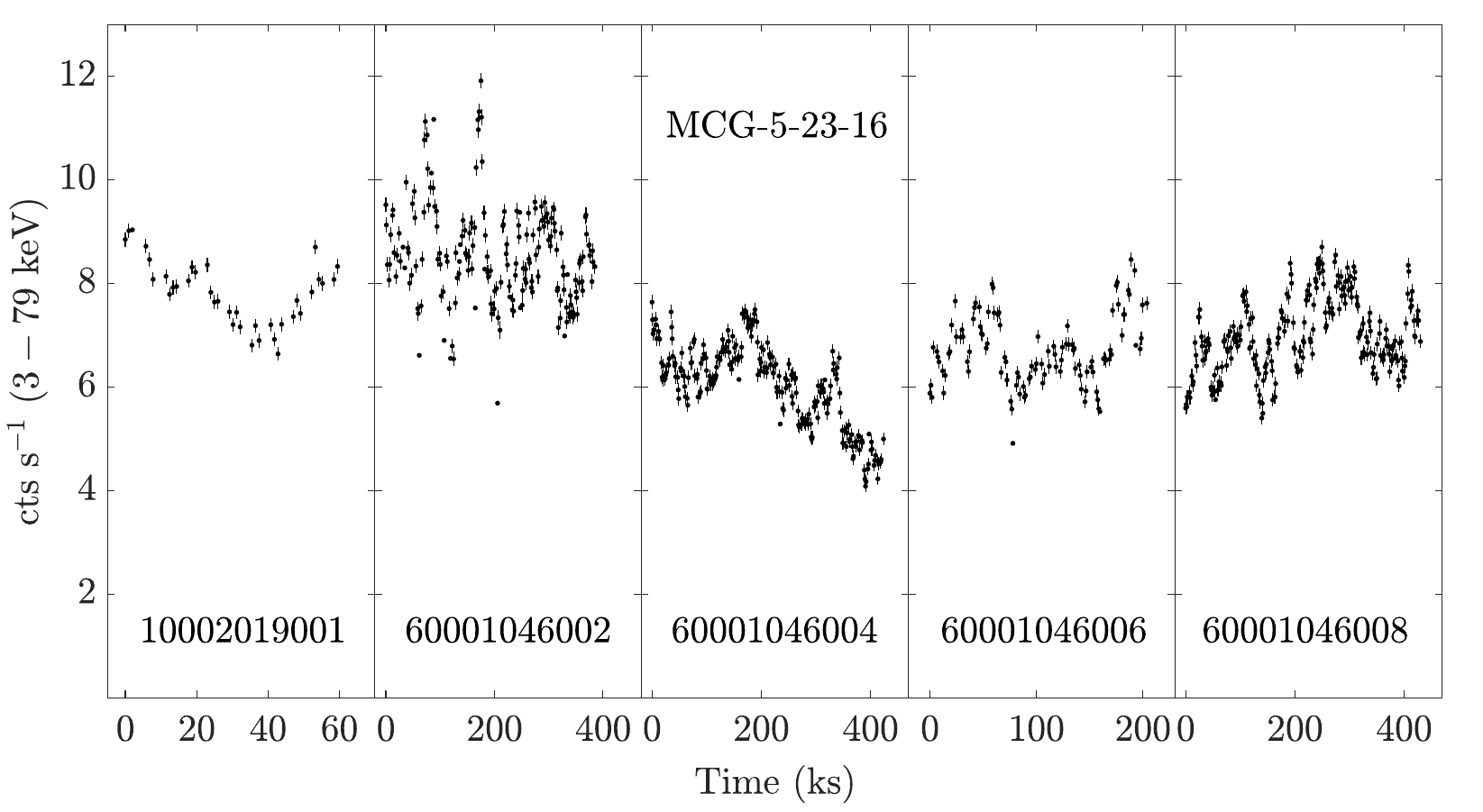}\\
\includegraphics[scale=0.32]{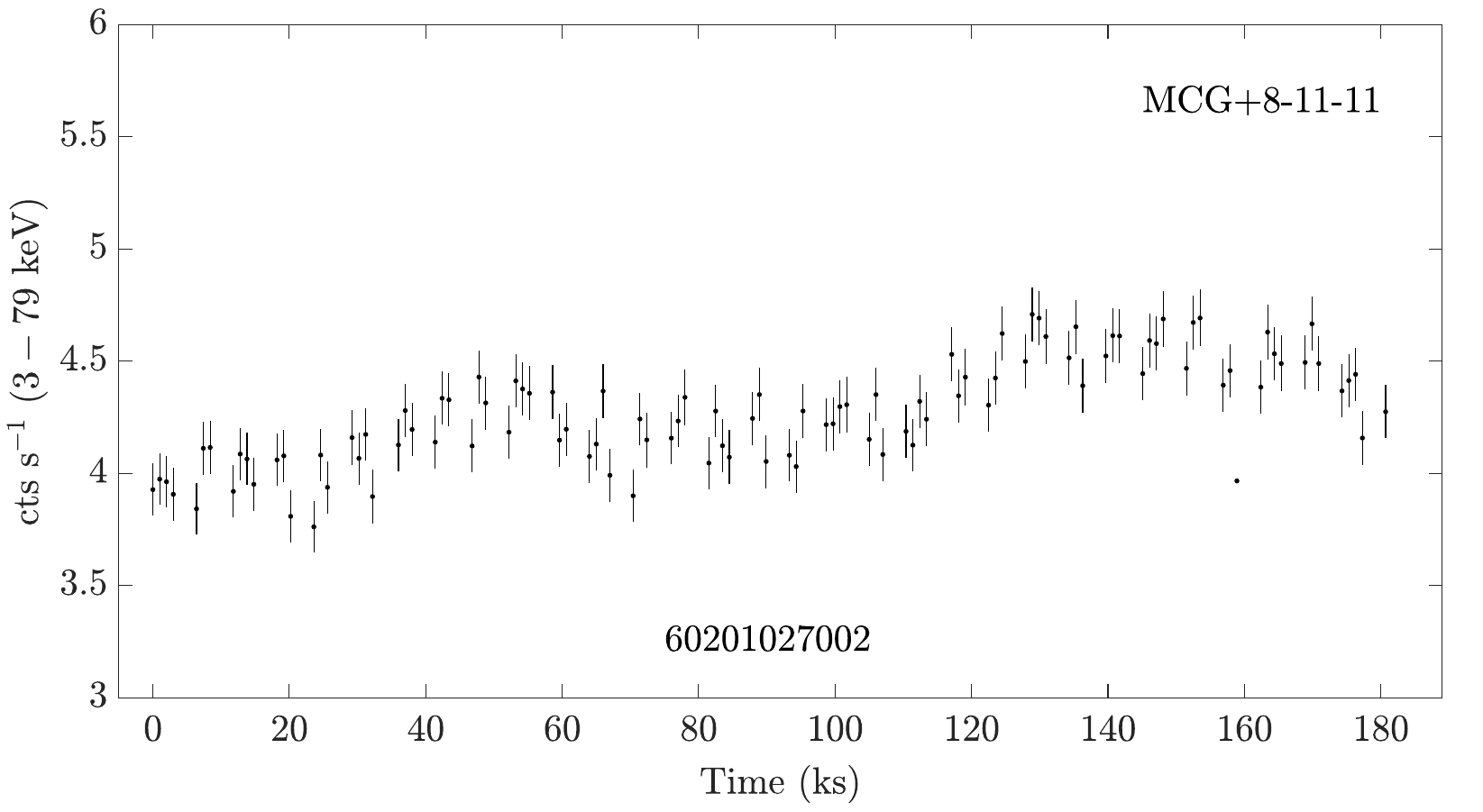}
\includegraphics[scale=0.32]{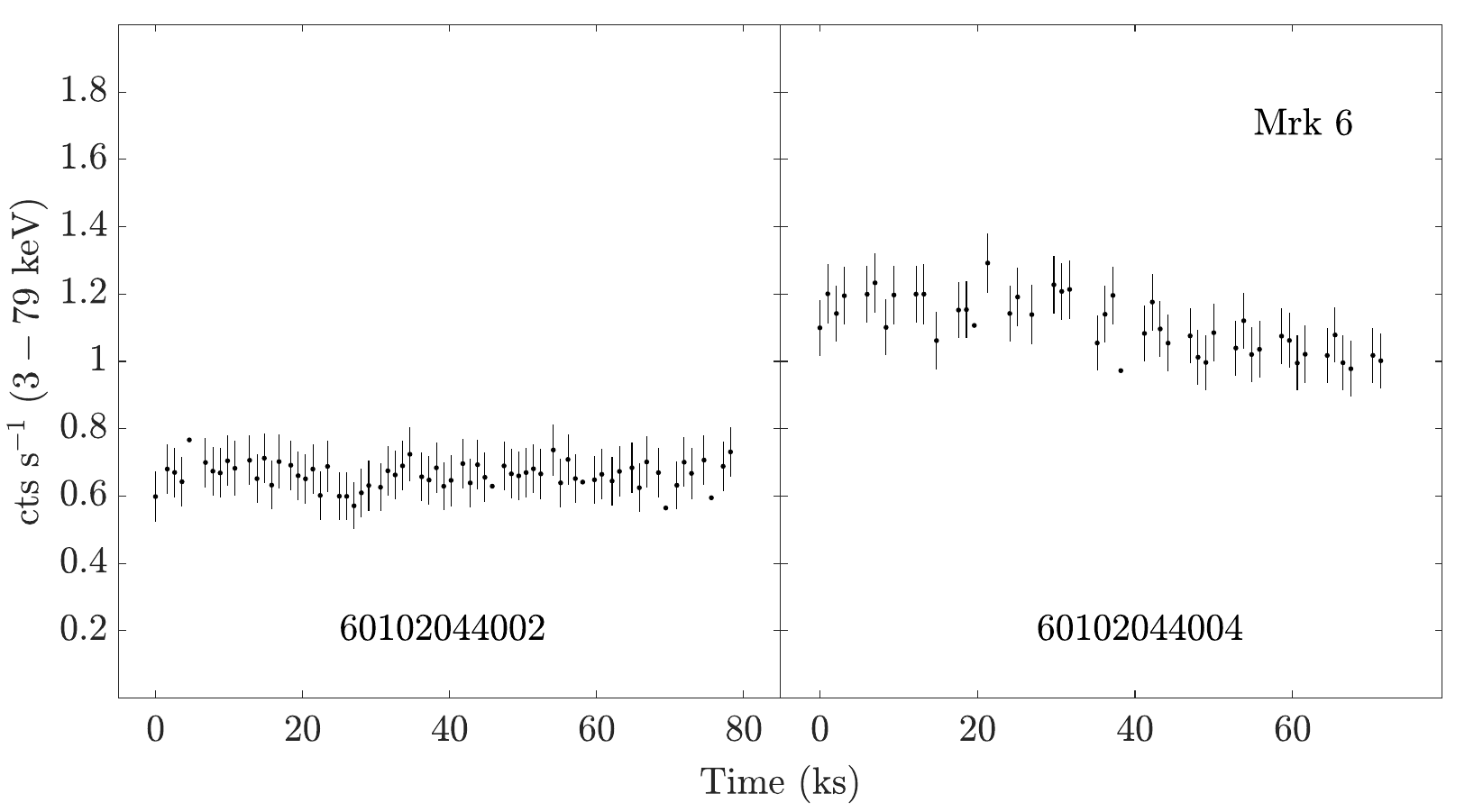}\\
\includegraphics[scale=0.32]{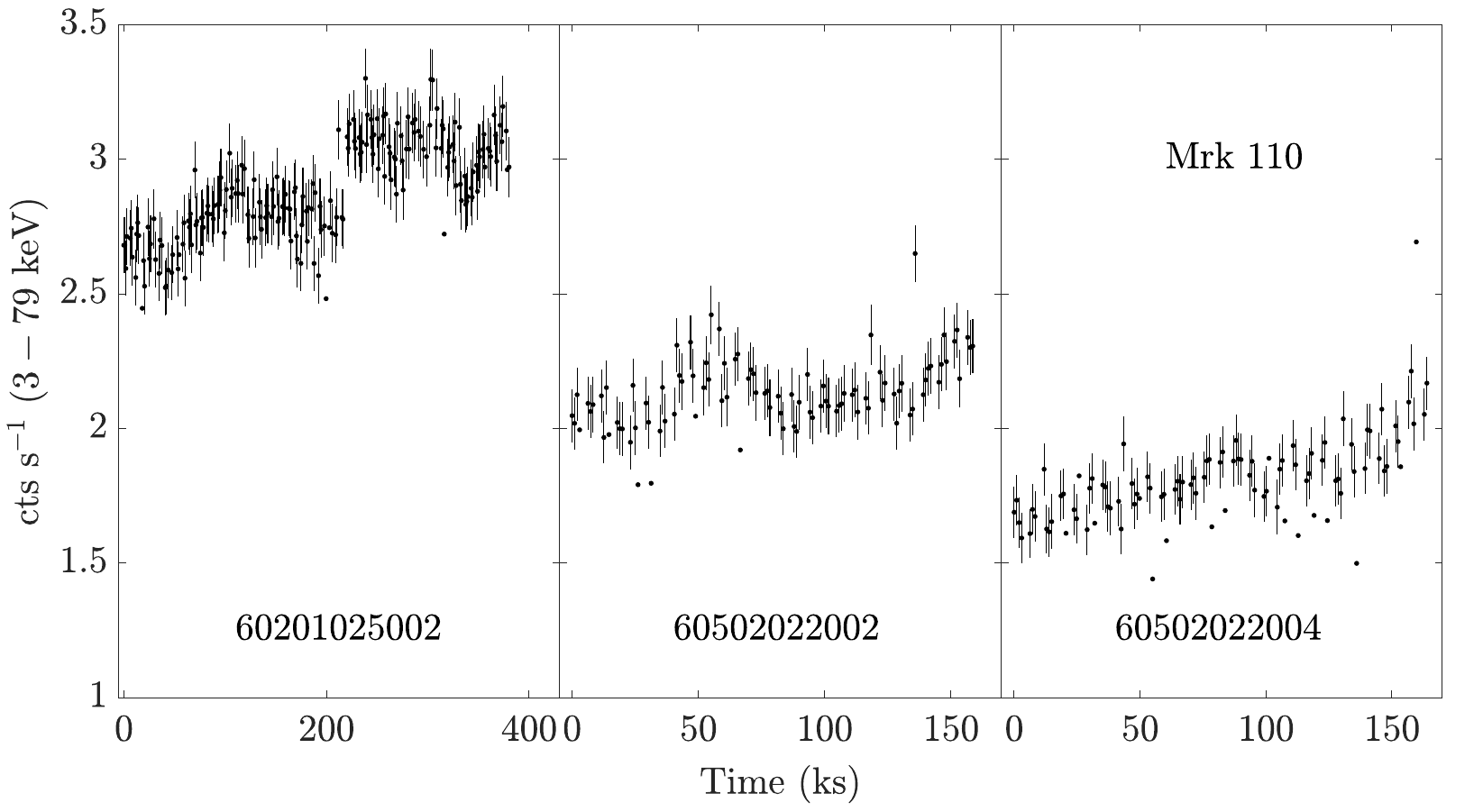}
\includegraphics[scale=0.32]{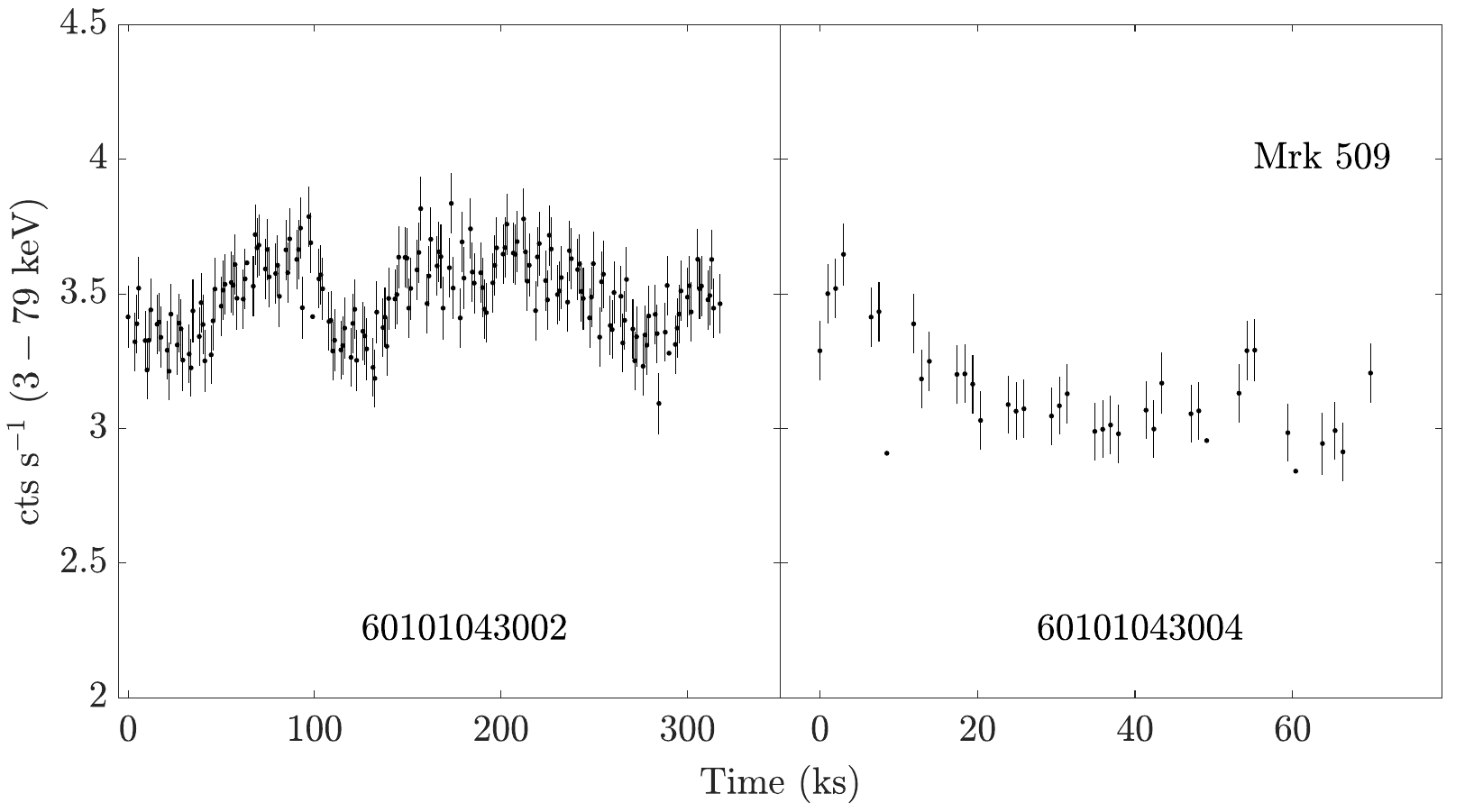}\\
\includegraphics[scale=0.32]{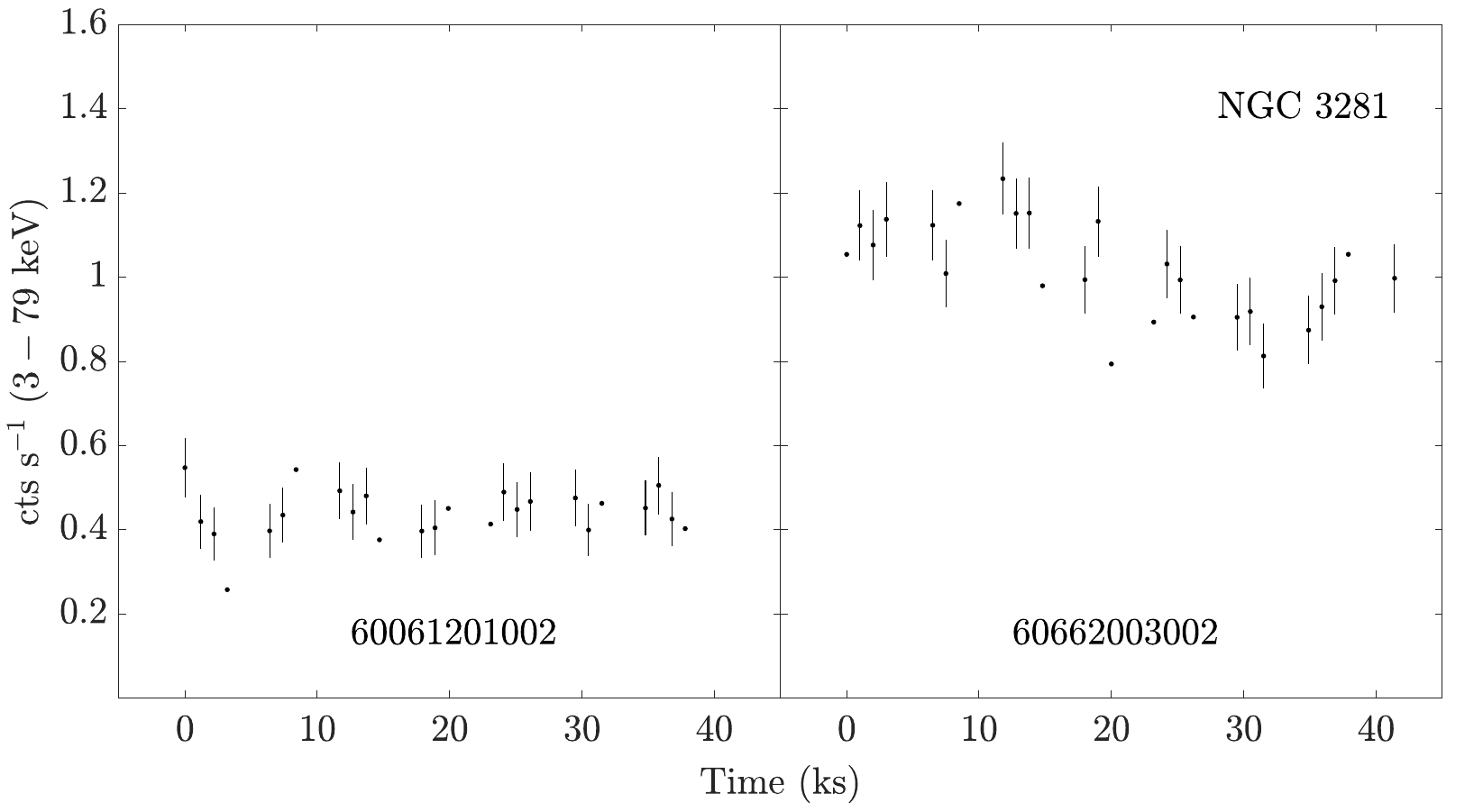}
\includegraphics[scale=0.32]{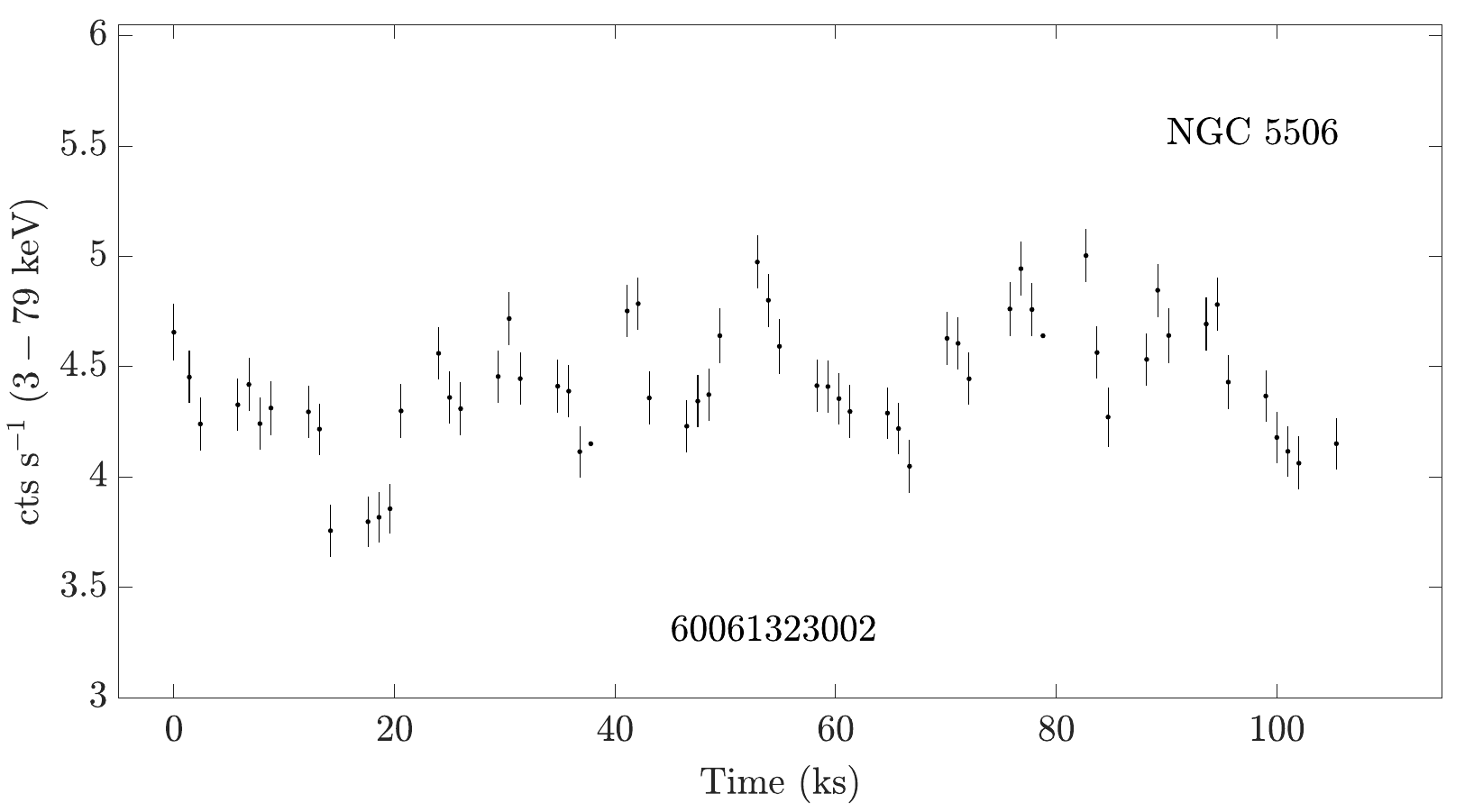}\\
\includegraphics[scale=0.32]{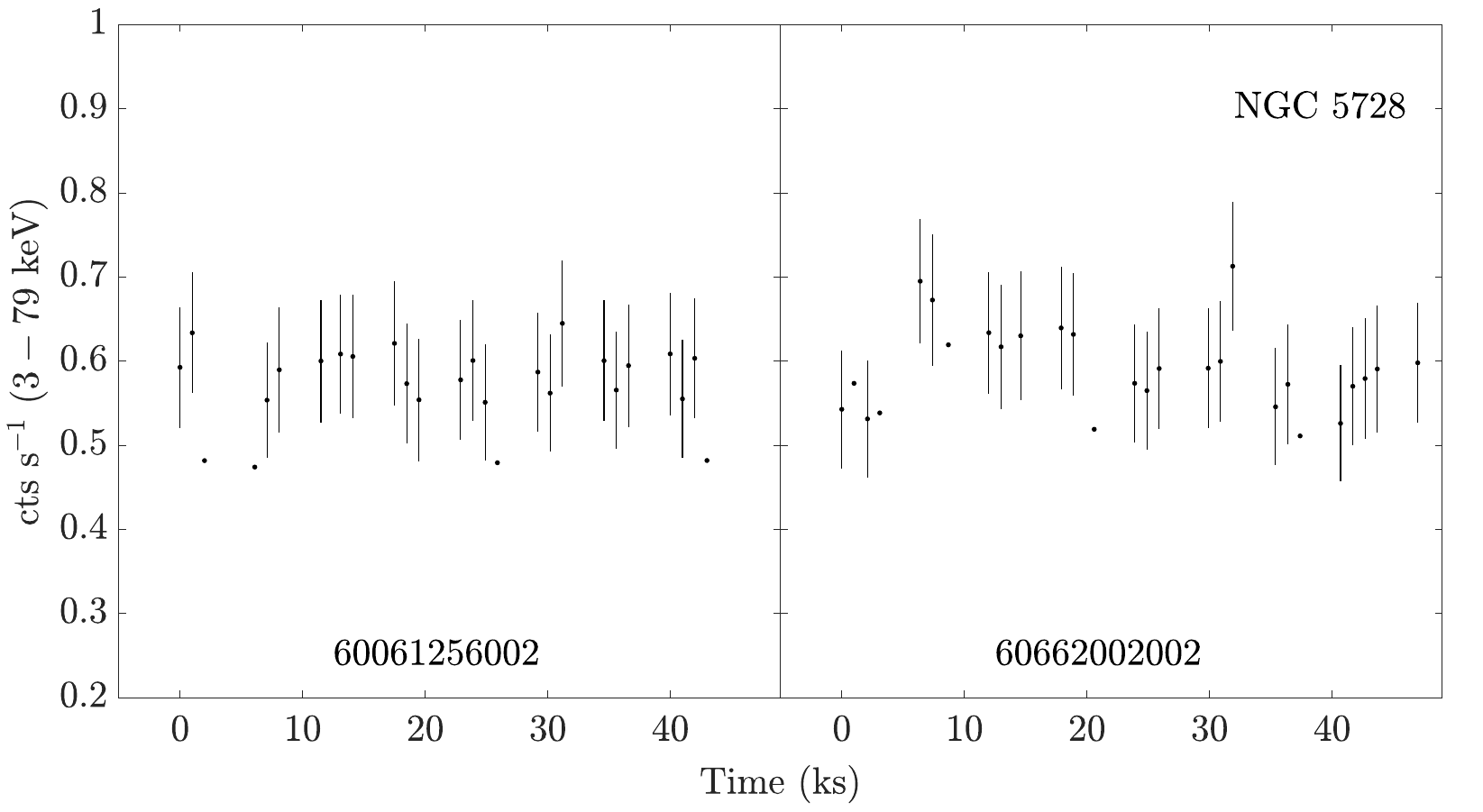}
\includegraphics[scale=0.32]{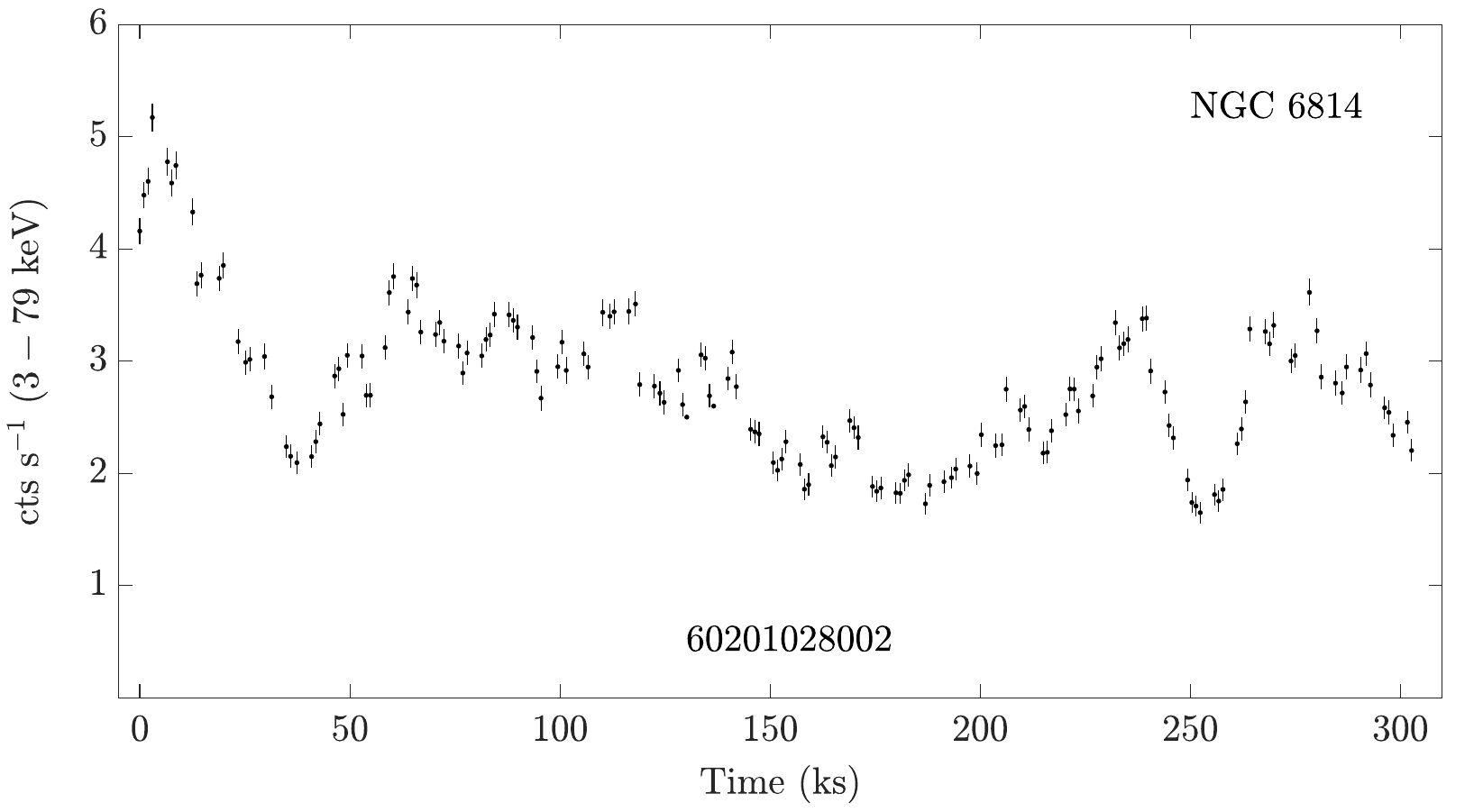}\\
\includegraphics[scale=0.32]{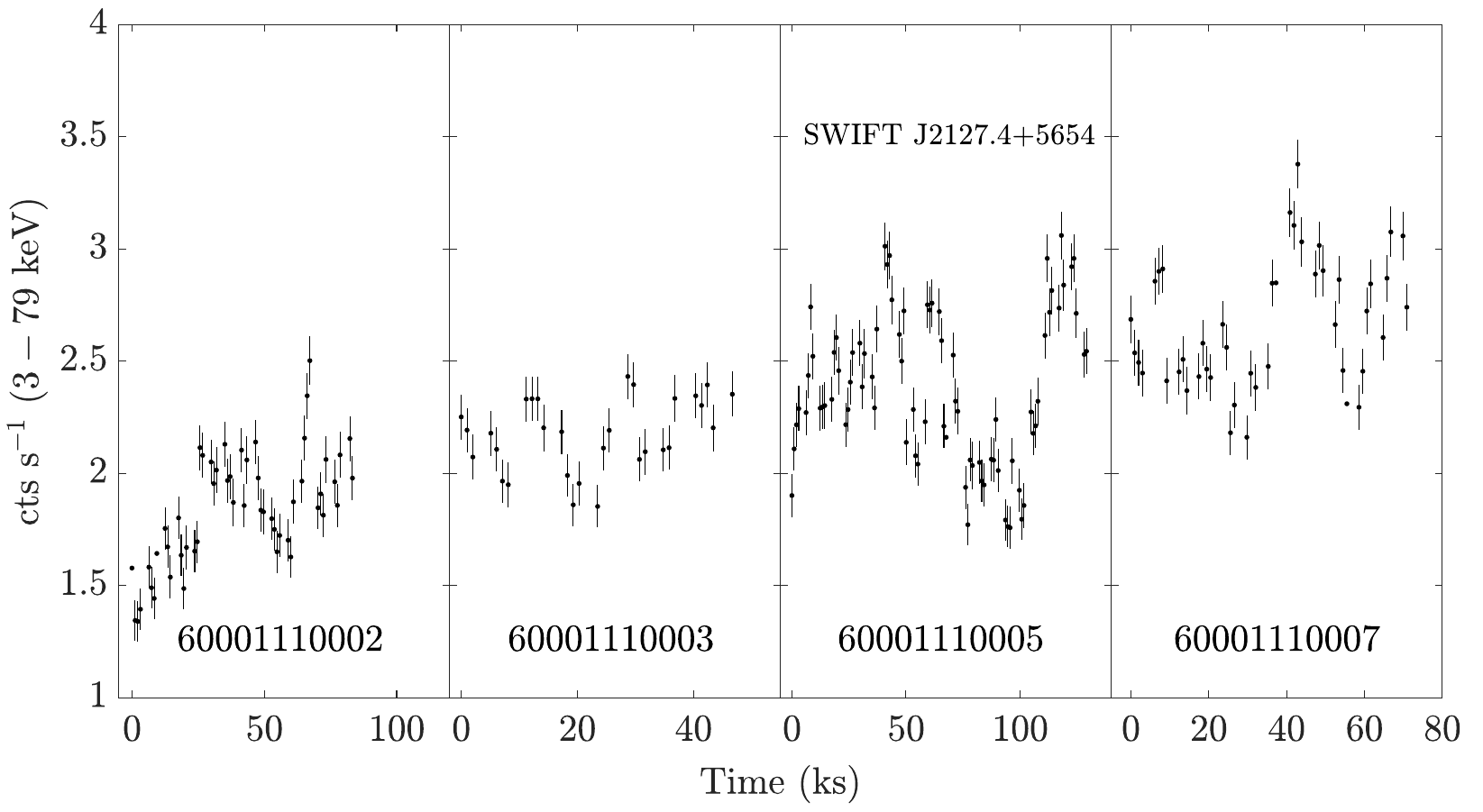}
\includegraphics[scale=0.32]{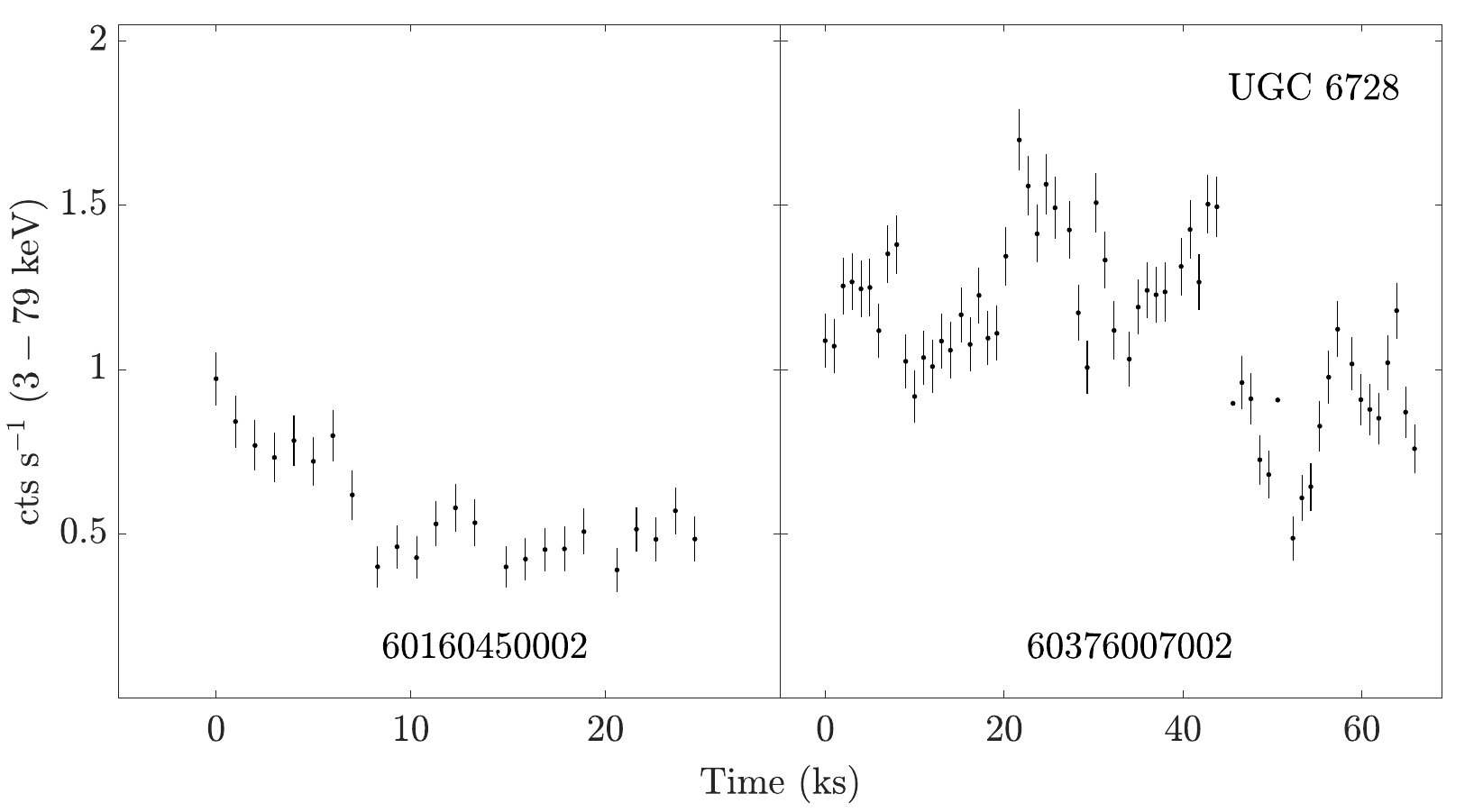}\\
\caption{\footnotesize Background-subtracted {\it NuSTAR} light curves. Each panel represents a source, while each sub-panel represents one epoch of each AGN. Source names and OBSIDs are printed in the figures. FPMA and FPMB light curves are combined together.}
\label{fig:appC}
\end{longfigure}





\end{appendix}

%
%

%
%
 




\end{document}